\documentclass[
footinbib,
a4paper,
aps,
prx,
superscriptaddress,
reprint,
twocolumn,
preprintnumbers,
nobalancelastpage,
floatfix,
10pt]{revtex4-2}

\usepackage{xcolor}
\usepackage{graphicx}
 	\graphicspath{ {Figures/} }
\usepackage{dcolumn}
\usepackage{tikz}
\usepackage{subfigure}
\usepackage{float}
\definecolor{cmblue}{rgb}{0.12156862745098039, 0.4666666666666667, 0.7058823529411765}
\definecolor{mygrey}{gray}{0.35}
\definecolor{myblue}{rgb}{0.2,0.2,0.8}
\definecolor{mygreen}{rgb}{0.2,0.8,0.5}
\definecolor{myzard}{cmyk}{0,0,0.05,0}
\definecolor{mywhite}{rgb}{1,1,1}
\definecolor{myred}{rgb}{1,0.,0.3}

\usepackage[utf8]{inputenc}
\usepackage{dsfont}
\usepackage[normalem]{ulem}
\usepackage{bbold}
\usepackage{soul}
\usepackage{slashed}

\usepackage{amsmath, amsthm, amsfonts, amssymb,amstext}
\usepackage{mathptmx}
\usepackage{bm}
\usepackage{xfrac}
\usepackage{mathrsfs}
\usepackage{latexsym}
\usepackage{array}

\usepackage{makecell}

 \def\ee{\mathord{\mathrm e}}
 
 \def\ii{\mathord{\mathrm i}}

\def\half{\textstyle\frac{1}{2}}

\newcommand{\diff}{\text{d}}
\def\beq{\begin{equation}}
\def\eeq{\end{equation}}
\def\barray{\begin{eqnarray}}
\def\earray{\end{eqnarray}}

\usepackage{physics}
\usepackage{braket}
\usepackage{tikz}

\usepackage[
colorlinks=true,
citecolor=myblue,
linkcolor=myred]
{hyperref}
\usepackage{cleveref} 
\usepackage{verbatim}

\usetikzlibrary{backgrounds,fit,decorations.pathreplacing,calc}
\usetikzlibrary{quantikz2}
\renewcommand{\ii}{{\mathrm i}}
\renewcommand{\ee}{{\mathrm e}}

\begin{document}
\author{P. Viñas}
\email{pablo.vinnas@estudiante.uam.es}
\affiliation{Instituto de F\'isica Te\'orica UAM-CSIC, Universidad Aut\'onoma de Madrid, Cantoblanco, 28049, Madrid, Spain}

\author{A. Bermudez}
\affiliation{Instituto de F\'isica Te\'orica UAM-CSIC, Universidad Aut\'onoma de Madrid, Cantoblanco, 28049, Madrid, Spain} 


\title{Context-aware gate set tomography: Improving the self-consistent characterization \\of trapped-ion universal gate sets by leveraging non-Markovianity}

\begin{abstract}
To progress in the characterization of noise for current   quantum computers, gate set tomography (GST) has emerged as a self-consistent tomographic protocol that can accurately estimate the complete set of noisy quantum gates, state preparations, and measurements. In its original incarnation, GST improves the estimation precision by applying the  gates sequentially,  provided that the noise makes them a set of fixed completely-positive and trace preserving (CPTP) maps independent of the  history of previous gates in the sequence. This ‘Markovian' assumption is sometimes in conflict with experimental evidence, as there might be time-correlated noise leading to non-Markovian dynamics or, alternatively,   slow drifts and cumulative calibration errors that lead to context dependence, such that the CP-divisible maps composed during a sequence actually change with  the circuit depth.
In this work, we address this issue for trapped-ion devices with phonon-mediated two-qubit gates. By a detailed microscopic modeling of high-fidelity light-shift gates, we tailor GST to  capture the  main source of context dependence: motional degrees of freedom. Rather than invalidating GST, we show that  context dependence can be incorporated in the parametrization of the gate set, allowing us to reduce the sampling cost of GST. Our results identify a promising research avenue that might be applicable to other platforms where microscopic modeling can be incorporated: the development of a context-aware GST.

\end{abstract}

\maketitle

\setcounter{tocdepth}{2}
\begingroup
\hypersetup{linkcolor=black}
\tableofcontents
\endgroup

\section{\bf {Introduction}}

Quantum information processors (QIPs) promise a speedup of certain computations relative to their classical counterparts~\cite{montanaro2016quantum}. To achieve this aim, however, several challenges still have to be overcome. One of these obstacles relates to the development of efficient methods for the characterization of noise. In the near term, detailed understanding and description of undesired interactions is central to advancing error mitigation techniques~\cite{PhysRevLett.119.180509,PRXQuantum.2.040326,PhysRevA.106.012423,RevModPhys.95.045005}, which might allow quantum computers to enter a utility stage prior to  fault tolerance (FT)~\cite{kim2023evidence}. In the long term, quantum error correction (QEC)~\cite{PhysRevA.54.1098,  PhysRevA.54.1098, PhysRevLett.77.793, RevModPhys.87.307} offers a clear direction towards FT.
When error rates of physical gates are small enough, one can obtain an exponential reduction  on logical errors by a polynomial increase of the physical qubit number~\cite{Nielsen_Chuang_2010}. From this perspective, 
a precise characterization of noise is crucial, not only to understand how to   reduce  error rates, but also to adapt and optimise the QEC protocols.

With this objective, the community working in quantum characterization, verification, and validation (QCVV) has developed a wide variety of methods~\cite{Eisert2020,PRXQuantum.2.010201,Gebhart2023}. Some approaches provide unified metrics to estimate average device performance, as exemplified by randomized benchmarking \cite{Emerson_2005,Knill_2008, Dankert_2009, PhysRevLett.106.180504,
PhysRevA.85.042311}. In contrast,  tomographic techniques~\cite{gill2004invitation, Banaszek} reconstruct the actual faulty operations in  noisy QIPs. While the  amount of information provided by the latter  makes them much more appealing, there is a considerable associated experimental cost, as epitomized by gate set tomography (GST)~\cite{Nielsen2021gatesettomography, 
greenbaum2015introduction,
blumekohout2013robust,
PhysRevA.89.052109,Merkel_2013}. What makes GST fundamentally different from the rest of techniques of the tomographic family~\cite{doi:10.1080/09500349708231894,PhysRevLett.78.390,PhysRevA.63.020101,PhysRevA.63.054104,PhysRevA.68.012305} is that it dispenses with the assumption of error-free state preparation and measurement (SPAM). Because of this, it is often referred to in the literature as a \textit{self-consistent} protocol. This is only possible by the simultaneous estimation of a reference initial state of the system $\rho_0$, a positive operator-valued measure (POVM) $M_0$ with $m$ possible outcomes~\cite{watrous2018theory}, and an informationally-complete (IC) set of operations $\big\{G_i:\,i\in \mathbb{G}\,\big\}$. The union of all these components is known as the gate set
\beq
\label{gate_set}
\mathcal{G}=\left\{\rho_0,\big\{G_i:\,i\in \mathbb{G}\,\big\},M_0\right\},
\eeq
and the goal of GST is to estimate it both accurately and with a high precision using  circuit-based measurement data.

Succinctly, GST provides an estimate of the gate set under study by fitting experimentally-measured frequencies to a model describing the gates. Optionally, some gate sequences might be chained several times to produce frequencies that enhance the precision of the estimation. This is the `long-sequence' version of GST \cite{Nielsen2021gatesettomography}.

{
Having outlined the main idea behind GST, we introduce now its two main drawbacks, which we address with novel approaches in this work. The more apparent one is the large number of experiments it requires. A generic two-qubit gate set description takes  $N_{\mathrm{param}}\sim10^3$ parameters \cite{ostrove2023nearminimalgatesettomography}. As a consequence, the number of different circuits required to run a given experimental GST design is $N_{\mathrm{GST}}\sim10^4$, with some variations depending on the circuit depth employed. This amount can substantially increase when crosstalk is modeled. Even with $N_{\mathrm{samples}}=10^3$ samples per circuit, which might already be too conservative given the minute errors of current high-fidelity QIPs one aims at estimating, one obtains a total of $N_{\mathrm{shots}}\sim10^7$ executions. Considering the gate speed and readout times of some of these QIPs, such as trapped-ion devices~\cite{bruzewicz2019trapped}, the total execution time of two-qubit GST is often impractical for a real time characterization. The second issue which restricts GST is more fundamental and platform-agnostic: it is unable to characterize context-dependent noise \cite{PhysRevX.9.021045}. GST relies entirely on the assumption that a noisy gate does not change with the number of times it is applied, nor with the specific previous history of operations along the circuit. When deviations from this assumption become significant, the  GST protocol  fails to provide reliable estimates. These violations of the constant-map behavior are often referred to in the GST literature as evidence of ``non-Markovianity'' 
\cite{Nielsen2021gatesettomography}. Because laboratory experiments across different platforms often suffer from Markovian deviations \cite{rebentrost2009optimal, wallman2016robust, rebentrost2009optimal, jing2018decoherence, piltz2014trapped}, the overhead required to run GST is not generally justified. This  highlights the importance of developing characterization techniques compatible with non-Markovianity. 

In this manuscript, we show that one can go beyond this limitation with a specific setup in mind: trapped-ion QIPs, focusing  on a geometric entangling gate known as \textit{light-shift} (LS) gate~\cite{leibfried2003experimental}. As we extensively discussed in our previous work for single-qubit GST~\cite{vinas2025microscopic}, a microscopically-motivated parametrization of GST can greatly reduce the number of parameters required to describe a gate set. This directly translates into a substantial reduction of the sampling complexity needed to achieve a desired GST precision. In the present  work, we use our microscopic parametrization to include noise in the LS gate, extending GST to the fully universal gate set that has demonstrated the highest  fidelities to date \cite{PhysRevLett.117.060504,PhysRevA.103.022427, PhysRevA.103.012603,PhysRevLett.127.130505, comment}. 

\begin{figure}
  \centering
  \includegraphics[width=1.02\columnwidth]{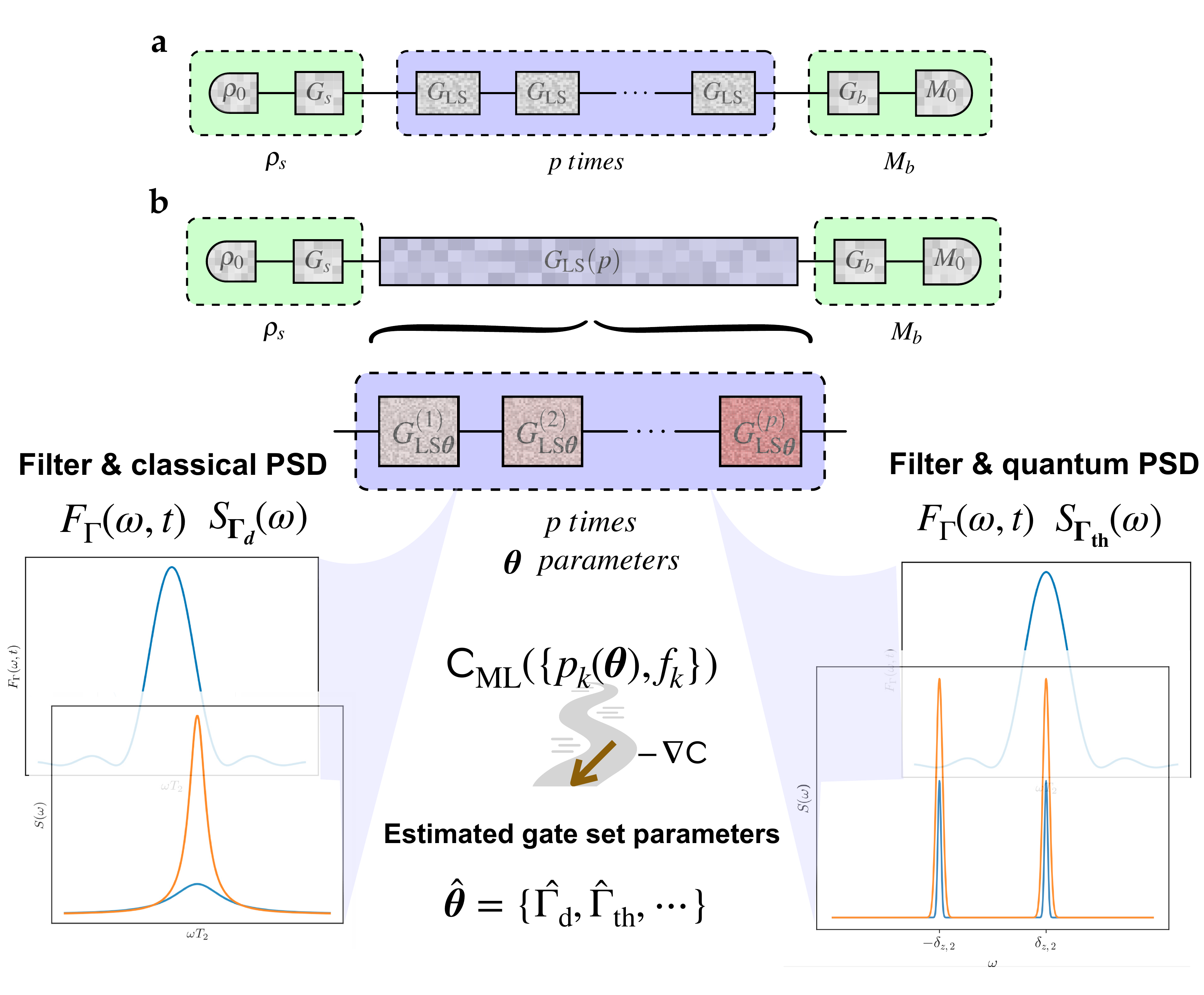}
  \caption{\textbf{Schematic representation of GST circuits for the LS germ. a:} circuit structure for context-independent GST with the microscopic parameterization. \textbf{b:} circuit structure for the context-dependent estimation of the $p$ LS gates with improved precision with respect to long-sequence GST. Different granularity levels represent precision of the estimations. In both schemes, long LS-gate sequences are executed to enhance precision. Hence, LS-gates are fine-grained depicted in comparison to the other circuit components. For the context-dependent estimation, we employ a microscopically-motivated parametrization of the gate set (see subsec.~\ref{analytical derivation}), where parameters are described in terms of filtered integrals~(Eqs.~(\ref{eq:deph_error},\ref{eq:thermal_error},\label{fig:circuits_scheme}\ref{eq:Gammas_app})) of the corresponding classical and quantum (\ref{quantum_psd}) power spectral densities. We then run the context-dependent tomography by solving a constrained optimization problem based on multi-parameter point estimation (\ref{MLE_GST}) with this parameterization.}
\end{figure}

We show that the number of resources required for GST can be reduced by several orders of magnitude, and we also find an alleviation of several technical challenges during the constrained non-linear optimization that underlies the GST statistical inference. 
This is explained in detail in Sec.~\ref{gate_map}, where we integrate the theory of LS trapped-ion gates with the GST formalism. This section also serves as the starting point for our generalization of GST to a context-dependent scenario, which addresses the second issue mentioned above. As we will further discuss, the microscopic map  for the LS gate  derived in Sec.~\ref{gate_map} can exhibit significant dependence on motional excitation of the ions, which introduces a specific history dependence on the  gate sequence as a whole due to the accumulated recoil of the ions by the laser fields that drive the LS gate. This prevents us from using the standard  long-sequence GST unless a cycle of sympathetic cooling~\cite{vsavsura2002cold, haffner2008quantum}, bringing the phonons to the same thermal state, can be applied prior to each LS gate. By employing two different ion species~\cite{PhysRevA.68.042302, schmidt2005spectroscopy}, this type of cooling reduces the ions' motional excitation near the ground-state while preserving the information stored in the qubit state, such that qubit-phonon entanglement is effectively erased prior to each gate, and the noisy LS gate becomes context independent. Unfortunately, this is too restrictive for many  trapped-ion devices that still operate with a single species, and would require  to keep track of the motional state during a sequence of LS gates. In Sec.~\ref{non-markovian_approach}, we upgrade the microscopic noise modeling resulting from $p$ successive LS gates, considering the cumulative  qubit-phonon dynamics  from one gate to the next. We are able to run tomography on the complete context-dependent channel, and actually show that the precision of the individual-gate GST estimates can improve with sequence depth $p$ faster than for standard GST, as the sensitivity can increase through the non-linear response of the qubit-phonon system. This idea is depicted in Figure~\ref{fig:circuits_scheme}. While Fig,~\ref{fig:circuits_scheme}\textbf{(a)} shows a schematic representation of the circuits that would be employed in a context-independent long-sequence GST, Fig.~\ref{fig:circuits_scheme}\textbf{(b)} reflects the idea behind the context-dependent estimation in section Sec.~\ref{non-markovian_approach}: doing tomography on the complete evolution after $p$ gates and then recovering each of the individual  maps with enhanced precision.

\section{\bf {Microscopic parametrization of trapped-ion  GST}}\label{gate_map}
The goal of this section is to provide a detailed description of our microscopic parametrization of  trapped-ion two-qubit GST, extending our work for single-qubit gates~\cite{vinas2025microscopic}  to a specific phonon-mediated entangling gate: the LS operation \cite{leibfried2003experimental, PhysRevA.76.040303, PhysRevA.103.012603}. Before discussing the leading error sources  considered in our work, let us briefly describe the ideal LS gate to set the notation and understand the role of the phonons as mediators of this gate, which turns out to be crucial to understand the noise sensitivity and the ultimate precision of GST.

Both the LS gate \cite{leibfried2003experimental} and the Mølmer-Sørensen  gate \cite{sorensen1999quantum} are variants of the so-called  geometric-phase gates. In contrast to the seminal  gate scheme by Cirac and Zoller \cite{PhysRevLett.74.4091}, these entangling  gates imprint a differential phase by exerting a force on the ions that  depends on the total spin state and, under specific conditions, gain a geometric robustness with respect to thermal fluctuations. The spin-dependent phase  depends on the area enclosed by the ion trajectories  in phase space, regardless of how fast the paths are traverse and, to a certain extend, on how much thermal fluctuations are present in the initial state. In contrast to the Cirac-Zolelr scheme,    ground-state cooling  is not a requisite to achieve high fidelities at leading order of the Lamb-Dicke parameter~\cite{sorensen1999quantum}. 

For the case of the LS gate, we consider a crystal of $N$ trapped ions, with the qubits of frequency $\omega_0$ encoded as e.g. two Zeeman sublevels of the $S_{1/2}$ ground-state of $^{40}{\rm Ca}^{+}$ \cite{PhysRevX.12.011032}. We note that the LS gate has also been implemented with other encodings \cite{PhysRevA.103.012603, Ballance_2016, Srinivas_2021, Sutherland_2019, Ospelkaus_2008}. In this work, we consider laser-driven LS gates by a pair of  non co-propagating  beams with frequencies $\omega_l$ for $l\in\{1,2\}$, where we define the beat-note laser frequency by $\Delta\omega_L=\omega_1-\omega_2$. The interaction of these fields with the ions produces two different AC-Stark shifts, given by the virtual absorption and emission of a photon either within the same beam or through a crossed-beam process \cite{PhysRevA.109.052417, Zhu_2006, leibfried2003experimental}. The later produces a state-dependent dipole force  through the recoil energy $E_R$ of photon-absorption in units of the corresponding normal mode energy. In the following, we assume that this recoil is oriented along the trap axis, and that the laser beams will couple to 2 out of $N$ ions in the crystal $j\in\{i_1,i_2\}$.
We move to the interaction picture with respect to 
\begin{equation}
\label{eq:ho}
H_0 = \frac{1}{2}\sum_{i}\omega_0Z_{i} + \sum_{m}\omega_{\alpha,m}a^{\dagger}_{\alpha,m}a_{\alpha,m},
\end{equation} 
where $\omega_{\alpha,m}$ are the normal-mode frequencies, and we have introduced the vibrational creation and annihilation operators $a^{\dagger}_{\alpha,m}a_{\alpha,m}$, as well as the qubit Pauli operators $Z_i,X_i,Y_i$. Within  the  Lamb-Dicke approximation, the light-matter  Hamiltonian describing a state-dependent force  along the trap axis reads
\begin{equation}
\label{eq:sdf}
H_{\mathrm{int}}(t)=\sum_{j,m} \mathcal{F}_{j,m}^z \bar{z}^0_ma_{z,m}\ee^{\ii(\Delta\boldsymbol{k}\cdot\boldsymbol{r}_j^0-\delta_mt+\phi_L)}Z_j+\mathrm{H.c.},
\end{equation}
where the detunings read $\delta_m=\omega_{z,m}-\Delta\omega_L$, and  the differential laser wave-vector (phase) $\Delta\boldsymbol{k}=\boldsymbol{k}_1-\boldsymbol{k}_2$ $(\phi_L=\phi_1-\phi_2)$.

The amplitude of this force is
\begin{equation}
\mathcal{F}_{i,m}^{z}=\ii\frac{\mid\tilde{\Omega}_{L,i}\mid}{2}\boldsymbol{\Delta k}\cdot\boldsymbol{e}_z\mathcal{M}_{i,m}^{z},
\end{equation}
where, using the notation of  \cite{PhysRevA.109.052417} ,  $\tilde{\Omega}_{L,i}$ is the crossed-beam ac Stark shift for the $i$-th ion, and $\mathcal{M}_{i,m}^{z}$ is the displacement of ion $i$ in the normal vibrational mode $m$ along the $z$ axis. In this expression $\bar{z}^0_m=1/\sqrt{2m\omega_{z,m}}$ is the ground-state oscillator width,  and $\eta_{z,m}=\boldsymbol{\Delta k}\cdot\boldsymbol{e}_z \bar{z}^0_m=\sqrt{E_R/\omega_{z,m}}\ll1 $ is the Lamb-Dicke parameter~\cite{RevModPhys.75.281}.

With this configuration of the force, the entangling gate will be driven by the `axial' motional modes of the gate, which describe trajectories in the corresponding phase space enclosing an area that will determine the geometric phase of the gate. The evolution operator for the LS interaction can be found after a second-order Magnus expansion \cite{magnus1954exponential, blanes2010pedagogical, Blanes_2009}
\begin{equation}
\label{evolution operator}
U(t)=\ee^{-\ii H_0t}\ee^{\sum_{j,m}(\phi_{j,m}(t)a^{\dagger}_{z,m}-\mathrm{H.c.})Z_{j}}\ee^{-\ii \mathcal{J}\!\!(t)\,Z_{i_1}Z_{i_2}},
\end{equation}
where $\mathcal{J}\!\!(t)$ controls the strength of a spin-spin coupling 
\begin{equation}
\label{coupling_strenght}
\mathcal{J}\!\!(t_g)=-\sum_{m}\frac{2\mathcal{F}_{i_1,m}^z\mathcal{F}_{i_2,m}^z}{\delta_m}\left(t_g\cos\phi^0_{i_1i_2}+\frac{1}{\delta_m}\sin(\delta_mt-\phi^0_{i_1i_2})\Big|_0^{t_g}\right)\!\!.
\end{equation}
and where we have introduced  $\phi^0_{i_1i_2}=\Delta\boldsymbol{k}(\boldsymbol{r}_{i_1}^0-\boldsymbol{r}_{i_2}^0)$.
We note that one could also include counter-rotating contributions to the phonon-mediated qubit-qubit interactions, such that the time $t_g$ for a maximally-entangling operation $\mathcal{J}\!\!(t_g)=\pi/4=:\theta_{LS}$ would be changed. {The ideal unitary describing this evolution reads $U_{\mathrm{LS}}={\rm exp}\{-\ii\theta_{LS}Z_{i_1}Z_{i_2}\}$, which generates Bell pairs  from  the product Hadamard basis $\ket{\sigma=\pm}=\frac{1}{\sqrt{2}}(\ket{0}\pm\ket{1})$, namely
\begin{equation}
U_{\mathrm{LS}}\ket{\pm,\pm\sigma}=\frac{1}{\sqrt{2}}(\ket{\pm,\pm\sigma}-\ii\ket{\mp,\mp\sigma}).
\end{equation}
}
In Eq.~(\ref{evolution operator}), we have also introduced 
\begin{equation}
\label{phi_small}
\phi_{j,m}(t_g)=\frac{\mathcal{F}_{j,m}^z\bar{z}^0_m}{\delta_m}{\ee^{-\ii(\Delta\boldsymbol{k}\cdot\boldsymbol{r}_j^0+\phi_L)}}\big(1-\ee^{\ii\delta_mt_g}\big),
\end{equation}
which describe the  trajectory in a complex  phase space of the corresponding driven modes subjected to the dipole forces. Similarly to the geometric phase, we note that counter-propagating terms can also be accounted for, modifying this expression in a way that they contribute to the trajectory.

The first two motional modes relevant for this gate implementation are the center-of-mass (com) mode, with frequency 
$\omega_{1,z}=\omega_z$, and the breathing mode, with frequency $\omega_{2,z}=\sqrt{3}\omega_z$ \cite{vsavsura2002cold}.
Typically, the com mode is used as the `active' mode, for which a circular phase-space trajectory is closed at the end of the LS gate, leading to the aforementioned geometric picture. In contrast, for unmodulated intensities of the laser beams, the breathing mode and other higher-energy normal modes will be driven along trajectories that cannot be simultaneously closed \cite{PhysRevA.109.052417,Oghittu_2024}. The unclosed phase-space trajectories introduce thermal noise, with each motional mode contributing in decreasing order of energy. This follows from the increasing frequency separation from 
the beat note $\Delta\omega_L$, which is close to the com-mode frequency. 
In the following section, we  derive a quantum channel  including this type of errors, focusing for simplicity on a $N=2$ ion crystal. Consequently, the breathing mode is the only contributor to this error. However, we note that for larger crystals and  2-out-of-$N$ LS gates, the error will include additional axial modes, which can be accounted for in an entirely analogous treatment.

\subsection{Analytical channel of the noisy light-shift gate}
\label{analytical derivation}

In this subsection, we discuss deviations from the ideal LS gates, which are  primarily affected by magnetic noise and motional errors from unclosed trajectories. As we anticipated in the ideal gate description, we assume that the LS gate is driven on the axial modes. Here, we take the com trajectory to be ideally closed by setting $t_g=2\pi/\delta_{\mathrm{com}}$. Thus, thermal noise from unclosed trajectories arises primarily from the unclosed breathing mode trajectory, which we now analyze in detail. 

It will prove convenient to write quantum channels on its vectorized or ``natural'' representation \cite{watrous2018theory, hashim2024practicalintroductionbenchmarkingcharacterization}, where density operators $\rho\in\mathcal{H}$ are written as vectors $\mathbf{vec}(\rho)=(\rho\otimes \mathbb{1}_{{\mathrm dim}(\mathcal{H})})\ket{\Phi}$ composed of the stacked columns (column-major order) of the matrix \cite{magnus1985matrix}. Here $\ket{\Phi}=\sum_{\boldsymbol{\sigma}}\ket{\boldsymbol{\sigma}}\otimes\ket{\boldsymbol{\sigma}}/2$ with $\boldsymbol{\sigma}=(\sigma_{i_1},\sigma_{i_2})\in\mathbb{Z}_2\times\mathbb{Z}_2$ refers to the maximally entangled state on a doubled Hilbert  space $\mathcal{H}\otimes \mathcal{H}$~\cite{Nielsen_Chuang_2010}.
To analyze gate performance, we are only interested in spin dynamics and we will consequently trace over the phononic degrees of freedom. In this subspace, all factors in Eq.~(\ref{evolution operator}) commute. As we will discuss later, magnetic noise introduces fluctuations of the characteristic frequency of ions $\omega_0$, hence, the resulting channel, which we will denote as $\mathcal{E}_{\Gamma_\mathrm{d}}$, will also commute with Eq.~(\ref{evolution operator}). As detailed below, the reduced quantum channel for the complete evolution can be divided into 
\begin{equation}
\label{complete_superoperator}
\mathcal{K}(\mathcal{E}_{\mathrm{LS}})=\mathcal{K}(\mathrm{tr}_{\mathrm{ph}}\{\mathcal{E}_{\mathrm{D}}\})\mathcal{K}(\mathcal{E}_{\Gamma_{\rm d}})\mathcal{K}(\mathcal{E}_{\omega_0})\mathcal{K}
(\mathcal{E}_{\mathrm{ZZ}}),
\end{equation}
{where $\mathrm{LS}$ and $\mathrm{D}$ refer to  `light-shift' and `(state-dependent-) displacement', respectively, whereas $\mathrm{ZZ}$ represents the spin-spin couplings. Note that, in this context, $\mathcal{E}_{\omega_0}$ only accounts for the spin evolution of Eq.~(\ref{eq:ho}). The channel thus represents the composition of the ideal LS gate followed by the respective error channels. We have taken the trace over phonons explicit in the state-dependent displacement channel, which is the only part of the evolution that generates entanglement between spins and phonons. In Eq.~(\ref{complete_superoperator}), as we will repeatedly do through the manuscript, we assumed evaluation at the gate time $t=t_g$ and omitted the explicit temporal dependence.}

Obtaining the superoperator representations for the unitary channels $\mathcal{E}_{\omega_0}$ and $\mathcal{E}_{ZZ}$ is trivial, as the natural representation of a unitary operator in column major order is given by $\mathcal{K}(\mathcal{E}_U) = \bar{U} \otimes U$, where $\bar{U}$ denotes the conjugate of $U$. This is a direct consequence of the vectorization rule of a matrix product, $\mathbf{vec}(ABC)=(C^{T}\otimes A)\mathbf{vec}(B)$ \cite{10.1145/3408039, magnus1985matrix}. We then focus on finding $\mathcal{K}(\mathrm{tr}_{\mathrm{ph}}\{\mathcal{E}_{\mathrm{D}}\})$.

\subsubsection{Motional errors and collective dephasing}
Let us now move to the discussion of the phonon induced motional errors. We begin by considering the contribution to expression~(\ref{evolution operator}) of the  state-dependent displacement
\begin{equation} 
\begin{split}
U_{\mathrm{D}}=\ee^{\sum_{j,m}(\phi_{j,m}(t_g)a^{\dagger}_m-{\rm H.c.})Z_j}=\sum_{\boldsymbol{\sigma}}\!\ket{\boldsymbol{\sigma}}\bra{\boldsymbol{\sigma}}\mathcal{D}(\boldsymbol{\sigma}),
\end{split}
\end{equation}
where we have introduced a product of displacement operators
\begin{equation}
\label{eq_prod_displacements}
\mathcal{D}(\boldsymbol{\sigma})=\bigotimes_{m}D_m\left(\textstyle\sum_{j}\phi_{j,m}(t_g)(-1)^{\sigma_{j}}\right)\ee^{\ii\Upsilon_m(\boldsymbol{\sigma})},
\end{equation}
with the following spin-dependent phase 
\begin{equation}
\Upsilon_m(\boldsymbol{\sigma})=(-1)^{\sigma_{i_1}\oplus\sigma_{i_2}}{\rm Im}\{\phi_{i_1,m}(t_g)\phi_{i_2,m}^*(t_g)\}.
\end{equation}
Here, $D_m(z)=\ee^{za_m^{\dagger}-z^*a_m}$ is the displacement operator \cite{scully1997quantum, schleich2015quantum} acting on the phase space of the $m$-th mode, and $\oplus$ denotes mod 2 addition. Also, we have used the relation \cite{leibfried2003experimental}
\begin{equation}
\label{eq:prod_disp}
\prod_{j}D(z_j)=D\left(\textstyle\sum_{j}z_j\right)\exp\left\{\frac{1}{2}\sum_{j<j'}^N(z_jz_{j'}^{*}-z_j^{*}z_{j'})\right\}.
\end{equation}
We now assume a separable initial state in which phonons follow a thermal distribution $\rho_{\mathrm ph}=\bigotimes_m\sum_{n_m}p_{m,n}\ket{n_m}\bra{n_m}$, where $p_{m,n}=\ee^{-\beta\omega_mn_m}/(1-\ee^{\beta\omega_m})$ \cite{gardiner2004quantum}.
Altogether, the unitary evolution produced by $U_{\mathrm D}$ reads
\begin{equation}
\label{sdd_evolution}
U_{\mathrm{D}}\rho {U_{\mathrm{D}}}^{\dagger}=\sum_{\boldsymbol{\sigma},\boldsymbol{\gamma}}\rho_{\boldsymbol{\sigma},\boldsymbol{ \gamma}}\ket{\boldsymbol{\sigma}}\bra{\boldsymbol{\gamma}}\bigotimes_m\sum_{n_m}p_{m,n}\mathcal{D}(\boldsymbol{\sigma})\ket{n_m}\bra{n_m}\mathcal{D}^{\dagger}(\boldsymbol{\gamma}),
\end{equation}
where $\rho_s=\sum_{\boldsymbol{\sigma},\boldsymbol{\gamma}}\rho_{\boldsymbol{\sigma},\boldsymbol{ \gamma}}\ket{\boldsymbol{\sigma}}\bra{\boldsymbol{\gamma}}$ represents any density operator describing the qubit state of the ions. Finally, we can trace out phonons and, using the expectation value of the displacement operator for a thermal state \cite{scully1997quantum}, we obtain
\begin{equation}
\label{trace2}
\mathrm{tr_{\mathrm{ph}}}\{\mathcal{E}_{\mathrm{D}}(\rho)\}=\sum_{\boldsymbol{\sigma},\boldsymbol{\gamma}}\rho_{\boldsymbol{\sigma},\boldsymbol{ \gamma}}\ket{\boldsymbol{\sigma}}\bra{\boldsymbol{\gamma}}\ee^{\sum_m\big( \ii\Upsilon_m^{\boldsymbol{\sigma},\boldsymbol{ \gamma}} - \mid\Phi_m^{\boldsymbol{\sigma},\boldsymbol{\gamma}}\mid^2(\bar{n}_m+1/2)\big)},
\end{equation}
where we have introduced  
\beq
\label{big_Phi}
\Phi_m^{\boldsymbol{\sigma},\boldsymbol{\gamma}}=2\sum_j(1-\delta_{\sigma_j, \gamma_j})\phi_{j,m}(t_g)(-1)^{\sigma_j},
\eeq
and where $\bar{n}_m$ is the mean number of phonons of the respective mode. We have also introduced
\begin{equation}
\Upsilon_m^{\boldsymbol{\sigma}, \boldsymbol{\gamma}}={\rm Im}\{\phi_{i_1,m}(t_g)\phi_{i_2,m}^*(t_g)\}\mathcal{S}_{\boldsymbol{\sigma},\boldsymbol{\gamma}},
\end{equation}
which depends on the following internal-state factor 
\begin{equation}
\label{eq:S}
\mathcal{S}_{\boldsymbol{\sigma},\boldsymbol{\gamma}}=(-1)^{\sigma_{i_1}\oplus\sigma_{i_2}}+(-1)^{\sigma_{i_1}\oplus\gamma_{i_2}}-(-1)^{\gamma_{i_1}\oplus\sigma_{i_2}}-(-1)^{\gamma_{i_1}\oplus\gamma_{i_2}}.
\end{equation}

In light of expression (\ref{phi_small}), we note that the phases $\Upsilon_m^{\boldsymbol{\sigma}, \boldsymbol{\gamma}}$ oscillate proportionally to $\sin\phi_{i_1,i_2}^0$. For simplicity, we assume equilibrium positions such that this function vanishes, although one could generalise easily.
Clearly, expression~(\ref{trace2}) represents a diagonal CPTP map.
When the gate duration is set to close the com-mode trajectory, we find $\Phi^{\sigma,\gamma}_{1}=0$, whereas for the breathing mode, the non-vanishing contribution to expression~(\ref{big_Phi}) leads to a map which corresponds to  spatially-anticorrelated dephasing with super-operator

\begin{equation}
\label{anticorrelated_dephasing}
\begin{split}
     \mathcal{K}(\mathrm{tr_{\mathrm{ph}}}\{\mathcal{E}_{\mathrm{D}}\})=\mathrm{diag}\big\{&1,\ee^{-\Gamma_{\mathrm{th}}},\ee^{-\Gamma_{\mathrm{th}}},1,\ee^{-\Gamma_{\mathrm{th}}},1,\ee^{-4\Gamma_{\mathrm{th}}},\ee^{-\Gamma_{\mathrm{th}}}, \\
    &\ee^{-\Gamma_{\mathrm{th}}},\ee^{-4\Gamma_{\mathrm{th}}},1,\ee^{-\Gamma_{\mathrm{th}}},1,\ee^{-\Gamma_{\mathrm{th}}},\ee^{-\Gamma_{\mathrm{th}}},1\big\},
\end{split}
\end{equation}
where $\mathrm{diag}\{\bullet\}$ refers to a diagonal matrix with the argument as elements, and we have introduced the thermal parameter 
\begin{equation}
\label{eq:thermal _rate}
\Gamma_{\mathrm{th}}=\left(\frac{\mid\tilde{\Omega}_{L}\mid\eta_{z,2}}{\delta_{z,2}}\right)^2\left(1-\cos\delta_{z,2}t_g\right)\left(\bar{n}_{\mathrm{2}}+\half\right),
\end{equation}
where we assumed $\tilde{\Omega}_{L,i_1}=\tilde{\Omega}_{L,i_2}=\tilde{\Omega}_{L}$.
Note that, in the event of closing the breathing mode trajectory instead of the com one, one would obtain a fully-correlated dephasing channel, which, as we now discuss, coincides with the magnetic noise channel.

\subsubsection{Green's functions,  classical and quantum spectral densities, and filter functions}
Let us now turn our discussion to the incorporation of magnetic noise fluctuations. We model the effect of such interactions by introducing a stochastic process $\tilde{\delta}(t)$ that modifies the characteristic qubit frequency  $\omega_0\to\omega_0+\tilde{\delta}(t)$ \cite{van1992stochastic, gardiner1985handbook, gardiner2004quantum, PhysRevResearch.7.013008}. To derive a quantum channel describing the effect of these fluctuations, as we introduced, we note that the ideal gate dynamics commute with the global magnetic noise. Thus, we can easily write a master equation for the noise channel
\begin{equation}
\label{time-convolutionless}
\frac{{\rm d}{\rho(t)}}{{\rm d}t} =-\sum_{i,j} \gamma(t)\big[{Z}_{i},[{Z}_j, \rho(t)]\big],
\end{equation}
where the time-dependent dephasing rate 
\beq
\gamma(t)=\frac{1}{4}\int_{0}^{t}\!\!{\rm d}t'\left(C(t,t')+C(t',t)\right),\hspace{1ex} C(t,t')=\mathbb{E}[ \tilde{\delta}(t) \tilde{\delta}(t')],
\eeq
depends on a symmetrized version of the auto-correlation function  of the stochastic process. For wide-sense stationary processes, this function is directly  symmetric $C(t,t')=C(|t-t'|)=C(t',t)$.
This leads, as we now argue,  to a spatially-correlated dephasing evolution with a dephasing rate that can be expressed in terms of the power spectral density (PSD) of the noise~\cite{RevModPhys.82.1155, varona2024lindbladlikequantumtomographynonmarkovian}, a common approach in different quantum noise analyses~\cite{PhysRevLett.87.270405, Biercuk_2011, Kofman_2001}. Because this equation does not involve any phonon evolution, it is straightforward to solve it in the qubit computational basis, yielding
\begin{equation}
\label{deph_evolution}
\mathbf{vec}(\rho(t_g)) = \mathcal{K}(\mathcal{E}_{{\Gamma}_{\rm d}}(t_g))\mathbf{vec}(\rho_0),
\end{equation}
where $\mathcal{K}(\mathcal{E}_{\Gamma_{\rm d}})$ denotes the superoperator describing the dephasing channel in the natural representation, and reads
\begin{equation}
\label{correlated_dephasing}
\begin{split}
    \mathcal{K}(\mathcal{E}_{\Gamma_{\rm d}}(t_g))= \mathrm{diag}\big\{&1,\ee^{-\Gamma_{\rm d}},\ee^{-\Gamma_{\rm d}},\ee^{-4\Gamma_{\rm d}},\ee^{-\Gamma_{\rm d}},1,1,\ee^{-\Gamma_{\rm d}}, \\
    &\ee^{-\Gamma_{\rm d}},1,1,\ee^{-\Gamma_{\rm d}},\ee^{-4\Gamma_{\rm d}},\ee^{-\Gamma_{\rm d}},\ee^{-\Gamma_{\rm d}},1\big\}.
\end{split}
\end{equation}
Here, we have introduced the dephasing parameter 
\begin{equation}
\Gamma_{\rm d}=2\int_{0}^{t_g}\!\!\mathrm{d}t\gamma(t),
\end{equation}
which can also be described  in terms of the noise PSD, making a direct connection with the filter function formalism~\cite{Kofman2000,PhysRevLett.93.130406,Gordon_2007,Uhrig_2008,PhysRevB.77.174509,Almog_2011}. The PSD, which is an even function in this case,  can be connected to  the auto-correlation function via  $C(t-t')=\int_{-\infty}^{\infty}\frac{{\rm d}\omega}{2\pi} S_\delta(\omega)\ee^{\ii\omega (t-t')}$, such that 
\begin{equation}
\label{eq:deph_error}
\Gamma_{\rm d}=\int_{-\infty}^{\infty}\mathrm{d}\omega S_\delta(\omega)F(\omega, t_g),
\end{equation}
where we have introduced $F(\omega, t)$ as the filter function
\begin{equation}
\label{eq:filter}
F(\omega,t_g)=\frac{1}{2\pi} \left(\frac{1-\cos(\omega t)}{\omega^2}\right)=:\frac{t}{2}\eta_{\frac{2
}{t}}(\omega).
\end{equation}
This is expressed in terms of  a nascent Dirac delta $\eta_\epsilon(x)=\frac{\epsilon}{\pi x^2}\sin^2(x/\epsilon)\to\delta(x)$ as $\epsilon\to0^+$,  showing that in the long-time Markovian limit, the filter  selects the  noise spectral density at zero frequency  to yield a constant dephasing  time $T_2=2/S_\delta(0)$. For shorter times, in particular those on the order of a characteristic correlation time of the noise $t\sim\tau_{\rm c}$, time correlations in the noise can affect the evolution. 

It is interesting to close this section by revisiting the motional errors in Eq.~\eqref{anticorrelated_dephasing} and~\eqref{eq:thermal _rate} from the perspective of a filtered quantum-mechanical PSD. The state-dependent force of Eq.~\eqref{eq:sdf} can be rewritten in a form that is similar to the stochastic qubit-frequency shift, but  the role of the classical process $\tilde{\delta}(t)$ is now played by the phononic operators $B_j(t)=\sum_m \mathcal{F}^z_{j,m}\tilde{\varphi}_{j,m}(t)$, where we have introduced the field operators $\tilde{\varphi}_{j,m}(t)=\bar{z}_m^0(a_m\ee^{\ii(\Delta\boldsymbol{k}\cdot\boldsymbol{r}_j^0-\delta_mt+\phi_L)}-{\rm H.c.})$. The effect of the autocorrelation function $C(t,t')$ can  be broader, and requires considering  various two-point functions of these  operators. In particular,   the spin-spin coupling~\eqref{coupling_strenght} is governed by 
\beq
\mathcal{J}\!(t_g)=\int_0^{t_g}\!\!{\rm d}t_1\int_0^{t_g}\!\!{\rm  d}t_2\sum_m\sum_{j\neq j'} \mathcal{F}^{z}_{j,m}G^{j,j'}_{{\rm R},m}(t_1-t_2)\mathcal{F}^z_{j',m},\eeq
where  the  retarded Green's function reads
\beq
G_{{\rm R},m}^{j,j'}(t)=-\ii\theta(t)\langle[\tilde{\varphi}_{j,m}(t),\tilde{\varphi}_{j',m}(0)]\rangle.
\eeq
This quantum correlation function is causal and temperature independent. We note that a  generalization of this expression, considering  time-dependent harmonic forces $\mathcal{F}^z_{j,m}\mathcal{F}^z_{j',m}\to\mathcal{F}^z_{j,m}(t_1),\mathcal{F}^z_{j',m}(t_2)$ as specific harmonic functions playing the role of sources, would allow one to also account  for  off-resonant contributions to the geometric phase. 

To account as well for  the motional gate errors  induced by the residual qubit-phonon entanglement~\eqref{trace2}, we need to consider the statistical  propagator
\beq
G^{j,j'}_{{\rm K},m}(t)=\frac{1}{2}\langle\{\tilde{\varphi}_{j,m}(t),\tilde{\varphi}_{j',m}(0)\}\rangle,
\eeq 
which  is also known as the Keldysh Green's function~\cite{Kamenev:341673}. Just as the  dephasing rate in Eq.~\eqref{time-convolutionless} depends on the symmetrized correlation function of the stochastic process $\tilde{\delta}(t)$, motional errors depend on the symmetrized fluctuations of the fields $\tilde{\varphi}(t)$ captured by $G_K$, and give contributions that are no longer causal and may actually  depend explicitly on temperature or, equivalently, on the mean phonon number. In fact, the motional dephasing rate in Eq.~\eqref{eq:thermal _rate} can be rewritten in terms of this statistical propagator or, alternatively, in terms of its Fourier transform. This leads to the equivalent of the noise PSD which, in this case $S_{\rm th}(\omega)=-\sum_m\sum_{j,j'} \mathcal{F}^{z}_{j,m}\mathcal{F}^z_{j',m}G^{j,j'}_{K,m}(\omega)$,   contains two peaks at the corresponding mode frequencies
\beq
\label{quantum_psd}
S_{\rm th}(\omega)=-2\pi\sum_m \sum_{j,j'}\mathcal{F}^{z}_{j,m}\mathcal{F}^z_{j',m}\ee^{\ii\Delta\boldsymbol{k}\cdot(\boldsymbol{r}_j^0-\boldsymbol{r}_{j'}^0)}W_m,
\eeq
where the delta peaks are contained in  
\beq
W_m=|\delta_m|(\bar{z}^0_m)^2(\bar{n}_m+\half)\delta(\omega^2-\delta_{m}^2).
\eeq
It is worth noting that one could also account for a motional heating of the phonons by exchanging the simple delta peaks by Lorentzians with a width set by the heating rate, which would actually lead to a PSD that is no longer even $S_{\rm th}(\omega)\neq S_{\rm th}(-\omega)$, a result of the quantum nature of the noise~\cite{RevModPhys.82.1155}. 

The motional error parameter appearing in Eqs.~\eqref{anticorrelated_dephasing} and~\eqref{eq:thermal _rate} can be expressed as follows
\begin{equation}
\label{eq:thermal_error}
\Gamma_{\rm th}=\int_{-\infty}^{\infty}\!\!\mathrm{d}\omega S_{\rm th}(\omega)F(\omega, t_g),
\end{equation}
with the same filter function as that of Eq.~\eqref{eq:filter}.
Since the function $F(\delta_1,t_g=2\pi/\delta_1)=0$ for the com-mode trajectory, it filters out the fluctuations of the  delta-shaped PSD stemming from this mode, such that the  motional dephasing is solely caused  by the breathing mode~\eqref{eq:thermal _rate}. Let us close by noting that the appearance  of the causal and statistical propagators connects to earlier works on the generating functional~\cite{PhysRevX.7.041012,PRXQuantum.3.020352,varona2024quantumcomputingfeynmandiagrams} and the finite-temperature response~\cite{Martinez2024thermalmasses,varona2024quantumcomputingfeynmandiagrams} through a Keldysh formalism. 

\subsubsection{Effective collective Kraus channel}

After the discussion in the previous sections, we have all the necessary ingredients to compute the LS-gate quantum channel~(\ref{complete_superoperator}). The corresponding super-operator  will be employed in next section to parametrize two-qubit GST, and show that one can indeed improve the precision in comparison to a fully-general approach. {The super-operator representation  will be convenient when dealing with the constrained optimization of GST, as the composition rule reduces to simple matrix multiplication. 
Let us, however, close  this section by providing a Kraus decomposition \cite{kraus1983states} of the quantum channel to give further physical intuition about the noisy dynamics. We have $\rho(t_g)=\mathcal{E}_N(\rho_0)=\sum_nK_n\rho_0K_n^{\dagger}$, where $\mathcal{E}_N\coloneq\mathcal{E}_N(\Gamma_{\mathrm{d}}, \Gamma_{\mathrm{th}})$ denotes the noise channel, and the Kraus operators $K_n$ can be expressed in terms of Pauli operators and the functions $\chi_{\mathrm{d}}=\ee^{-\Gamma_{\mathrm{d}}}$, $\chi_{\mathrm{th}}=\ee^{-\Gamma_{\mathrm{th}}}$ as follows
\begin{equation}
\begin{split}
&K_1=\frac{1}{2}(\sqrt{\chi_{\mathrm{d}}}+\sqrt{\chi_{\mathrm{th}}})\mathbb{1}+\frac{1}{2}(\sqrt{\chi_{\mathrm{d}}}-\sqrt{\chi_{\mathrm{th}}})Z_{i_1}Z_{i_2},\\
&K_2=\frac{1}{2}\sqrt{\chi_{\mathrm{th}}(1-\chi_{\mathrm{th}})}(Z_{i_1}-Z_{i_2}),\\
&K_3=\frac{1}{2}\sqrt{\chi_{\mathrm{d}}(1-\chi_{\mathrm{d}})}(Z_{i_1}+Z_{i_2}),\\
&K_4=\frac{1}{4}(1-\chi_{\mathrm{th}})(\mathbb{1}+Z_{i_1})(\mathbb{1}-Z_{i_2}),\\
&K_5=\frac{1}{4}(1-\chi_{\mathrm{th}})(\mathbb{1}-Z_{i_1})(\mathbb{1}+Z_{i_2}),\\
&K_6=\frac{1}{4}(1-\chi_{\mathrm{d}})(\mathbb{1}+Z_{i_1})(\mathbb{1}+Z_{i_2}),\\
&K_7=\frac{1}{4}(1-\chi_{\mathrm{d}})(\mathbb{1}-Z_{i_1})(\mathbb{1}-Z_{i_2}).\\
\end{split}
\end{equation}

The last four operators clearly account for the (anti-) correlated  nature of the dephasing induced by the (motional) frequency noise.
We can use this representation to describe the complete process (\ref{complete_superoperator}) as a composition of the unitary gate followed by the noise channel
\begin{equation}
\label{gate_krauss}
\mathcal{E}_{\mathrm{LS}}(\bullet)=\mathcal{E}_{\mathrm{N}}(\Gamma_{\mathrm{d}}, \Gamma_{\mathrm{th}})\circ \ee^{-\ii\frac{\omega_0}{2}t_g(Z_{i_1}+Z_{i_2})}\ee^{-\ii\theta_{LS}Z_{i_1}Z_{i_2}}(\bullet).
\end{equation}
}
where we recall that the maximally-entangling phase area is $\theta_{LS}=\pi/4$ and the unitary part is understood to act by conjugation $U(\bullet)=U\bullet U^{\dagger}$.

Having discussed the thermal and magnetic noise contributions, we note  that additional sources of noise  could be incorporated in future work, including off-resonant drivings, spectator modes, anomalous heating, intensity fluctuations or residual photon scattering. The effect of laser phase fluctuations,   which are relevant in single-qubit pulses or Mølmer-Sørensen entangling gates, become  negligible in  the LS gate scheme.  We expect this other noise sources to be  residual as compared to the magnetic and thermal errors  hereby considered \cite{PhysRevA.109.052417}. Let us emphasize that, while the aim of this work is to adapt GST to the dominant error sources in trapped-ion QIPs, a  ``tailored-to-experiment'' noise modeling that also parameterizes these effects is left for future investigation.

\subsection{Trapped-ion two-qubit GST:  A  comparison of fully-general and parametrized approaches}
\label{context-independent-gst}
In this section, we employ the previous LS-gate modelling to parameterize a two-qubit GST, and show that it allows for an improved precision at fixed resources or, alternatively, lower resources for a target precision.
We note that the previous LS-gate channel requires an initially separable spin-phonon state. As we mentioned during the introduction, under the assumption that sympathetic cooling~\cite{vsavsura2002cold, haffner2008quantum} is applied prior to each LS gate~ \cite{DeCross_2025, Pino_2021}, the ions would be cooled near to the motional ground-state, and this assumption would hold. In addition to separability, this leads to context independence for a sequence of noisy LS gates. 
We also assume context independence for the dephasing channel due to  magnetic noise. For this, we need to assume that the characteristic correlation time that governs magnetic noise $\tau_c$ is sufficiently small compared to the time in between consecutive gates $\Delta t_{i,j}$, such that one can neglect correlations between different gates. In contrast, this  time can be  larger in comparison to the gate time itself  $\tau_c\sim t_g$, leading to  non-Markovian dynamics during each gate.

Under these assumptions, the composition of the corresponding  collective dephasing channels for a sequence of $p$ gates will result in a map with a rate $p\Gamma_{\mathrm{d}}(t_g)$. This connects to an implicit assumption  in our previous work \cite{vinas2025microscopic}, enabling us to find a  informationally complete set of gates that can be efficiently parametrized. Building on those results~\cite{vinas2025microscopic}, we now include the context-independent entangling  LS gate, and thus  parametrize a  two-qubit GST. {Our ideal gate set reads
\begin{equation}
\label{eq:gate_set_GST}
\begin{split}
\mathcal{G}&=\Bigl\{\rho_{0},\hspace{1ex}\big\{ G_{LS}(\theta_{LS});\ G_i(\theta_i,\phi_i), \,i={1,\cdots,5}\big\},\hspace{1ex}M_0\Bigr\}, 
\end{split}
\end{equation}
where $G_{\mathrm{LS}}$ denotes the LS gate, and $G_i(\theta_i,\phi_i)$ refers to the $i$-th single-qubit gate. Ideally, those are defined as
\beq
\label{eq:ideal_gate}
G^{\rm id}_i(\theta_i,\phi_i)={\rm exp}\left\{-\ii\frac{\theta_i}{2}\big(\cos\phi_i X_i-\sin\phi_i Y_i\big)\right\},
\eeq
and an IC set corresponds to  the angles $\theta_{LS}=\pi/4$ and
\beq
\label{eq:gate_ste_angles}
(\theta_i,\phi_i)\in\left\{(\pi,0),\left(\frac{\pi}{2},0\right),\left(\frac{\pi}{2},\frac{3\pi}{2}\right),\left(\frac{\pi}{2},\frac{\pi}{2}\right),\left(\frac{\pi}{2},\pi\right)\right\}.
\eeq
For the microscopic parameterization of the noise gate set,  we assume single-qubit gates  affected by phase and frequency fluctuations. Optionally, intensity noise can also be considered \cite{vinas2025microscopic}. This leads to the definition of additional noise parameters and effective shifts  $\Gamma_1,\,\Gamma_2, \, \Delta_1,\,\Delta_2$ and $\Delta\Gamma$, such that the noisy version of the gate $G(\theta,\phi)$ reads $\tilde{G}(\theta,\phi)=\tilde{G}(\theta,\phi;\,\Gamma_1, \Gamma_2, \Delta_1,\Delta_2, \Delta\Gamma),\,{\rm for\,}\,n=1,2$.  These noise parameters can also be expressed in terms of filtered noise   integrals, albeit in this case all of the PSDs have a classical origin 
\beq
\label{eq:Gammas_app}
\begin{split}
	\Gamma_n = & \int_{-\infty}^\infty \!\!{\diff\omega}\,S_{{\delta}}(\omega)\,F_{\Gamma_n}(\omega,\Omega,t_g),\\
	\Delta_n = & \int_{-\infty}^\infty \!\!{\diff\omega}\,S_{{\delta}}(\omega)\,F_{\Delta_n}(\omega,\Omega,t_g),\\
    \Delta\Gamma_1 = & \int_{-\infty}^\infty \!\!\!{\rm d}\omega {S}_{\Omega}(\omega)F_\Omega(\omega,t_g),
\end{split}
\eeq
where, $S_{\delta}(\omega)$ and $S_\Omega(\omega)$ are the corresponding PSDs of the frequency and amplitude noise stochastic processes, which in this case incorporate fluctuations of the laser phase and intensity used to drive the single-qubit gates. In addition, since such drivings are also transversal, the specific filters $F_{\Gamma_n}(\omega,\Omega,t_g)$, $F_{\Delta_n}(\omega,\Omega,t_g)$ differ with respect to the free-induction decay form of Eq.~\eqref{eq:filter}, and also depend on the Rabi frequency $\Omega$ of the driving.  In contrast,   $F_{\Omega}(\omega,t)$ filters the intensity fluctuations in a dressed-state basis and has the same functional form,  see \cite{PhysRevResearch.7.013008,vinas2025microscopic}. As noted in these works,  $\Gamma_2$ and $\Delta_2$  become negligible when non-Markovian effects in the single-qubit gates remain small.
Finally,  the noisy version of the LS gate $G_{LS}(\theta_{LS})$ is described by  $\tilde{G}_{LS}(\theta_{LS}; \Gamma_{\rm d},\Gamma_{\rm th})$ in Eq.~ (\ref{gate_krauss}), where the error parameters $\Gamma_{\mathrm{d}}$ and  $\Gamma_{\mathrm{th}}$ depend on the classical~\eqref{eq:deph_error} and quantum~\eqref{eq:thermal_error} PSDs.

Considering also the faulty SPAM $\tilde{\rho}_0,\tilde{M}_0$, for the gate set under study (\ref{eq:gate_set_GST}) and our microscopically-motivated  parameterization, we find $N_{\mathrm{param}}\in(24,36)$ parameters to be determined via GST, depending on the Markovianity assumptions and the inclusion of intensity noise fluctuations. This contrasts with the fully-general approach to GST~\cite{Nielsen2021gatesettomography},  which would require  $N_{\mathrm{param}}=423$  to account for the same gate set. This difference in the number of parameters would considerably increase  if one includes crosstalk.

While we already presented GST during the introduction, let us further discuss its main aspects, which will be relevant for the estimation of gate set (\ref{eq:gate_set_GST}) with the microscopic parameterization. In a nutshell, GST provides an estimate of the gate set under study by fitting experimentally-measured frequencies to a model describing the gates. These frequencies are nothing but the outcomes of carefully designed circuits obeying a fixed structure~\cite{Nielsen2021gatesettomography}: $i)$ \textit{state preparation}: the reference state $\rho_0$ undergoes an evolution produced by elements extracted from $\mathcal{G}$
$ii)$ \textit{base circuit}: a `germ', i.e., a particular combination of gates, is executed to provide sensitivity with respect to some of the parameters of the gate set. Optionally, this sequence of gates may be repeated several times to enhance precision in what it is called the long-sequence GST protocol. $iii)$ \textit{measurement}: followed by some transformation such as those in state preparation, the measurement operation $M_0$ is applied and the results are collected.

It is important to mention that stages $1)$ and $3)$ are designed to generate the informationally complete~\cite{scott2006tight} sets $\big\{G_{s}\rho_0=\rho_{s}:\,s\in \mathbb{S},\,|\mathbb{S}|=d^2\ \big\}$ and $\big\{G_{b}M_0=M_{b}:\ {b} \in \mathbb{M}_{b} \big\}$, where the Hilbert space  dimension  is $d$. Each POVM $M_b$ has $m_b$ possible outcomes, such that the measurements are labeled by $\mu=(b,m_b)\in\mathbb{M}=\mathbb{M}_b\times\mathbb{M}_{m_b}$ with 
$|\mathbb{M}_{b}|=d^2$. This means that the \textit{fiducial pairs} $\big\{\rho_{s},M_{b}\big\}$ are sufficient to obtain a unique representation of any quantum channel $\hat{\mathcal{E}}$ from the probabilities $p_{\mu,s}=\mathrm{tr}\{M_{\mu}\mathcal{E}(\rho_s)\}$. Similarly, long-sequence GST requires the set of germs to amplify every non-redundant parameter used to describe it. This condition is known as amplification completeness (AC). Once the outputs of these circuits have been collected, experimental frequencies are constructed as $f_{\gamma_c,\mu} = N_{\gamma_c,\mu}/N_{\gamma_c}$, where $N_{\gamma_c,\mu}$ denotes the number of times that outcome $\mu$ has been obtained from circuit $\gamma_c$, and $N_{\gamma_c}$ the number of times that circuit has been executed. Finally, one estimates the gate set by fitting this data to the theoretical probabilities $p_{\gamma_c,\mu}$ that describe those same circuits. In principle, this could be achieved by means of a simple linear inversion algorithm. Finite sampling noise, however, can lead to  un-physical solutions, yielding estimates that do not satisfy the completely positive and trace preserving (CPTP) conditions of quantum maps. To solve this issue, the linear inversion estimate is often only used as an initial guess of non-convex nonlinear optimization problem, in which quantum maps can be enforced to be physical by imposing certain constraints. Different cost functions and optimization routines have been refined to circumvent local minima, and give a precise estimate at a certain resource cost~\cite{Nielsen2021gatesettomography}. More precisely, GST provides an estimate $\hat{\boldsymbol{\theta}}$ to a constrained multi-parameter point-estimation problem in which, typically, a negative log-likelihood cost function $\mathcal{C}_{\rm ML}(\boldsymbol{\theta})$ or a weighted least-squares cost function $\mathcal{C}_{\rm LS}(\boldsymbol{\theta})$ is built from the finite-frequency outcomes and then minimized
\beq
\label{MLE_GST}
\begin{split}
&\hat{\boldsymbol{\theta}}=\texttt{argmin}\hspace{1ex} \begin{cases}
\mathcal{C}_{\rm ML}(\boldsymbol{\theta})=-\sum\limits_{\boldsymbol{\gamma}}\sum\limits_{m_b}f_{\boldsymbol{\gamma}}(m_b)\log p_{\boldsymbol{\gamma}}(m_b),\\
\mathcal{C}_{\rm LS}(\boldsymbol{\theta})=\sum\limits_{\boldsymbol{\gamma}}\sum\limits_{m_b}\frac{1}{p_{\boldsymbol{\gamma}}(m_b)}\big(f_{\boldsymbol{\gamma}}(m_b)-p_{\boldsymbol{\gamma}}(m_b)\big)^2\!.
\end{cases}\\ 
\end{split}
\eeq
here, we have omitted the parametric dependence of the probabilities. Of course, the vector $\boldsymbol{\theta}$ contains all the parameters describing the gate set, in our case $\boldsymbol{\theta}=(\Gamma_{\mathrm{th}},\Gamma_{\mathrm{d}};\Gamma_1^1,\Delta_1^1,\dots)$.

Other technical details relevant to GST are gauge freedom and optimization~\cite{Nielsen2021gatesettomography}, as well as Fiducial Pair Reduction (FPR)~\cite{rudinger2023twoqubitgatesettomography, ostrove2023nearminimalgatesettomography}. Gauge freedom describes a potential ambiguity of the estimated gate set $\mathcal{\hat{G}}$. Given $\mathcal{\hat{G}}$ compatible with some experimental frequencies, the similarity-transformed gate set $\mathcal{\hat{G}}^M=M\mathcal{\hat{G}}M^{-1}$ is equally consistent with experimental results. The similarity transformation is understood to act independently over each gate set component. To set a common reference and compare results, usually done by means of gauge-dependent distances, the closest-to-ideal gate set out of all the similar ones will be chosen. The process of finding this `privileged' point in the space of sets compatible with experimental data is known as gauge optimization.
On the other hand, fiducial pair reduction aims to cut down the experimental overhead of long-sequence GST. Once an AC set has been found, sandwiching each of its germs with all the fiducial pairs results on many redundant circuits, which provide sensitivity to shared parameters. Different FPR techniques identify and discard many of those unneeded experiments, resulting on more efficient experimental designs.

Having further discussed GST, let us come back to the analysis of our physically-parameterized gate set. Remarkably, under the microscopic parameterization employed, such a gate set provides amplification completeness with no need to combine the bare gates into more complex germs. Also, as we discuss at the end of this subsection, the experimental designs that we obtain are minimal. In a similar line to what we exposed in our previous work, the inclusion of the LS gate in the gate set does not introduce any gauge freedom to the already gauge-free single-qubit gate set. Therefore, gauge optimization techniques remain unnecessary with our approach. This is a direct consequence of the spectrum of the gates being incompatible with more than one set of parameters. 

It is also notable that, on par with our single-qubit GST analysis~\cite{vinas2025microscopic}, physicality constraints are also readily imposed for the entire two-qubit  gate set. Complete positivity can indeed be imposed simply by restricting to positive error parameters $\Gamma_{\rm d}, \Gamma_{\rm th}\geq0$, which guarantees the mapping of the superoperator elements to probabilities. Similarly, $\Gamma_1>0$ can be imposed to guarantee the complete positivity of single-qubit gates. We note, nonetheless, that CP constraints lead to a potentially biased estimation \cite{Nielsen2021gatesettomography}. In particular, we find that estimating a high-fidelity gate set with designs that are not sufficiently precise  to capture such small deviations results in the optimization converging to the ideal gates. This is not so problematic, as this bias eventually disappears when employing enough samples and sufficiently large depths of the long-sequence GST sequences. 
However, with the aim of observing the expected precision at any depth or at any $N_{\mathrm{shots}}$, we will relax the CP conditions for the optimizations in the following figures.} This contrasts the general approach~\cite{Nielsen2021gatesettomography}, in which complete positivity of the estimated channels cannot be easily imposed in the super-operator formalism, which is preferred for performance over other channel representations. No additional condition is needed for trace preservation. This is easily checked by moving to the Pauli Transfer Matrix representation under a simple change of basis. There, we verify that the first row of the superoperator~(\ref{anticorrelated_dephasing}) is $(1,0,\dots,0)$, which is the trace preservation condition.

In order to generate noisy data to run two-qubit GST both with our model and with the fully-general approach, we focus on frequency noise, and consider the Ornstein-Uhlenbeck stochastic process~\cite{gardiner1985handbook, 10.1119/1.18210}. This choice is motivated by the facility with which we can tune the correlation time $\tau_{\rm c}$, and the diffusion constant $c$ of this process, but also  because it is a standard choice for Gaussian colored noise that serves as an idealized version for various processes~\cite{PhysRevA.21.1289, avan1977two}.

For the simulation of the entangling LS gate, in which equation~(\ref{time-convolutionless}) is exact within a Lamb-Dicke approximation, we  directly compute the parameter describing magnetic noise by means of the PSD of the OU process, which is a simple Lorentzian,
\beq
\label{eq:PSD_OU}
    S_{\delta}(\omega)=\frac{ c\tau_{\rm c}^2}{1+(\omega\tau_{\rm c})^2}.
\eeq
The depahsing parameter can then be expressed as
\begin{equation}
\label{gammaOU}
\Gamma_{\rm d}= \frac{c\tau_c^2}{2}\left(t_g+\tau_c(\ee^{-\frac{t_g}{\tau_c}}-1)\right).
\end{equation}
As evident from previous expression,
for sufficiently long gate times compared to $\tau_c$, the linear term in Eq.~(\ref{gammaOU}) dominates. Hence, under the assumption that time between gates is long enough, we can safely assume $\Gamma(pt_g)=p\Gamma_{\mathrm{d}}$, as previously anticipated in connection to context-dependence. 
At the same time, we can easily generate realistic simulation values for $\Gamma_{\mathrm{th}}$, as it does not have a stochastic dependence, and is completely determined by the filtered quantum PSD of the discrete vibrational modes, depending on microscopic parameters $\phi_{i,m}(t_g)$ and $\bar{n}_m$.

With all of this, we generate simulated data for each circuit $\gamma_c$ and run two-qubit GST, which provides an estimate $\hat{\boldsymbol{\theta}}$ as a solution to the constrained minimization problem Eq.~(\ref{MLE_GST}). For this, we rely on the open-source python library \texttt{pyGSTi} \cite{Nielsen_2020, pygsti}. We then compare the estimates with the actual gate set used for the numerical simulations by means of the diamond distance \cite{aharanov_kitaev}
\begin{equation}
\label{diamond}
\mid\mid \hat{G}-G_0\mid\mid_{\diamond} = \mathrm{max}_{\rho}\mid\mid (\hat{G}\otimes\mathbb{1})\rho-(G_0\otimes\mathbb{1})\rho\mid\mid_{1},
\end{equation}
where we maximize over all density matrices with dimension ${ dim}(\mathcal{H})^2$, $\hat{G},G_0$ denote the estimated and the `true' versions of each noisy gate $G$ respectively, and $\mid\mid\bullet\mid\mid_1$ is the usual trace norm. Unless  specified differently, we compute this norm for each element of the gate set and then average.

\subsubsection{Scaling of long-sequence GST:  single-depth estimates  versus logarithmic-depth spacing }

We now present the results of Figure~\ref{fig: td_vs_p_mic_noisy}, showing the scaling of precision with the maximum depth employed for long sequence (red line and data-points), using the microscopically-modelled gate set. The results follow a $m=-1$ slope tendency in log-log axis (black dashed line), consistent with the expected GST results obtained in other works~\cite{Nielsen2021gatesettomography}. As discussed in \cite{vinas2025microscopic}, and also noted in prior GST treatments~\cite{Nielsen2021gatesettomography, ostrove2023nearminimalgatesettomography}, we appreciate how this scaling saturates for sufficiently large depths, as noise begins to dominate the dynamics and we can no longer improve the estimations as the signal to noise ratio no longer improves with circuit depth.

Figure~\ref{fig: td_vs_p_mic_noisy} motivates the discussion of GST scaling. Long-sequence GST produces estimates whose error asymptotically scales with $O(1/p\sqrt{N_{samples}})$, where $p$ is the maximum number of repetitions for each germ in the GST circuits. As maximum-likelihood  is an asymptotically unbiased estimator, this leads to a linear increase in precision with $p$. Because only gate parameters are amplified, we do not include SPAM elements in the computation of the diamond norm (\ref{diamond}). This scaling is often referred to as a \textit{Heisenberg-like} scaling. We note, however, that the concept of a Heisenberg limit arises in quantum metrology and frequency estimation~\cite{degen2017quantum, haase2016precision, smirne2016ultimate}, and  gives a fundamental bound for the minimal estimation error one can achieve with   with a fixed  number of qubits,  scaling  as $1/N_{\rm qb}$. Reaching this bound requires using entanglement, and points to the fact that one can obtain a more precise estimate 
if, instead of performing the frequency estimation with an ensemble of qubits in a product state leading to a $1/\sqrt{N_{\rm qb}}$ scaling, one exploits entanglement to boost the estimation precision  linearly $1/{N_{\rm qb}}$. In this context, it is preferable to sample directly the entangled qubit probes $N_{\rm samples}=N_{\rm qb}$. 
In the context of long-sequence GST, entanglement does not play any major role and, in fact,  the Heisenberg-like scaling mentioned above   can be viewed as a depth-dependent prefactor  upon the standard quantum limit   $O(1/\sqrt{N_{samples}})$, which is just the usual quantum projection or shot noise uncertainty ~\cite{PhysRevA.47.3554}. To avoid confusion, we will just employ $1/p$ notation to refer to this scaling at the present work. The standard long-sequence GST allocates sampling resources logarithmically over deeper circuits, which is a preferable strategy compared to simply increasing the number of samples of the shorter ones in light of the linear $1/p$ amplification.

\begin{figure}
  \centering
  \includegraphics[width=1\columnwidth]{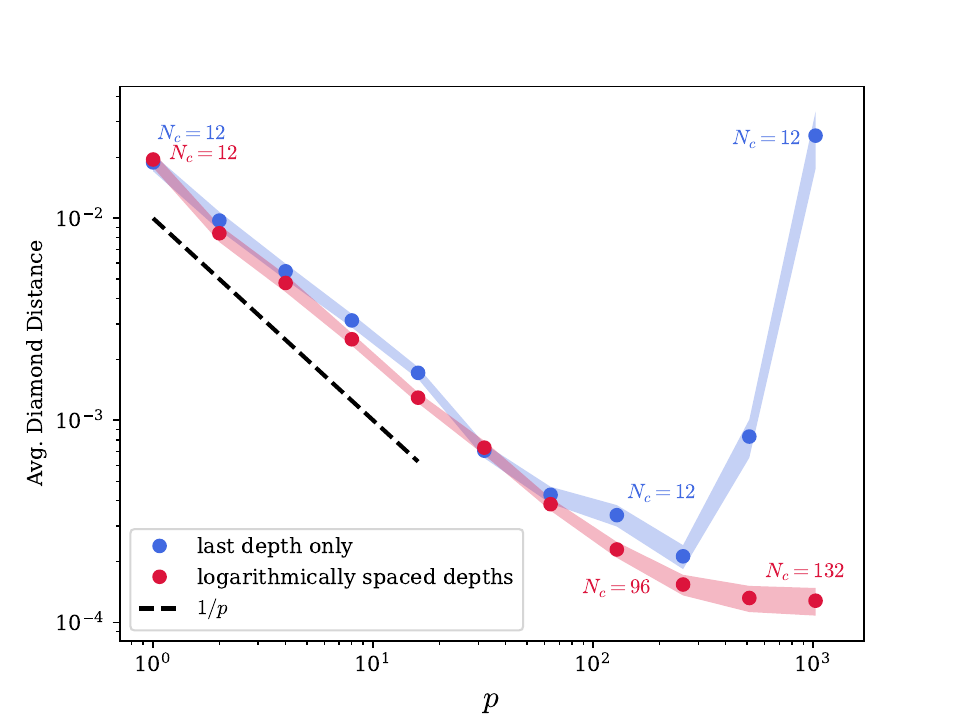}
  \caption{\textbf{Average diamond distance as a function of circuit depth for the microscopic model.} We run GTS on simulated data for the usual logarithmically-spaced depths GST scheme (red) and for a last-depth only scheme (blue). The number of circuits required are $N_{\mathrm{c}}=12$ for the later scheme and $N_{\mathrm{c}}=12\times\log_2p$ for the first scheme. We observe $1/p$ tendencies for both schemes (gray guide to the eye). We note that the noise-dominated region in which precision plateaus for the logarithmic scheme results in a degradation of precision for the last-depth only one. Parameters used for the simulations are $c=2\times 10^9s^{-3}$, $\tau_c=5\times 10^{-4}s$ and $\bar{n}_{\mathrm{b}}=5$. Each circuit is sampled $N_{\rm samples}=10000$ times. Light shaded regions represent $95\%$ confidence intervals after averaging over $100$ GST estimations for each depth.}
  \label{fig: td_vs_p_mic_noisy}
\end{figure}

The aforementioned logarithmically-spaced allocation of circuits in long-sequence GST serves to avoid branch ambiguities and local minima in the complex non-linear optimization problem (\ref{MLE_GST}). However, as shown by the quantitative comparison in Fig.~\ref{fig: td_vs_p_mic_noisy}, we find that we can obtain comparable precisions by only including the circuit with the largest depth (blue line and data-points). This already saves a logarithmic amount of resources \cite{Nielsen2021gatesettomography}, as we are not forced to dedicate measurement samples to the intermediate-depth circuits, and can instead allocate them at the more precise larger-depth circuit directly. We believe that this difference is rooted in the fact that other non-linear minimization for our microscopic parametrization is much simpler, and not afflicted by the complications of the standard fully-general GST. 

Let us also note that, in the region where the GST precision cannot be further improved, the absence of shorter circuits leads to a faster degradation of precision when using a scheme with only the  largest-depth circuit, which contrasts the plateau displayed by the standard scheme based on circuits with logarithmically-spaced depths. This difference is expected, as the logarithmically-spaced scheme
still includes the data from shorter-depth circuits which are not affected by the loss of the signal-to-noise ratio, maintaining the precision of the estimations even if the larger depths do not add further information due to the accumulated noise. On the other hand, in a regime which is not dominated by noise, these results show that one can save considerable resources by using the microscopic parameterization and focusing instead on  single large-depth circuits.

\subsubsection{Sample complexity of the microscopic GST }

Let us now explore the sampling complexity of our parametrized GST, comparing the total number of shots needed to run long-sequence GST
with those of the fully-general scheme. We choose a fixed number of shots per circuit $N_{\rm samples}=10^4$, and compare the estimates obtained with both models, which will end up requiring a very different number of circuits to achieve a target precision. These results are presented in  Fig.~\ref{fig: td_vs_p_mic_vs-general}. As can be observed, the reduced number of circuits required to achieve high precision in the estimation of our microscopic model improves dramatically faster with the total number of shots $N_{\rm shots}=N_c\times N_{samples}$ than the fully-general scheme. For instance, using $N_{\rm samples}=10^4$  per circuit, achieving a target average diamond distance $d_{\diamond} = 10^{-3}$ requires $N_{\mathrm{shots}}\approx5\times10^5$ for our microscopically-motivated model, whereas a fully general approach demands resources ranging in the interval $N_{\mathrm{shots}}\in(10^8,10^9)$, depending on the efficacy of the FPR strategy. In the figure, theoretical tendency predictions for different FPR schemes are plotted, indicating the ratio of circuits employed with respect to the original design. For this, we assume that the $1/p$ tendency can be attained regardless of FPR. While depending on the FPR strategy this assumption might not be guaranteed in practice, the predictions shown serve as lower bounds of the actual FPR results. We also note that the AC set used in the fully general design is not necessarily optimal, as alternative germ sets might yield equivalent results at a slightly lower cost. This stems from the fact that finding an AC set is a computationally demanding task, and we were not capable of finding a more compact one.

\begin{figure}
  \centering
  \includegraphics[width=1\columnwidth]{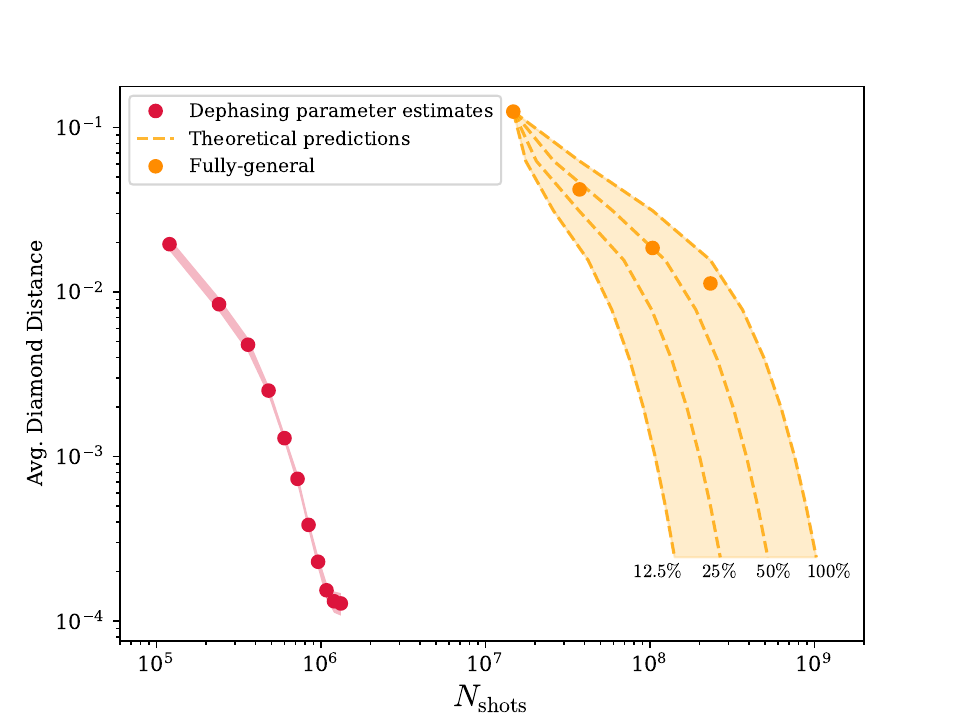}
  \caption{\textbf{Average diamond distance as a function of the total number of shots $N_\mathrm{shots}$ for both the microscopic model and the fully general one.} We run GTS on simulated data for both models, showing a dramatic improvement in the precision of the estimates for the microscopic model at a fixed number of shots. Parameters used for the simulations are $c=2\times 10^9s^{-3}$, $\tau_c=5\times 10^{-4}s$ and $\bar{n}_{\mathrm{b}}=5$. Each circuit is sampled $N_{\rm samples}=10000$ times. Light shaded regions for the microscopic model represent $95\%$ confidence intervals after averaging over $100$ GST estimations for each depth. We complement the fully general estimates with theoretical tendency predictions for different FPR schemes, indicating the ratio of circuits employed with respect to the original design.}
  \label{fig: td_vs_p_mic_vs-general}
\end{figure}

Let us finally discuss the choice of the GST experimental designs employed with more detail. Under a general parameterization of quantum channels, near-minimal experimental designs for GST can be found using recent FPR techniques and relaxing the AC conditions \cite{rudinger2023twoqubitgatesettomography, ostrove2023nearminimalgatesettomography}. This may produce a slight degradation of the $1/p$ tendency expected for a `robust' design. Luckily, this is usually  compensated by the reduction in the number of circuits required to run the protocol. This results in a much more favorable scaling with the total number of resources when opting for the near-minimal design. 

In contrast to this, our physically-motivated approach directly provides a minimal experimental design, as the number of circuits needed to run long-sequence GST equals the theoretical lower bound. Nonetheless, to get a more robust design for the optimization, we supplement this minimal set with a few circuits, obtaining a total of just $N_{\mathrm{c}}=12$ two-qubit circuits. We do not appreciate a degradation of the $1/p$ tendency when employing this (near-) minimal set, as we do not need to relax AC conditions. This compact design is a direct consequence of the channel structure we derived in previous subsection. First, it is straightforward to verify that the parameters in our model are linearly amplified with gate repetition. Thus, we conclude that the LS gate on its own is a reliable germ for long-sequence GST, in the same way as we proved for single-qubit gates in our previous work. Hence, the base circuits containing the LS germ are 
described by the channel
\begin{equation}
\label{gate_krauss_to_p}
\mathcal{E}_{\mathrm{LS}}^{p}(\bullet)=\mathcal{E}_{\mathrm{N}}(p\Gamma_{\mathrm{d}}, p\Gamma_{\mathrm{th}})\circ \ee^{-\ii p\frac{\omega_0}{2}t_{g}(Z_{i_1}+Z_{i_2})}\ee^{-\ii p\theta_{LS}Z_{i_1}Z_{i_2}}(\bullet).
\end{equation}

The ultimate check for this enhancement of precision is the $1/p$ scaling of precision with depth which we reported in Fig.~\ref{fig: td_vs_p_mic_noisy}.

To introduce the idea which motivates next section, let us close the present one by pointing out a crucial aspect which we already mentioned in the introduction. Long-sequence GST assumes constant quantum channels for the gates, which are repeated a large number of times within the GST circuits  in order to  amplify all possible  error channels. Nonetheless, it is clear that the state resulting from the evolution in eq.~(\ref{sdd_evolution}) exhibits spin-phonon entanglement, unless trajectories from all phononic modes driving the gate are perfectly closed after the pulse. As we are assuming thermal motional errors in the LS gate, the condition required for the channel derivation, i.e., that the initial state is separable, cannot be satisfied if the entangling gate is applied sequentially. 
Therefore, if sympathetic cooling cannot be applied in between consecutive LS gates, the map that we derived in the previous subsection would only provide a correct description for the first  LS gate, making it unsuitable for long-sequence GST. Note that this is not a problem specific to our microscopic parameterization: any description of the LS gate assuming a constant map, even a fully general parameterization, will equally fail to correctly describe the successive applications of the phonon-mediated gate when the state of the phonon auxiliary degrees of freedom is not reset prior to each entangling gate. This connects to a limitation of
GST: it is unable to capture context-dependent dynamics \cite{PhysRevX.9.021045}, referred to as `non-Markovian' in the context of GST \cite{Nielsen2021gatesettomography}. In this work, however, we prefer to avoid this terminology, as non-constant or context-dependent quantum maps may also be described by a purely Markovian evolution. We note that a rigorous criterion for non-Markovianity of quantum evolutions  is formalized in terms of CP-divisibility of the quantum dynamical maps~\cite{PhysRevLett.103.210401, PhysRevLett.105.050403}, and that there can be context-dependent gate sequences that are still   CP-divisible, hence qualifying as Markovian. Aside from terminology, it is clear that the LS gate thermal error is context-dependent, as it depends on the LS history during the circuits. Therefore, standard GST may result in model violation and unprecise estimates if sympathetic cooling is not applied prior to each LS gate. However,  mid-circuit sympathetic cooling is still not available for many trapped-ion processors. We propose a second solution that does not require any  hardware-based improvement, but instead treats the map resulting from the successive gate applications as a whole. This leads to a central result of this work: using microscopic modeling, one can modify  the long-sequence protocol 
to allow for context-dependence in the gate sequences, such that one not only avoids model violation, but actually can achieve precisions surpassing standard long-sequence GST.

\section{\bf Microscopic modeling  of context-dependent GST}
\label{non-markovian_approach}

If no mid-circuit sympathetic cooling can be applied in between  pairs of LS gates, residual spin-phonon entanglement will accumulate, such that the gate error will depend on the previous LS-gate history, making the  quantum channel of the GST circuit context dependent. The phonons that mediate the gate are not a rapidly thermalizing environment, but instead a quantum data bus that can accumulate  motional fluctuations   leading to a progressive decay of the gate fidelities. This contrast the frequency noise, in which correlations of the magnetic-field fluctuations across different gates can be neglected for sufficiently large time between the gates, making the dephasing contributions effectively constant. 

\subsection{Analytical channel of sequential  light-shift gates}
\label{analytical derivation2}

In this section, we show that the microscopic modeling can be extended to  the composition of $p$ LS entangling gates, obtaining  a parametrized context-dependent  quantum channel. For the superoperators in Eq.~(\ref{complete_superoperator}) derived from evolutions acting trivially over the phononic space, the respective counterparts after $p$ gates simply read
\begin{equation}
\Gamma_{\rm d}\to p\Gamma_{\rm d},\quad
\mathcal{J}\!(t_g)\to p\!\mathcal{J}\!(t_g),\quad
\omega_0\to p\omega_0.
\end{equation}
In contrast, the  super-operator that accounts for motional errors $\mathcal{K}(\mathrm{tr}_{\mathrm{ph}}\{\mathcal{E}^{(p)}_{\mathrm{D}}\})$ associated to  the phase-space trajectories of the normal modes after $p$ consecutive gates, will depend  on the history of motional excitations during the concatenated qubit-phonon couplings. Let us revisit our previous derivation and extend it to this new scenario, which requires a microscopic evaluation of  
\begin{equation}
\mathcal{K}(\mathrm{tr}_{\mathrm{ph}}\{\mathcal{E}^{(p)}_{\mathrm{D}}(\rho)\})=\mathrm{tr}_{\mathrm{ph}}\{({U_{\mathrm{D}}})^p\rho {(U_{\mathrm{D}}}^{\dagger})^{p}\}.
\end{equation}
Following similar steps as in our discussion  of Sec.~\ref{evolution operator}, we find
\begin{equation}
\mathrm{tr_{\mathrm{ph}}}\{\mathcal{E}_{\mathrm{D}}^{(p)}(\rho)\} =\sum_{\boldsymbol{\sigma},\boldsymbol{\gamma}}\rho_{\boldsymbol{\sigma}, \boldsymbol{\gamma}}\ket{\boldsymbol{\sigma}}\bra{\boldsymbol{\gamma}}\times\mathcal{A},
\end{equation}
where we have introduced
\begin{equation}
\mathcal{A}=\prod_m\sum_{n_m}p_{m,n}\bra{n_m}\left(\mathcal{D}^{\dagger}(\boldsymbol{\gamma})\ee^{\ii H_0t_g}\right)^{\!\!p}\left(\ee^{-\ii H_0t_g}\mathcal{D}(\boldsymbol{\sigma})\right)^{\!\!p}\ket{n_m}.
\end{equation}
and where we recall that the operators $\mathcal{D}^{\dagger}(\boldsymbol{\sigma})$ involve a product of phase-space spin-dependent  displacements~\eqref{eq_prod_displacements}.
As evident from this expression, we now need to consider the accumulation of rotations induced by the
free evolution, as it does not commute with the sequential displacements, and will affect the concatenated trajectory in phase space.
Using  the relation for displacement operators $\ee^{-\ii H_0t_g}D_m(\alpha)\ee^{\ii H_0 t_g}= D_m(\alpha\ee^{-\ii\omega_{z,m} t_g})$~\cite{gerry2023introductory}, and the previous expression for their respective products~\eqref{eq:prod_disp},  we obtain
\begin{equation}
\label{expectation_displacement}
\mathcal{A}=\prod_m\ee^{\eta_m^{\boldsymbol{\sigma}, \boldsymbol{\gamma}}(p)}\sum_{n_m}p_{m,n}\bra{n_m}D_m(\Phi^{\boldsymbol{\sigma},\boldsymbol{\gamma}}_m\textstyle\sum_{l=0}^{p-1}\ee^{\ii l\omega_{z,m}t_g})\ket{n_m},
\end{equation}
where we have introduced

\begin{equation}
\begin{split}
&\eta_m^{\boldsymbol{\sigma}, \boldsymbol{\gamma}}(p)=\ii p\Upsilon_m^{\boldsymbol{\sigma}, \boldsymbol{\gamma}}-\ii2\phi_{i_1,m}(t_g)\phi^*_{i_2,m}(t_g)\times\\&\times((-1)^{\sigma_{i_1}\oplus\sigma_{i_2}}-(-1)^{\gamma_{i_1}\oplus\gamma_{i_2}})\textstyle\sum_{l=1}^{p-1}(p-l)\sin(l\omega_{z,m}t_g),
\end{split}
\end{equation}
and where we made use of Eq.~\eqref{eq:S}.
Again, we can restrict ourselves to the case $N=2$ where the unclosed trajectory is that of the breathing  mode, such that the thermal error channel after $p$ consecutive LS gates can be expressed in the natural representation as follows
\begin{widetext}
\begin{equation}
\mathcal{K}(\mathrm{tr_{\mathrm{ph}}}\{\mathcal{E}_{\mathrm{D}}^{(p)}\}) =\mathrm{diag}\big\{1,\ee^{\ii\zeta_{\mathrm{b}}^p-\Gamma_{\mathrm{th}}^p},\ee^{\ii\zeta_{\mathrm{b}}^p-\Gamma_{\mathrm{th}}^p},1,\ee^{-\ii\zeta_{\mathrm{b}}^p-\Gamma_{\mathrm{th}}^p},1,\ee^{-4\Gamma_{\mathrm{th}}^p},\ee^{-\ii\zeta_{\mathrm{b}}^p-\Gamma_{\mathrm{th}}^p}, \ee^{-\ii\zeta_{\mathrm{b}}^p-\Gamma_{\mathrm{th}}^p},\ee^{4\Gamma_{\mathrm{th}}^p},1,\ee^{-\ii\zeta_{\mathrm{b}}^p-\Gamma_{\mathrm{th}}^p},1,\ee^{\ii\zeta_{\mathrm{b}}^p-\Gamma_{\mathrm{th}}^p},\ee^{\ii\zeta_{\mathrm{b}}^p-\Gamma_{\mathrm{th}}^p},1\big\}.
\end{equation}
\end{widetext}

In this expression, we observe that the $p=1$ thermal rates~\eqref{eq:thermal _rate} are now rescaled by an oscillatory factor  due to the relative orientation of the consecutive state-dependent forces
\begin{equation}
\label{effective_gamma}
\begin{split}
\Gamma_{\mathrm{th}}^p=\Gamma_{\mathrm{th}}\frac{\cos(p\omega_{z,2}t_g)-1}{\cos(\omega_{z,2}t_g)-1}.
\end{split}
\end{equation}
In addition, we get the additional term
\begin{equation}
\label{eq:phase_imt}
\zeta_{\mathrm{b}}^p = -\frac{1}{2}\left(\frac{\mid\tilde{\Omega}_{L}\mid\eta_{\alpha,m}}{\delta_m}\right)^2\frac{p\sin(\omega_{z,2}t_g)-\sin(p\omega_{z,2}t_g)}{\cos(\omega_{z,2}t_g)-1}.
\end{equation}

For the complete quantum channel describing the LS sequence, we find that the equivalent of Eq.~(\ref{gate_krauss}) reads
\begin{equation}
\label{gate_krauss_sequence}
\mathcal{E}_{\mathrm{LS}}^{(p)}(\bullet)= \mathcal{E}_{\mathrm{N}}\left(p\Gamma_{\mathrm{d}}, \Gamma_{\mathrm{th}}^p\right)\circ \ee^{-\ii p\frac{\omega_0}{2}t_g(Z_{i_1}+Z_{i_2})}\ee^{-\ii \left(p\theta_{\rm LS} + \frac{\zeta_{\mathrm{b}}^p}{2}\right)Z_{i_1}Z_{i_2}}(\bullet).
\end{equation}

This evolution may also be written in terms of the intermediate individual  LS maps as follows
\begin{equation}
\label{gate_sequence}
\mathcal{E}_{\mathrm{LS}}^{(p)} = \mathcal{E}_{\mathrm{LS}}^{(p-1, p)}\circ\dots\circ\mathcal{E}_{\mathrm{LS}}^{(r-1, r)}\circ\dots\circ\mathcal{E}_{\mathrm{LS}}^{(0,1)},\,\,\,\,r<p,
\end{equation}
where we have defined the intermediate maps
\begin{equation}
\label{intermediate}
\mathcal{E}_{\mathrm{LS}}^{(r-1, r)}=\mathcal{E}_{\mathrm{LS}}^{(r)}\circ(\mathcal{E}_{\mathrm{LS}}^{(r-1)})^{-1}.
\end{equation}

There are some relevant points to note here. The first is that the phase $\zeta_{\mathrm{b}}^p$~\eqref{eq:phase_imt} breaks the symmetries that lead to an anti-correlated dephasing channel for $p=1$, introducing a new contribution to   the effective coupling strength 
$\mathcal{J}\!(pt_g)\to\tilde{\mathcal{J}}\!(pt_g)=\mathcal{J}\!(pt_g)+\zeta_{\mathrm{b}}^p/2$.
As a consequence, as $p$ increases, the gate deviates periodically from the maximally-entanglement conditions. Second, we observe that the rescaling factor of the thermal rate  is periodic in $p$, maximizing at the values $p=kp_{max}$ with $k \in \mathbb{N}$, such that the quantity 
\begin{equation}
g(p)=\frac{1}{2}\left(\frac{p\omega_{z,2}t_g}{\pi}-1\right)
\end{equation}
is as close as possible to an integer, i.e. $p_{max}=\mathrm{argmin}_p(|g(p)-\mathrm{round}(g(p))|$.} This contrasts the linear scaling described for the context-independent version of the map Eq.~(\ref{gate_krauss_to_p}), which enabled us to perform long-sequence GST and obtain the usual $1/p$ scaling of precision with depth (see Fig.~\ref{fig: td_vs_p_mic_noisy}). From this perspective, one may naively discard  the concatenated context-dependent LS gate as a valid germ  that can increase precision in long-sequence GST. Nonetheless, upon a closer inspection, we note that the periodic rescaling factor can actually behave as a monotonically amplification function for a range of depths $p$,  provided one chooses a specific laser beat-note $\Delta\omega_L$, and thus a certain gate time $t_g$. Moreover, one can exploit the non-linear dependence in this regime to get a super-linear amplification of the thermal noise parameters and get estimations which, as we will now show, surpass the  precision of the standard long-sequence GST. 

\begin{figure}
  \centering
  \includegraphics[width=1\columnwidth]{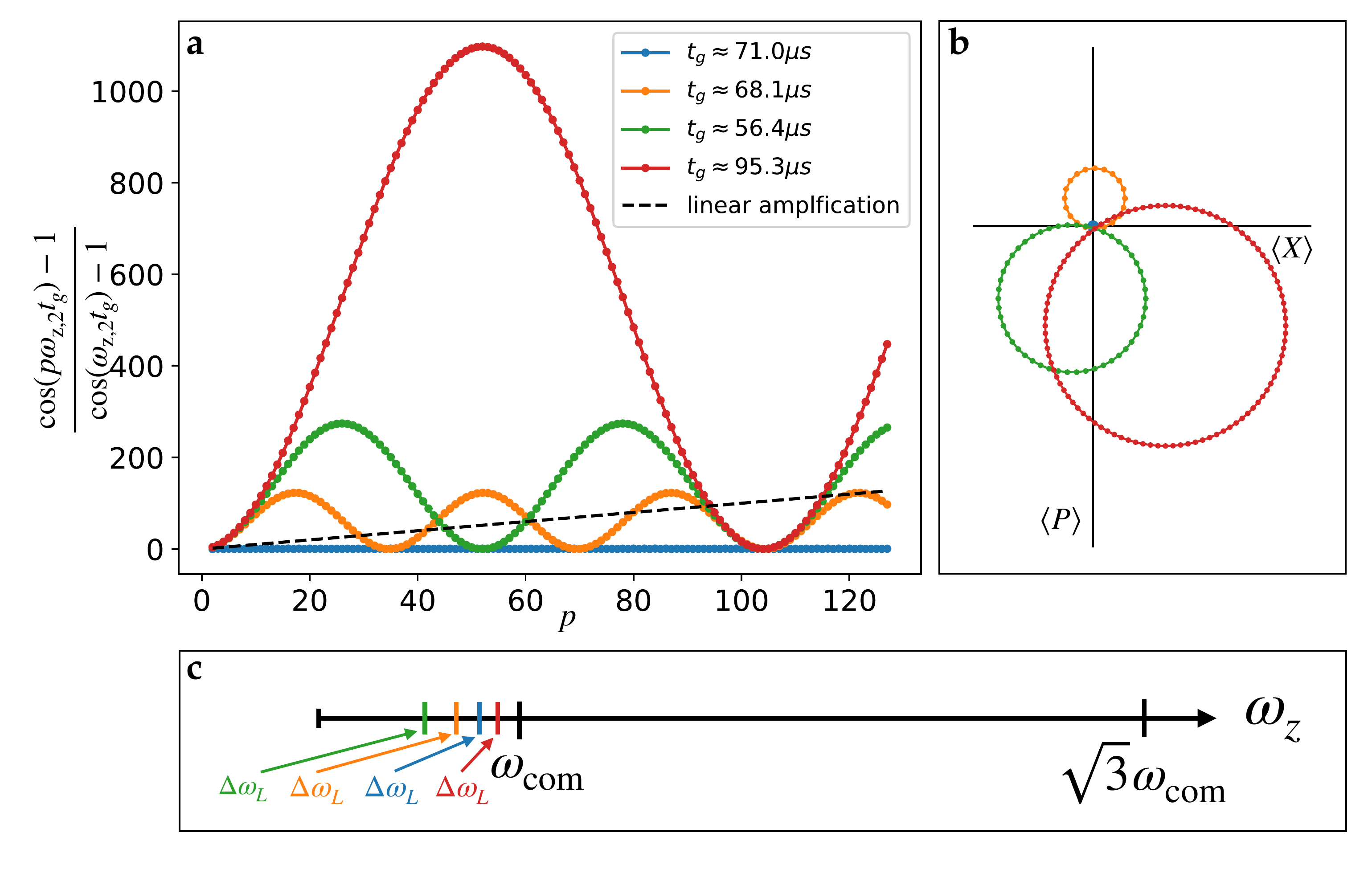}
  \caption{\textbf{Adjustment of the gate time $t_g$ for super-linear error amplification. a: } amplification factor for different gate times. While the thermal parameter $\Gamma_{\mathrm{th}}^p$ exhibits a periodic behavior, by adjusting the laser beat-note $\Delta\omega_L$, an effective super-linear amplification is observed in the range of depths relevant to long-sequence GST. \textbf{b:} phase-space trajectory precession. Residual displacement accumulation at the unclosed breathing-mode phase-space trajectory induces precession. This periodic accumulation of the gate's geometric phase enables the amplification of thermal error. Each point represents the final value of the trajectory for each of the $p$ gate executions. \textbf{c:} Different laser beat-note values correspond to the different gate times considered. Gate time is determined by the detuning $\delta_{\mathrm{com}}^z$, such that $t_g=2\pi/\delta_{\mathrm{com}}^z$ and the com-mode trajectory is always closed. Different $\Delta\omega_L$ values produce different amplifications of the thermal error with
  $p$, $\Delta\omega_L$ values are not at scale.}
  \label{fig: amplifications}
\end{figure}

The aforementioned super-linear amplification of the thermal parameter can be appreciated in Fig.~\ref{fig: amplifications}\textbf{(a)}, in which different amplifications are shown for different gate times at increasing circuit depths. Given the value of the breathing mode $\omega_{z,2}$, one can adjust the gate time to amplify the thermal parameter in a range of depths that are not dominated by noise, and are therefore useful to the long-sequence protocol. The periodic behavior of $\Gamma_{\mathrm{th}}^p$ is consistent with the precession of the composite breathing-mode phase-space trajectory, which is a consequence of the residual displacement of the ion after each gate repetition. This precession is represented in Fig.~\ref{fig: amplifications}\textbf{(b)} for the different gate times considered. Fig.~\ref{fig: amplifications}\textbf{(c)} represents the different laser beat-note values chosen to achieve the amplifications at their respective gate times, which are completely determined by the relation $t_g = 2\pi/\delta_{\mathrm{com}}$ when the com-mode trajectory is closed. Notice that the selection of $\Delta\omega_L$ for optimal amplification is not in conflict with the conditions for the  maximally-entangling gate, as Eq.~(\ref{coupling_strenght}) is also sensitive to changes in the laser intensity that fixes the forces $\mathcal{F}^z_{i,m}$.

\subsection{Non-linear amplification, estimation sensitivity and the Fisher information matrix}

Let us now characterize quantitatively this sensitivity of the context-dependent GST. The Cramér-Rao bound \cite{eadie1973statistical} defines a lower bound for the error in the estimate of a set of parameters  $\hat{\boldsymbol{\theta}}$ with respect to their true value $\bar{\boldsymbol{\theta}}$, considering that one uses an unbiased estimator. This bound is expressed in terms of the Fisher Information (FI) matrix $I(\boldsymbol{\theta})$ as 
\begin{equation}
\label{CRB}
\Sigma(\hat{\boldsymbol{\theta}})\geq I^{-1}(\bar{\boldsymbol{\theta}}),
\end{equation}
where $\Sigma(\hat{\boldsymbol{\theta}})$ quantifies the precision by the covariance matrix of the estimate~\cite{rossi2018mathematical}. As noted in \cite{rudinger2023twoqubitgatesettomography}, using the negative log-likelihood $\mathcal{C}_{\rm ML}$ defined in (\ref{MLE_GST}), the FI matrix for each of the circuits employed for the long-sequence GST is
\begin{equation}
\label{FI_per_circuit}
I_c=N_{\mathrm{samples}}\sum_{\mu}^{\mathbb{M}}\left(\frac{1}{p_{\gamma_c, \mu}}(\nabla_{\boldsymbol{\theta}} p_{\gamma_c, \mu})({\nabla_{\boldsymbol{\theta}}} p_{\gamma_c, \mu})^T-H_{\gamma_c, \mu}\right),
\end{equation}
where $\nabla_{\boldsymbol{\theta}} p_{\gamma_c, \mu}$ and $H_{c,\gamma_c}$ denote the gradient and the Hessian of the observed probabilities $p_{\gamma_c, \mu}$ for each circuit 
\begin{equation}
\begin{split}
(\nabla_{\boldsymbol{\theta}} p_{\gamma_c, \mu})_i = \frac{\partial p_{\gamma_c, \mu}}{\partial\theta_i},\hspace{2ex}
(H_{c,\gamma_c})_{i,j} = \frac{\partial^2 p_{\gamma_c, \mu}}{\partial\theta_i\partial\theta_j}.
\end{split}
\end{equation}
It is also relevant to mention that, as the cost function $\mathcal{C}_{\rm ML}$ is additive with respect to the GST circuits, the FI is also additive~\cite{rossi2018mathematical}, meaning that its value for the entire GST design is described in terms of all circuits used $C$ by
\begin{equation}
I_{C}=\sum_{c\in C}I_c.
\end{equation}

\begin{figure}
  \centering
  \includegraphics[width=1\columnwidth]{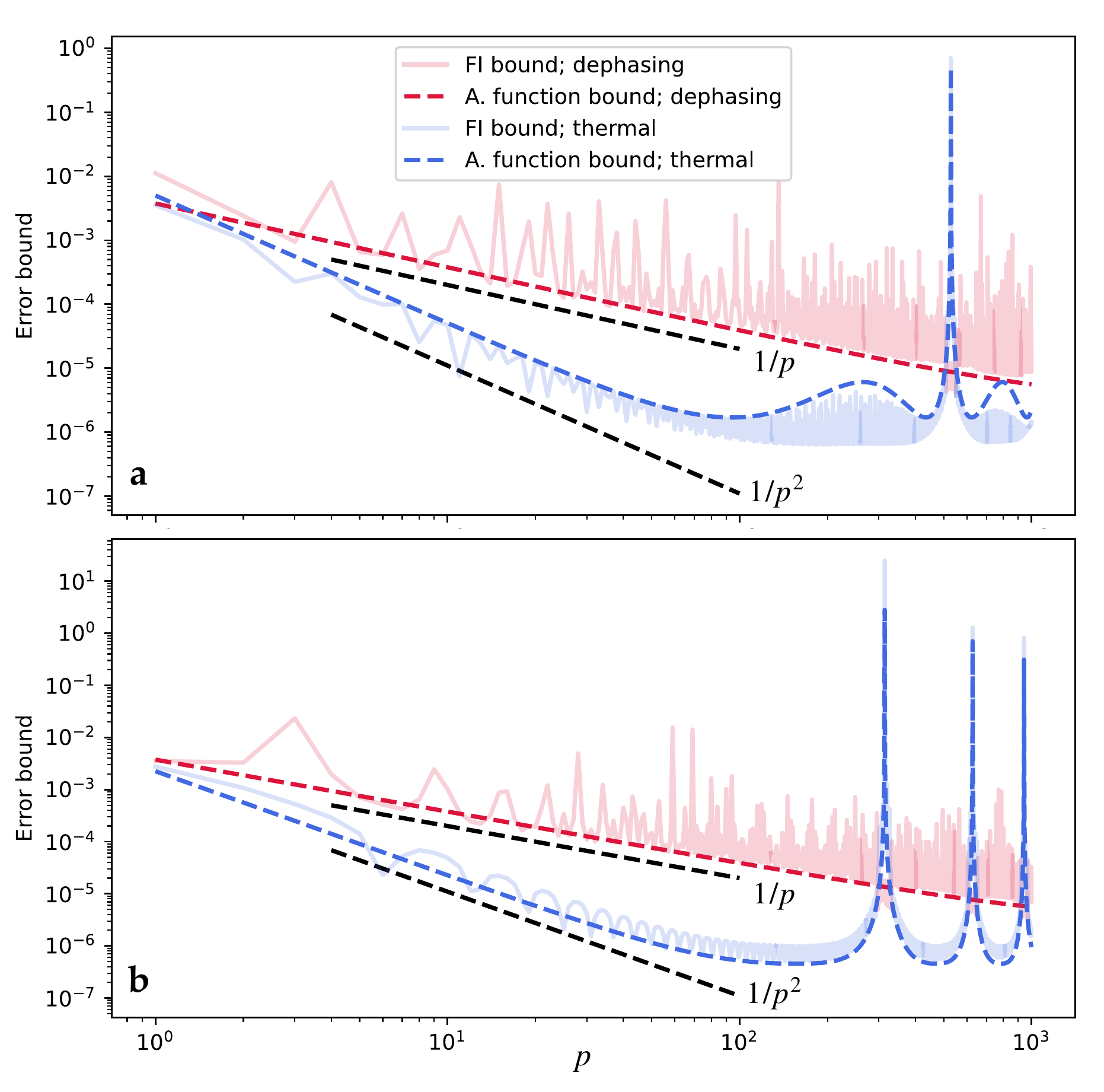}
  \caption{\textbf{Error bound estimations for GST.} We compute the Fisher Information matrix for the LS gate, assuming the initial state $\rho_0=\ket{++}\bra{++}$ and measuring on the same basis, $\ket{\pm}=\frac{1}{\sqrt{2}}(\ket{0}+\ket{1})$. This permits us to obtain an estimated scaling for error in both the thermal parameter (blue) and dephasing parameter (red). In lighter colors we plot the estimates resulting from the FI matrix spectrum, while dashed lines represent the estimations obtained from inverting the derivative of the corresponding amplification functions. We modify $\Delta\omega_L$ to observe different scaling behaviors, appreciating closing points for the breathing-mode trajectory at different depths. Guides to the aye showcasing different scaling tendencies are plotted in black-dashed lines, demonstrating $1/p$ and $1/p^2$ scalings for dephasing and thermal parameter estimates respectively. For the upper panel \textbf{a}, we set $t_g\approx97\mu s$, while for the lower panel \textbf{b} $t_g\approx95\mu s$. At the same time, we adjust $\mathcal{F}_{i,m}^{z}$ to satisfy the maximal-entanglement condition $J_{12}(t_g)= \pi/4$. Other parameters used are $c=2\times 10^7s^{-3}$, $\tau_c=5\times 10^{-4}s$ and $\bar{n}_{\mathrm{b}}=5$. }
  \label{fig: FI_error_bounds}
\end{figure}

To understand the non-linear amplification due to the consecutive phase-space displacements, let us momentarily switch off the SPAM  and single-qubit gate  errors, and then compute the FI matrix.
Motivated by the expected collective dephasing and Ramsey-like experiments~\cite{PhysRev.78.695, PhysRevB.72.134519, ramsey2, ramsey3}, we can focus on GST fiducial pairs defined by the initial state $\rho_0=\ket{++}\bra{++}$ and projective measurement on the basis $\ket{\pm}=(\ket{0}\pm\ket{1})/\sqrt{2}$. This constitutes a reduced experimental design which, as we show below, contains nonetheless all the information required to  enable a prediction of the GST precision.
We first compute the spectrum of the Fisher information matrix for this scheme. This can be achieved analytically by means of expression (\ref{FI_per_circuit}), after the calculation of the model probabilities. Then, motivated by the fact that $\mathrm{det}\Sigma(\hat{\boldsymbol{\theta}})$ is proportional to the volume enclosed by the covariance elliptical region \cite{varona2024lindbladlikequantumtomographynonmarkovian}, we use the spectrum of this matrix as an error bound for the two parameters under estimation, in this case $\Gamma_{\rm d}$ and $\Gamma_{\rm th}$. If we plot the square root of each of these eigenvalues as a function of the number of times $p$ the LS gate is repeated, we obtain the error bounds displayed in Fig.~\ref{fig: FI_error_bounds} with lighter colors. As can be concluded from this figure, the thermal parameter can be estimated with a precision that grows with  $1/p^2$, which is a substantial gain compared to the $1/p$ gain expected from long-sequence GST. On the other hand, the dephasing parameter $·\Gamma_{\rm d}$ does not show any  context dependence or non-linear sensitivity,  and leads to the typical $1/p$ scaling in the estimation precision.

Let us now present a different perspective that allows us to obtain even simpler analytical bounds for the GST precision. For a given point estimation problem in which one aims to estimate a single parameter $\bar{\theta}$, the estimate $\hat{\theta}$ will present an error $\epsilon_\theta$. We assume the existence of a function $f(\theta)$ such that the inverse function $f^{-1}\circ f(\theta)=\theta$ exists in a vicinity of the true value $[\bar{\theta}-\delta\theta,\bar{\theta}+\delta\theta]$. Then, if one can estimate $f(\bar{\theta})$ with error $\epsilon_f$, an estimate for the target parameter can be obtained by a simple propagation of errors~\cite{2016QMQM....5....2H}, namely 
\begin{equation}
\label{error_prop}
\theta = f^{-1}(f(\bar{\theta})+\epsilon_f)\approx\bar{\theta}+\frac{\epsilon_f}{f'(\theta)|_{\bar{\theta}}},
\end{equation}

For the long-sequence GST protocol of the previous section, we can identify a linear function amplifying the dependence of the signal with the noise parameter $\theta=\Gamma_{\rm th}$   linearly with the depth $f(\Gamma_{\rm th})=p\Gamma_{\rm th}$ , which translates into the $1/p$ scaling of the GST precision. Since, in the standard GST approach, one distributes measurement shots with  circuits of increasing depths to aid the non-linear minimization, this scaling translates into a linear improvement of the estimation precision which, as already noted, is sometimes referred to as a Heisenberg-like scaling.
On the other hand, from the perspective of quantum metrology, this scaling  with measurement shots is not fixed by the statistical inference, but by the way in which one has decided to face the non-linear minimization and distribute the measurement shots. As we have shown in Fig.~\ref{fig: td_vs_p_mic_noisy}, with our parametrization, there is actually no need to devote resources to the intermediate-depth circuits in the region where the signal-to-noise ratio is improved. Accordingly, the scaling with $1/p$ gets decoupled from the scaling with the number of shots, and the statistical estimation obeys the standard quantum limit (SQL) \cite{PhysRevLett.116.120801} consistent with the $\sqrt{N_{\rm samples}}$ scaling of the standard deviation  from Eqs.~(\ref{CRB})-(\ref{FI_per_circuit}).

For our context-dependent version of the GST, the corresponding function for the thermal parameter is
\begin{equation}
\label{amplification_function}
f(\Gamma_{\mathrm{th}})=\exp\left(\Gamma_{\mathrm{th}}\frac{\cos(p\omega_{z,2}t_g)-1}{\cos(\omega_{z,2}t_g)-1}\right).
\end{equation} 

The derivative of this function, governing the GST precision, is depicted  with a blue dashed line in Fig.~\ref{fig: FI_error_bounds}, showing a clear agreement with the FI predictions (dashed blue lines), and serving as a tight lower bound for the error when the sequences are sufficiently deep. As evident from these results, we emphasize again that the $1/p^2$ scaling of precision with depth arises in a regime in which the cumulative breathing-mode trajectories do not close. On the contrary, for sufficiently deep sequences, the precision of the end  point displayed in Fig.~\ref{fig: amplifications}{\bf (b)} can result in an eventual  closing of the breathing-mode trajectory. In this case, the signal becomes insensitive to the thermal noise parameter, and the estimation error  dramatically increases,  leading  to the  consecutive peaks displayed in Fig.~\ref{fig: FI_error_bounds}. 

For the dephasing parameter, a similar approximate treatment yields the corresponding precision estimates displayed by a light red  line in Fig.~\ref{fig: FI_error_bounds}. These estimates  lead to the usual $1/p$ scaling expected from the linear amplification, and again capture accurately the results obtained by calculating the full Fisher information matrix for each of the employed circuits (dashed red lines), serving as a tight lower bound when the sequence depth is sufficiently large. We note that the high variability of the FI estimates compared to the ones obtained with the amplification function is not surprising, as the FI matrix (\ref{FI_per_circuit}) incorporates crossed-parameter terms from second-order derivatives, and the complete operator in Eq.~(\ref{complete_superoperator}) presents oscillations from different rotation terms.

\subsection{Context-aware increase of  GST precision}

After having discussed the expected error bounds for the  LS-gate error parameters in this idealized situation in which the single-qubit gates and the SPAM are perfect, let us go back to the more realistic GST protocol where all components of the gate set~(\ref{gate_set}) are faulty. To proceed with our parametrised GST, we  first need to select the  range of depths required  to amplify the thermal parameter, which can be done by tuning the values of $\Delta\omega_L$ (see Fig.~\ref{fig: amplifications}). An important point of the estimation is that, instead of using long-sequence GST, which would require expressing all the intermediate channels as constant maps, we perform a single GST estimation of the complete channel after applying the $p$ gates, which only requires being able to efficiently parametrize the complete quantum channel of the faulty sequence Eqs.~(\ref{gate_krauss_sequence}-\ref{gate_sequence}). This results in the estimation of $\Gamma_{\mathrm{th}}^p$ with the standard shot-noise  precision $O(1/\sqrt{N_{\mathrm{samples}}})$. Then, we invert the amplification function (\ref{amplification_function}) to obtain $\Gamma_{\mathrm{th}}$ with error $O(1/(f'(\Gamma_{\mathrm{th}})\sqrt{N_{\mathrm{samples}}})$, which, as we just discussed, can yield an effective $1/p^2$ scaling.

\begin{figure}
  \centering
  \includegraphics[width=1\columnwidth]{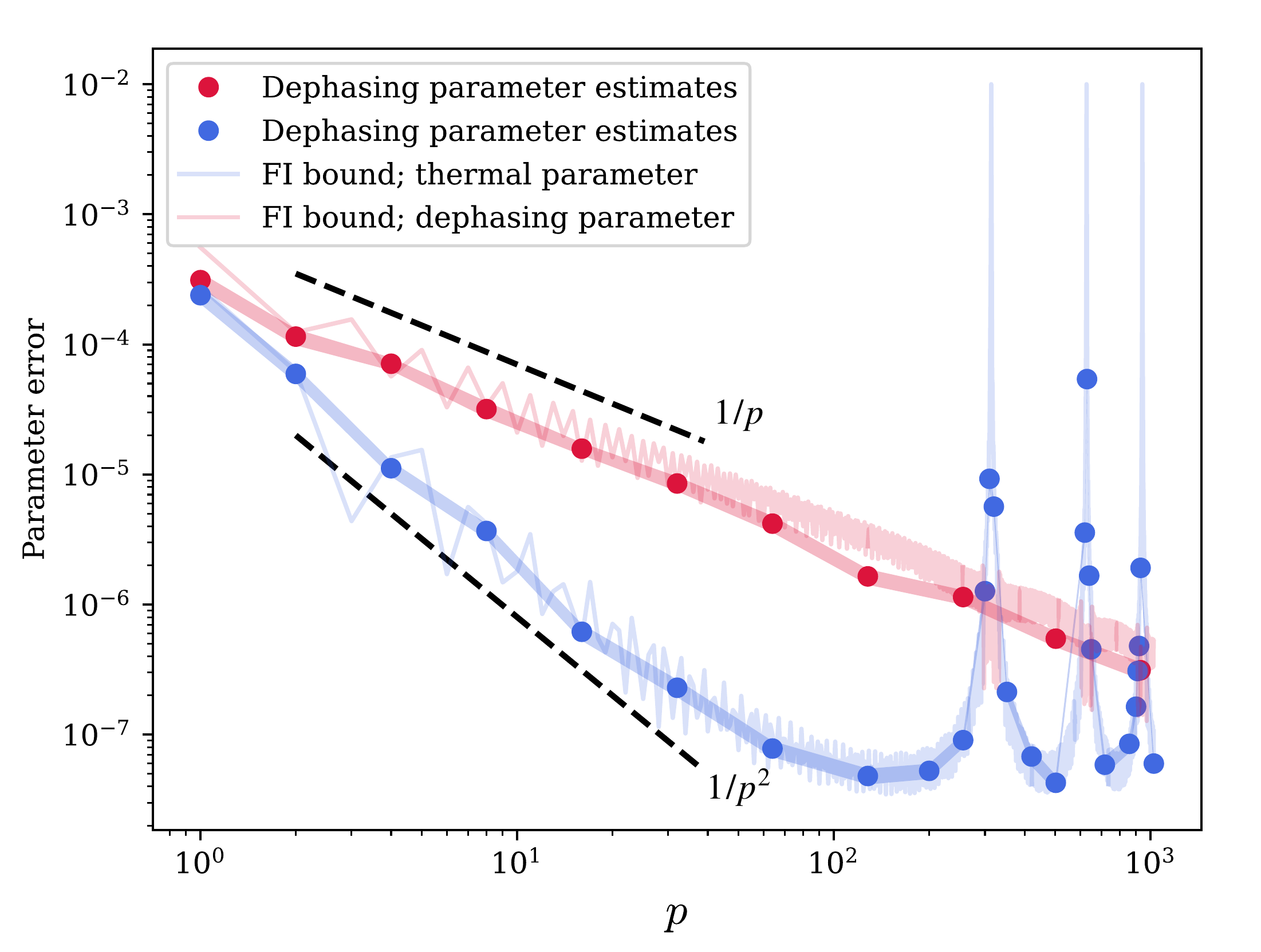}
  \caption{\textbf{Error of parameter estimates as a function of depth for the non-markovian approach.} Scaling of the error for both the dephasing parameter (red) and thermal parameter (blue) with depth, computed for the first LS gate execution of the sequence. Guides to the eye with the expected scalings for short depths are displayed for both parameters. The thermal parameter error exhibits $1/p^2$ scaling of precision with depth, while respecting the SQL with the number of samples $N_{\mathrm{samples}}$. After the accumulation of free rotations, this tendency is reverted, to the point that no information from the thermal parameter can be obtained when the phase-space trajectory closes. This manifests as divergences of the scaling, given by inverse of the derivative of eq.~(\ref{amplification_function}). With the aim of obtaining a reduced variance, for this plot, we sandwich the LS germ with all possible fiducial pairs. In shaded colors, we display the numerical FI error bounds obtained for these fiducial pairs. Parameters used for the simulations are $c=2\times 10^7s^{-3}$, $\tau_c=5\times 10^{-4}s$ and $\bar{n}_{\mathrm{b}}=5$. Each circuit is sampled $N_{\rm samples}=10^4$ times. Light shaded regions represent $95\%$ confidence intervals after averaging over $100$ GST estimations for each depth.}
  \label{fig:d_vs_p_non_markovian_parameteres}
\end{figure}

The previous scheme allows us to get an estimation of the quantum channel for a single noisy LS gate within the whole sequence with enhanced precision. {We note here that, even if the whole sequence must be described as a CPTP dynamical quantum map to be a physically-admissible evolution,  many of the intermediate quantum channels associated to a single gate in the sequence are actually not CP. Hence, attending to the CP-divisibility criteria ~\cite{PhysRevLett.103.210401, PhysRevLett.105.050403}, the complete evolution is not only context-dependent, but actually  non-Markovian. This can be easily checked by noting that the intermediate map $\mathcal{E}_{LS}^{(r-1, r)}$~\eqref{intermediate} is just a single-gate map with thermal parameter $\Gamma_{\mathrm{th}}^{r-1,r}=\Gamma_{\mathrm{th}}^r-\Gamma_{\mathrm{th}}^{r-1}$ and phase $\zeta_{\mathrm{b}}^{r-1,r}=\zeta_{\mathrm{b}}^r-\zeta_{\mathrm{b}}^{r-1}$. Given the oscillating nature of Eq.~(\ref{effective_gamma}), it is clear that the thermal parameter can attain negative values, such that it is not CP. We analyze this in terms of a non-Markovianity measure \cite{PhysRevLett.105.050403} in Appendix \ref{appendix_1}.} For single-qubit gates, which we treat as context-independent, we repeat each germ $p$ times to get long-sequence precision. In Fig.~\ref{fig:d_vs_p_non_markovian_parameteres}, we plot the estimation error for the thermal and dephasing parameters using  our context-dependent GST. In agreement with  the simplified SPAM-error free setting, the estimated dephasing parameter $\hat{\Gamma}_{\rm d}$ 
shows an error that scales with sequence depth as $1/p$ (red dots for GST data, red shaded line for the numerical FI estimates, and black dashed line for the inverse linear scaling). In contrast, the estimated thermal noise parameter $\hat{\Gamma}_{\rm th}$ shows an error with a faster decrease with sequence length (blue dots for GST data), showing an inverse quadratic scaling that agrees very well with the predicted error propagation formula (\ref{error_prop}) (blue dots for GST data, blue shaded line for the numerical FI estimates, and black dashed line for the inverse quadratic scaling). This scaling approximates the expected $1/p^2$ behavior for short to intermediate depths, and would stick to this limit in the absence of the cumulative free rotations $\ee^{\ii H_0 t_g}$ that change the directionality of the state-dependent forces. In fact, the periodic accumulation of these rotations eventually reverts this tendency, to the point that no information can be obtained about the thermal parameter when the phase-space trajectory of the breathing mode closes after $p$ gates, and the sensitivity to the thermal parameter is lost. This corresponds to the peaks already observed in Fig.~\ref{fig:d_vs_p_non_markovian_parameteres}, for which the exponent in function (\ref{amplification_function}) vanishes, and highlights the fact that a careful microscopic understanding of the phonon-mediated gates and the context-dependent noise is required to identify the optimal regime in which amplification yields lower estimation error.

Let us remark that  the estimation of the single-LS gate channel discussed above can be done for any specific gate in the sequence, as we know the explicit $p$-dependence of the map. Therefore, even if we get estimates bounded by the standard shot-noise scaling when aiming for the complete evolution comprising all $p$ gates, we can infer any of the $p$ individual operations with improved precision --it is only the concatenation of all $p$ evolutions that amplifies the overall error--. Consequently, we can extract how the effective LS channel evolves with the number of gates $p$. This is something beyond the reach of general GST, where context-dependent effects lead to model violation and, when significant, can even lead to a failure of the parametric estimation \cite{Nielsen2021gatesettomography}.
This results in a paradoxical scenario when aiming for full generality. While such an analysis is designed to avoid relying on specific physical assumptions —maintaining an `agnostic' perspective— the framework inherently depends on a stringent requirement: context-independence. This assumption undermines the generality it pursues. Contrary to this, our microscopic parameterization is based on making physical assumptions about a specific trapped-ion processor, while enabling us to capture these commonly-observed effects.

\begin{figure}
  \centering
  \includegraphics[width=1\columnwidth]{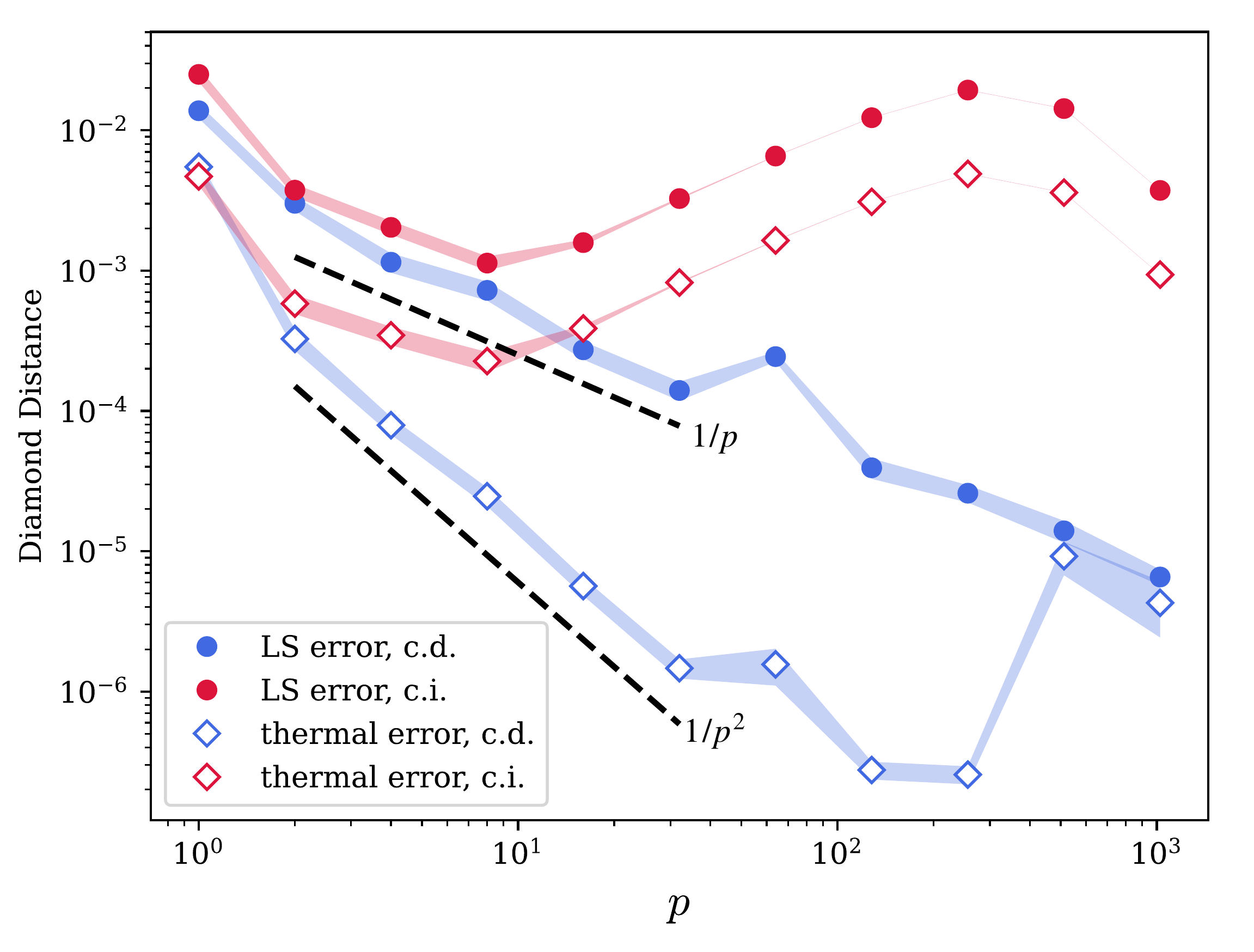}
  \caption{\textbf{Diamond distance for the light-shift channel as a function of depth for the first LS gate execution.} We run GST on simulated data and plot the diamond distance of the estimates as a function of depth. As we are mainly interested in the LS gate scaling, we do not include single-qubit gates in the computation of the diamond distances. We employ an experimental design in which only 12 two-qubit circuits are used in total. We plot the estimations obtained using both the context-dependent version of GST (blue solid markers) and the context-independent one (red solid markers). The $1/p$ scaling is depicted by a black guide to the eye, and is consistent with the results obtained for the context-dependent estimations. On the other hand, context-independent estimations only provide reliable results for short circuits, in which context-dependent effects are still small. Because context-dependence is encoded in the thermal parameter, we also plot the error in the estimation of this parameter with both GST models (empty blue and red markers). For the context-dependent estimation, the thermal parameter error scales as $1/p^2$ at short depths. This scaling is also depicted with a black guide to the eye. Parameters used for the simulations are $c=2\times 10^7s^{-3}$, $\tau_c=5\times 10^{-4}s$ and $\bar{n}_{\mathrm{b}}=5$. Each circuit is sampled $N_{\rm samples}=10^4$ times. Light shaded regions represent $95\%$ confidence intervals after averaging over $100$ GST estimations for each depth.}
  \label{fig:td_vs_p_non_markovian}
\end{figure}

Let us finish this section with some results that exhibit the scaling of the diamond distance with depth. This integrates the estimations of the two parameters previously discussed on a single plot. Fig.~\ref{fig:td_vs_p_non_markovian} contains these estimations for the context-dependent GST and the standard GST version employed in subsec.~\ref{context-independent-gst}, using the same experimental design with $N_{\mathrm{c}}=12$ circuits. As can be deduced from this figure, the context-dependent description allows for an estimation that is bounded by the usual $1/p$ scaling. Contrary, standard GST does only provide correct estimates for short circuits, where context-dependent effects are still small enough to provide a precise characterization by a fixed map. Nonetheless, as depth increases and these effects accumulate, long-sequence GST in its original incardination no longer provides accurate results. We note that this behavior is periodic, as when long enough depths are reached so that breathing mode trajectory closes, context-dependent effects are again small. Because context-dependent effects are fully encoded in the thermal parameter, we use empty markers to plot the error of the estimates for this parameter with both GST models.

\section{\bf{Conclusions and outlook}}
\label{conclusions}

In this work, we have presented a microscopic parameterization of entangling light-shift gates in trapped-ion devices which, in conjunction with our previous single-qubit work \cite{vinas2025microscopic}, allows as to construct a more efficient  GST  design at a significantly-reduced sampling cost as compared to a device-independent approach. Our results open the door to applying GST as a tool for real-time characterization and adaptive calibration of trapped-ion processors.

We have  demonstrated that a microscopic understanding of QIPs  paves the way for a promising new  direction in characterization: the self-consistent tomography of context-dependent dynamics, which are inaccessible to standard GST. We identified the primary source of context-dependence in trapped-ion LS gates as the accumulation of residual displacement from unclosed trajectories in phase space, which introduces thermal motional  noise. This accumulation depends on the residual entanglement between motional and internal degrees of freedom. Because such correlations are dependent on the LS gate history, also is thermal noise. We noted that the  LS sequences are not only context-dependent, but also non-Markovian, as intermediate maps are generally not CP.

Finally, we conducted a detailed analysis to illustrate the scaling of the estimation precision with the depth of the sequences used $p$. Notably, we showed that context-dependent effects can enhance precision beyond the usual $1/p$ scaling of long-sequence GST. This shows that memory effects, often seen as a limitation, can in fact be used as a resource. As an outlook, we believe it would be of great interest to validate these results on real trapped-ion hardware. For the purpose of this work, which is mainly focused on the LS gate characterization, we have not considered crosstalk affecting single-qubit gates. The study of different crosstalk models in GST \cite{PRXQuantum.2.040338} applied to our LS gate set --which might be determined by the specific processor design-- is left as the subject of future work. 

Another interesting avenue is related to the design of efficient error mitigation strategies for the noise studied in this work. {Probabilistic error cancellation} \cite{PhysRevLett.119.180509, PhysRevX.8.031027}, for instance, allows for the recovery of unbiased estimates of ideal expectation values by combining the results of noisy ones, provided that the noise model is learned in advance. However, the number of noisy circuit executions grows exponentially with circuit depth and the number of qubits. It has been shown that this overhead can be greatly reduced when an underlying noise structure can be assumed \cite{van_den_Berg_2023}. As our microscopic modelling offers such a great simplification over a maximal parameterization, we believe it might provide a useful structure for such mitigation routines. The estimation of error-free expectation values could also serve as a validation for the characterization itself, which is relevant because the true gate set would of course not be accessible.

\section{Code availability}
The code to reproduce our findings can be found in GitHub \cite{myrepo}.

\acknowledgements
The project leading to this publication has received funding from the US Army Research Office through Grant
No. W911NF-21-1-0007. We acknowledge support from
PID2021-127726NB-I00 (MCIU/AEI/FEDER, UE), from
the Grant IFT Centro de Excelencia Severo Ochoa CEX2020-001007-S, funded by MCIN/AEI/10.13039/501100011033, 
from the CSIC Research Platform on Quantum Technologies
PTI-001, and from the European Union’s Horizon Europe research and innovation programme under grant agreement No 101114305 (“MILLENION-SGA1” EU Project). Views and
opinions expressed are, however, those of the author(s) only
and do not necessarily reflect those of the European Union or
the European Commission. Neither the European Union nor
the granting authority can be held responsible for them.
\appendix
\section{Non-Markovianity measure for intermediate maps}
\label{appendix_1}
As  mentioned in the main text, our context-dependent LS-gate sequences, which are described a physical CPTP map,  can only be divided into the composition of intermediate maps that are, as we no show,  not necessarily CP. In view of the CP-divisibility criteria~\cite{PhysRevLett.103.210401, PhysRevLett.105.050403}, the  time evolution of the entire gate sequence can thus  be interpreted as a non-Markovian quantum dynamical map. For a given evolution represented by the channel $\mathcal{E}_{t,0}$, one can actually quantify the degree of non-Markovianity of the channel  by accounting for each infinitesimal non-CP contribution~\cite{PhysRevLett.105.050403}, leading to the definition of the non-Markovianity measure $\mathcal{N}_{\rm CP}=\int_0^t{\rm d}t'g(t')$, and
\beq
g(t')=\lim_{\Delta t\to 0}\frac{1}{\Delta t}\bigg(\big\lVert\mathcal{E}_{t'+\Delta t,t'}\otimes\mathbb{1}(\ket{\Phi}\!\bra{\Phi})\big\rVert_1-1\bigg),
\eeq
where, again, $\ket{\Phi}$ refers to the maximally entangled state on a doubled Hilbert space $\mathcal{H}\otimes\mathcal{H}$, and $\mathcal{E}_{t'+\Delta t,t'}=\mathcal{E}_{t'+\Delta t,0} \mathcal{E}^{-1}_{t',0}$ is the instantaneous intermediate map, which might not be completely positive. Typically, we will only be interested in the CP-divisibility of the intermediate LS gates with duration $t_g$, and not other partitions of the complete evolution. Thus, it is convenient to define a `discrete' version of the non-Markovianty measure. In analogy to $\mathcal{N}_{\rm CP}$, for a sequence of $p$ LS gates, we define $\mathcal{N}^d_{\rm CP}=\sum_{k=1}^pg^d(k)$, where
\begin{equation}
g^d(k)=\big\lVert\mathcal{E}_{LS}^{(k-1,k)}\otimes\mathbb{1}(\ket{\Phi}\!\bra{\Phi})\big\rVert_1-1.
\end{equation}

\begin{figure}
  \centering
  \includegraphics[width=1\columnwidth]{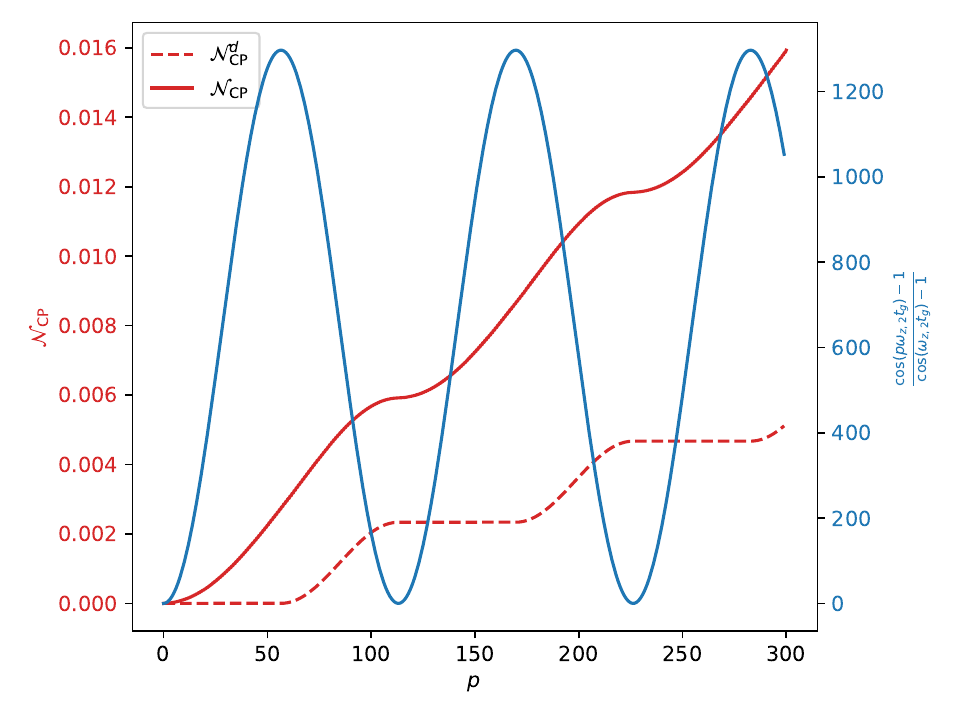}
  \caption{\textbf{Discrete and continuous non-Markovianity measures as a function of $p$.} We plot the non-Markovianity measures $\mathcal{N}^d_{\rm CP}$ (red dashed line) and $\mathcal{N}_{\rm CP}$ (red solid line) as a function of the sequence length. We also plot the amplification factor of the thermal parameter (blue line) in the same $p$ axis. The discrete measure accounts for the non-CP contributions stemming from the intermediate LS maps, increasing when the amplification factor decreases. On the contrary, the continuous version of the measure increases at each $p$, revealing that even those LS maps which are CP are non-Markovian.}
  \label{fig: non-markovianity}
\end{figure}

This quantity is useful to get an intuition on how the non-Markovianity of the individual context-dependent gates accumulates with the sequence length, augmenting the non-Markovian behavior of the complete evolution. In Figure~\ref{fig: non-markovianity}, we plot $\mathcal{N}^d_{\rm CP}$ with a dashed red line as a function of the number of gates $p$. In the same $p$ axis, we plot with a blue solid line the thermal-noise parameter enhancement in Eq.~(\ref{effective_gamma}). As one can  appreciate, when this effective parameter decreases, the intermediate maps Eq.~(\ref{intermediate}) display  an effective negative  rate, and the non-Markovianity measure increases. In contrast, for the regions in which $\Gamma_{\mathrm{th}}^{p}$ increases, intermediate maps are physical and the non-Markovianity remains constant. 

The coarse-grained measure $\mathcal{N}^d_{\rm CP}$, however,  does not quantify entirely how non-Markovian the complete evolution is, and  serves as a tool to identify which intermediate LS gates  are CP-divisible. This is because even those LS maps that are CP can be non-Markovian, as they may also contain non-CP  contributions from a finer-grain instantaneous evolutions. Thus, we now use $\mathcal{N}_{\rm CP}$ to rigorously measure non-Markovianity. This is plotted with a solid red line in Fig.~\ref{fig: non-markovianity}, which is now monotonically  increasing for  every $p$ value, except for those where $\Gamma_{\mathrm{th}}^{p}$ vanishes, corresponding to the closure in phase-space of the breathing-mode trajectory. Hence, the whole  LS-gate sequence evolution is non-Markovian, even if some of the intermediate LS gates are actually CP. 

This is consistent with the fact that, for an infinitesimal time increment $\Delta t$, the effective rate of the corresponding intermediate channel $\Gamma_{\mathrm{th}}({t,t+\Delta t})$ directly depends on $\phi_{i_1,\mathrm{2}}(t)$. Specifically, we assume a time $t$ such that $t=pt_g+\tau$ and $\tau<t_g$, so that the last LS gate has not been completed. Then, the total displacement for the $m$-th mode  $D_m$, which we obtained in Eq.~(\ref{expectation_displacement}) for the case of an integer number of gates, now reads
\begin{equation}
\label{total_displacement_appendix}
D_m\coloneq D_m\big(\Phi^{\boldsymbol{\sigma},\boldsymbol{\gamma}}_m(t_g)\textstyle\sum_{l=0}^{p-1}\ee^{\ii l\omega_{z,m}t_g}+\Phi^{\boldsymbol{\sigma},\boldsymbol{\gamma}}_m(\tau)\ee^{\ii p\omega_{z,m}t_g}\big).
\end{equation}

Hence, the effective intermediate rate reads
\begin{equation}
\label{int_eff_rate}
\Gamma_{\mathrm{th}}^{t,t+\delta t}=4\left( \bar{n}_{\mathrm{b}}+\half\right)\left(\xi(t+\delta t)-\xi(t)\right),
\end{equation}
wherein we have introduced 
\begin{equation}
\label{int_eff_rate_factor}
\xi(t)=|\textstyle\phi_{i_1,2}(t_g)\sum_{l=0}^{p-1}\ee^{\ii l\omega_{z,m}t_g}-\textstyle\phi_{i_1,2}(\tau)\ee^{\ii p\omega_{z,m}t_g}|^2.
\end{equation}

The function (\ref{int_eff_rate_factor}) oscillates  in the interval $0\leq\tau\leq t_g$, yielding negative rates for intermediate maps inside each LS gate. 

We emphasize that non-Markovian gates in which the non-Markovianity only takes place inside a gate pulse are perfectly accessible through GST. The differential aspect of the present work is the ability to capture context-dependent dynamics, i.e., the history-dependent behavior of the gates, and even a degree of strict non-Markovianity between different gates.

\newpage
\bibliographystyle{apsrev4-1}
\bibliography{biblio}

\begin{thebibliography}{119}%
\makeatletter
\providecommand \@ifxundefined [1]{%
 \@ifx{#1\undefined}
}%
\providecommand \@ifnum [1]{%
 \ifnum #1\expandafter \@firstoftwo
 \else \expandafter \@secondoftwo
 \fi
}%
\providecommand \@ifx [1]{%
 \ifx #1\expandafter \@firstoftwo
 \else \expandafter \@secondoftwo
 \fi
}%
\providecommand \natexlab [1]{#1}%
\providecommand \enquote  [1]{``#1''}%
\providecommand \bibnamefont  [1]{#1}%
\providecommand \bibfnamefont [1]{#1}%
\providecommand \citenamefont [1]{#1}%
\providecommand \href@noop [0]{\@secondoftwo}%
\providecommand \href [0]{\begingroup \@sanitize@url \@href}%
\providecommand \@href[1]{\@@startlink{#1}\@@href}%
\providecommand \@@href[1]{\endgroup#1\@@endlink}%
\providecommand \@sanitize@url [0]{\catcode `\\12\catcode `\$12\catcode `\&12\catcode `\#12\catcode `\^12\catcode `\_12\catcode `\%12\relax}%
\providecommand \@@startlink[1]{}%
\providecommand \@@endlink[0]{}%
\providecommand \url  [0]{\begingroup\@sanitize@url \@url }%
\providecommand \@url [1]{\endgroup\@href {#1}{\urlprefix }}%
\providecommand \urlprefix  [0]{URL }%
\providecommand \Eprint [0]{\href }%
\providecommand \doibase [0]{http://dx.doi.org/}%
\providecommand \selectlanguage [0]{\@gobble}%
\providecommand \bibinfo  [0]{\@secondoftwo}%
\providecommand \bibfield  [0]{\@secondoftwo}%
\providecommand \translation [1]{[#1]}%
\providecommand \BibitemOpen [0]{}%
\providecommand \bibitemStop [0]{}%
\providecommand \bibitemNoStop [0]{.\EOS\space}%
\providecommand \EOS [0]{\spacefactor3000\relax}%
\providecommand \BibitemShut  [1]{\csname bibitem#1\endcsname}%
\let\auto@bib@innerbib\@empty
\bibitem [{\citenamefont {Montanaro}(2016)}]{montanaro2016quantum}%
  \BibitemOpen
  \bibfield  {author} {\bibinfo {author} {\bibfnamefont {A.}~\bibnamefont {Montanaro}},\ }\href {\doibase 10.1038/npjqi.2015.23} {\bibfield  {journal} {\bibinfo  {journal} {npj Quantum Information}\ }\textbf {\bibinfo {volume} {2}} (\bibinfo {year} {2016}),\ 10.1038/npjqi.2015.23}\BibitemShut {NoStop}%
\bibitem [{\citenamefont {Temme}\ \emph {et~al.}(2017)\citenamefont {Temme}, \citenamefont {Bravyi},\ and\ \citenamefont {Gambetta}}]{PhysRevLett.119.180509}%
  \BibitemOpen
  \bibfield  {author} {\bibinfo {author} {\bibfnamefont {K.}~\bibnamefont {Temme}}, \bibinfo {author} {\bibfnamefont {S.}~\bibnamefont {Bravyi}}, \ and\ \bibinfo {author} {\bibfnamefont {J.~M.}\ \bibnamefont {Gambetta}},\ }\href {\doibase 10.1103/PhysRevLett.119.180509} {\bibfield  {journal} {\bibinfo  {journal} {Phys. Rev. Lett.}\ }\textbf {\bibinfo {volume} {119}},\ \bibinfo {pages} {180509} (\bibinfo {year} {2017})}\BibitemShut {NoStop}%
\bibitem [{\citenamefont {Nation}\ \emph {et~al.}(2021)\citenamefont {Nation}, \citenamefont {Kang}, \citenamefont {Sundaresan},\ and\ \citenamefont {Gambetta}}]{PRXQuantum.2.040326}%
  \BibitemOpen
  \bibfield  {author} {\bibinfo {author} {\bibfnamefont {P.~D.}\ \bibnamefont {Nation}}, \bibinfo {author} {\bibfnamefont {H.}~\bibnamefont {Kang}}, \bibinfo {author} {\bibfnamefont {N.}~\bibnamefont {Sundaresan}}, \ and\ \bibinfo {author} {\bibfnamefont {J.~M.}\ \bibnamefont {Gambetta}},\ }\href {\doibase 10.1103/PRXQuantum.2.040326} {\bibfield  {journal} {\bibinfo  {journal} {PRX Quantum}\ }\textbf {\bibinfo {volume} {2}},\ \bibinfo {pages} {040326} (\bibinfo {year} {2021})}\BibitemShut {NoStop}%
\bibitem [{\citenamefont {Yang}\ \emph {et~al.}(2022)\citenamefont {Yang}, \citenamefont {Raymond},\ and\ \citenamefont {Uno}}]{PhysRevA.106.012423}%
  \BibitemOpen
  \bibfield  {author} {\bibinfo {author} {\bibfnamefont {B.}~\bibnamefont {Yang}}, \bibinfo {author} {\bibfnamefont {R.}~\bibnamefont {Raymond}}, \ and\ \bibinfo {author} {\bibfnamefont {S.}~\bibnamefont {Uno}},\ }\href {\doibase 10.1103/PhysRevA.106.012423} {\bibfield  {journal} {\bibinfo  {journal} {Phys. Rev. A}\ }\textbf {\bibinfo {volume} {106}},\ \bibinfo {pages} {012423} (\bibinfo {year} {2022})}\BibitemShut {NoStop}%
\bibitem [{\citenamefont {Cai}\ \emph {et~al.}(2023)\citenamefont {Cai}, \citenamefont {Babbush}, \citenamefont {Benjamin}, \citenamefont {Endo}, \citenamefont {Huggins}, \citenamefont {Li}, \citenamefont {McClean},\ and\ \citenamefont {O'Brien}}]{RevModPhys.95.045005}%
  \BibitemOpen
  \bibfield  {author} {\bibinfo {author} {\bibfnamefont {Z.}~\bibnamefont {Cai}}, \bibinfo {author} {\bibfnamefont {R.}~\bibnamefont {Babbush}}, \bibinfo {author} {\bibfnamefont {S.~C.}\ \bibnamefont {Benjamin}}, \bibinfo {author} {\bibfnamefont {S.}~\bibnamefont {Endo}}, \bibinfo {author} {\bibfnamefont {W.~J.}\ \bibnamefont {Huggins}}, \bibinfo {author} {\bibfnamefont {Y.}~\bibnamefont {Li}}, \bibinfo {author} {\bibfnamefont {J.~R.}\ \bibnamefont {McClean}}, \ and\ \bibinfo {author} {\bibfnamefont {T.~E.}\ \bibnamefont {O'Brien}},\ }\href {\doibase 10.1103/RevModPhys.95.045005} {\bibfield  {journal} {\bibinfo  {journal} {Rev. Mod. Phys.}\ }\textbf {\bibinfo {volume} {95}},\ \bibinfo {pages} {045005} (\bibinfo {year} {2023})}\BibitemShut {NoStop}%
\bibitem [{\citenamefont {Kim}\ \emph {et~al.}(2023)\citenamefont {Kim}, \citenamefont {Eddins}, \citenamefont {Anand}, \citenamefont {Wei}, \citenamefont {van~den Berg}, \citenamefont {Rosenblatt}, \citenamefont {Nayfeh}, \citenamefont {Wu}, \citenamefont {Zaletel}, \citenamefont {Temme},\ and\ \citenamefont {Kandala}}]{kim2023evidence}%
  \BibitemOpen
  \bibfield  {author} {\bibinfo {author} {\bibfnamefont {Y.}~\bibnamefont {Kim}}, \bibinfo {author} {\bibfnamefont {A.}~\bibnamefont {Eddins}}, \bibinfo {author} {\bibfnamefont {S.}~\bibnamefont {Anand}}, \bibinfo {author} {\bibfnamefont {K.~X.}\ \bibnamefont {Wei}}, \bibinfo {author} {\bibfnamefont {E.}~\bibnamefont {van~den Berg}}, \bibinfo {author} {\bibfnamefont {S.}~\bibnamefont {Rosenblatt}}, \bibinfo {author} {\bibfnamefont {H.}~\bibnamefont {Nayfeh}}, \bibinfo {author} {\bibfnamefont {Y.}~\bibnamefont {Wu}}, \bibinfo {author} {\bibfnamefont {M.}~\bibnamefont {Zaletel}}, \bibinfo {author} {\bibfnamefont {K.}~\bibnamefont {Temme}}, \ and\ \bibinfo {author} {\bibfnamefont {A.}~\bibnamefont {Kandala}},\ }\href {\doibase 10.1038/s41586-023-06096-3} {\bibfield  {journal} {\bibinfo  {journal} {Nature}\ }\textbf {\bibinfo {volume} {618}},\ \bibinfo {pages} {500} (\bibinfo {year} {2023})}\BibitemShut {NoStop}%
\bibitem [{\citenamefont {Calderbank}\ and\ \citenamefont {Shor}(1996)}]{PhysRevA.54.1098}%
  \BibitemOpen
  \bibfield  {author} {\bibinfo {author} {\bibfnamefont {A.~R.}\ \bibnamefont {Calderbank}}\ and\ \bibinfo {author} {\bibfnamefont {P.~W.}\ \bibnamefont {Shor}},\ }\href {\doibase 10.1103/PhysRevA.54.1098} {\bibfield  {journal} {\bibinfo  {journal} {Phys. Rev. A}\ }\textbf {\bibinfo {volume} {54}},\ \bibinfo {pages} {1098} (\bibinfo {year} {1996})}\BibitemShut {NoStop}%
\bibitem [{\citenamefont {Steane}(1996)}]{PhysRevLett.77.793}%
  \BibitemOpen
  \bibfield  {author} {\bibinfo {author} {\bibfnamefont {A.~M.}\ \bibnamefont {Steane}},\ }\href {\doibase 10.1103/PhysRevLett.77.793} {\bibfield  {journal} {\bibinfo  {journal} {Phys. Rev. Lett.}\ }\textbf {\bibinfo {volume} {77}},\ \bibinfo {pages} {793} (\bibinfo {year} {1996})}\BibitemShut {NoStop}%
\bibitem [{\citenamefont {Terhal}(2015)}]{RevModPhys.87.307}%
  \BibitemOpen
  \bibfield  {author} {\bibinfo {author} {\bibfnamefont {B.~M.}\ \bibnamefont {Terhal}},\ }\href {\doibase 10.1103/RevModPhys.87.307} {\bibfield  {journal} {\bibinfo  {journal} {Rev. Mod. Phys.}\ }\textbf {\bibinfo {volume} {87}},\ \bibinfo {pages} {307} (\bibinfo {year} {2015})}\BibitemShut {NoStop}%
\bibitem [{\citenamefont {Nielsen}\ and\ \citenamefont {Chuang}(2010)}]{Nielsen_Chuang_2010}%
  \BibitemOpen
  \bibfield  {author} {\bibinfo {author} {\bibfnamefont {M.~A.}\ \bibnamefont {Nielsen}}\ and\ \bibinfo {author} {\bibfnamefont {I.~L.}\ \bibnamefont {Chuang}},\ }\href@noop {} {\emph {\bibinfo {title} {Quantum Computation and Quantum Information: 10th Anniversary Edition}}}\ (\bibinfo  {publisher} {Cambridge University Press},\ \bibinfo {year} {2010})\BibitemShut {NoStop}%
\bibitem [{\citenamefont {Eisert}\ \emph {et~al.}(2020)\citenamefont {Eisert}, \citenamefont {Hangleiter}, \citenamefont {Walk}, \citenamefont {Roth}, \citenamefont {Markham}, \citenamefont {Parekh}, \citenamefont {Chabaud},\ and\ \citenamefont {Kashefi}}]{Eisert2020}%
  \BibitemOpen
  \bibfield  {author} {\bibinfo {author} {\bibfnamefont {J.}~\bibnamefont {Eisert}}, \bibinfo {author} {\bibfnamefont {D.}~\bibnamefont {Hangleiter}}, \bibinfo {author} {\bibfnamefont {N.}~\bibnamefont {Walk}}, \bibinfo {author} {\bibfnamefont {I.}~\bibnamefont {Roth}}, \bibinfo {author} {\bibfnamefont {D.}~\bibnamefont {Markham}}, \bibinfo {author} {\bibfnamefont {R.}~\bibnamefont {Parekh}}, \bibinfo {author} {\bibfnamefont {U.}~\bibnamefont {Chabaud}}, \ and\ \bibinfo {author} {\bibfnamefont {E.}~\bibnamefont {Kashefi}},\ }\href {\doibase 10.1038/s42254-020-0186-4} {\bibfield  {journal} {\bibinfo  {journal} {Nature Reviews Physics}\ }\textbf {\bibinfo {volume} {2}},\ \bibinfo {pages} {382} (\bibinfo {year} {2020})}\BibitemShut {NoStop}%
\bibitem [{\citenamefont {Kliesch}\ and\ \citenamefont {Roth}(2021)}]{PRXQuantum.2.010201}%
  \BibitemOpen
  \bibfield  {author} {\bibinfo {author} {\bibfnamefont {M.}~\bibnamefont {Kliesch}}\ and\ \bibinfo {author} {\bibfnamefont {I.}~\bibnamefont {Roth}},\ }\href {\doibase 10.1103/PRXQuantum.2.010201} {\bibfield  {journal} {\bibinfo  {journal} {PRX Quantum}\ }\textbf {\bibinfo {volume} {2}},\ \bibinfo {pages} {010201} (\bibinfo {year} {2021})}\BibitemShut {NoStop}%
\bibitem [{\citenamefont {Gebhart}\ \emph {et~al.}(2023)\citenamefont {Gebhart}, \citenamefont {Santagati}, \citenamefont {Gentile}, \citenamefont {Gauger}, \citenamefont {Craig}, \citenamefont {Ares}, \citenamefont {Banchi}, \citenamefont {Marquardt}, \citenamefont {Pezz{\`e}},\ and\ \citenamefont {Bonato}}]{Gebhart2023}%
  \BibitemOpen
  \bibfield  {author} {\bibinfo {author} {\bibfnamefont {V.}~\bibnamefont {Gebhart}}, \bibinfo {author} {\bibfnamefont {R.}~\bibnamefont {Santagati}}, \bibinfo {author} {\bibfnamefont {A.~A.}\ \bibnamefont {Gentile}}, \bibinfo {author} {\bibfnamefont {E.~M.}\ \bibnamefont {Gauger}}, \bibinfo {author} {\bibfnamefont {D.}~\bibnamefont {Craig}}, \bibinfo {author} {\bibfnamefont {N.}~\bibnamefont {Ares}}, \bibinfo {author} {\bibfnamefont {L.}~\bibnamefont {Banchi}}, \bibinfo {author} {\bibfnamefont {F.}~\bibnamefont {Marquardt}}, \bibinfo {author} {\bibfnamefont {L.}~\bibnamefont {Pezz{\`e}}}, \ and\ \bibinfo {author} {\bibfnamefont {C.}~\bibnamefont {Bonato}},\ }\href {\doibase 10.1038/s42254-022-00552-1} {\bibfield  {journal} {\bibinfo  {journal} {Nature Reviews Physics}\ }\textbf {\bibinfo {volume} {5}},\ \bibinfo {pages} {141} (\bibinfo {year} {2023})}\BibitemShut {NoStop}%
\bibitem [{\citenamefont {Emerson}\ \emph {et~al.}(2005)\citenamefont {Emerson}, \citenamefont {Alicki},\ and\ \citenamefont {Życzkowski}}]{Emerson_2005}%
  \BibitemOpen
  \bibfield  {author} {\bibinfo {author} {\bibfnamefont {J.}~\bibnamefont {Emerson}}, \bibinfo {author} {\bibfnamefont {R.}~\bibnamefont {Alicki}}, \ and\ \bibinfo {author} {\bibfnamefont {K.}~\bibnamefont {Życzkowski}},\ }\href {\doibase 10.1088/1464-4266/7/10/021} {\bibfield  {journal} {\bibinfo  {journal} {Journal of Optics B: Quantum and Semiclassical Optics}\ }\textbf {\bibinfo {volume} {7}},\ \bibinfo {pages} {S347} (\bibinfo {year} {2005})}\BibitemShut {NoStop}%
\bibitem [{\citenamefont {Knill}\ \emph {et~al.}(2008)\citenamefont {Knill}, \citenamefont {Leibfried}, \citenamefont {Reichle}, \citenamefont {Britton}, \citenamefont {Blakestad}, \citenamefont {Jost}, \citenamefont {Langer}, \citenamefont {Ozeri}, \citenamefont {Seidelin},\ and\ \citenamefont {Wineland}}]{Knill_2008}%
  \BibitemOpen
  \bibfield  {author} {\bibinfo {author} {\bibfnamefont {E.}~\bibnamefont {Knill}}, \bibinfo {author} {\bibfnamefont {D.}~\bibnamefont {Leibfried}}, \bibinfo {author} {\bibfnamefont {R.}~\bibnamefont {Reichle}}, \bibinfo {author} {\bibfnamefont {J.}~\bibnamefont {Britton}}, \bibinfo {author} {\bibfnamefont {R.~B.}\ \bibnamefont {Blakestad}}, \bibinfo {author} {\bibfnamefont {J.~D.}\ \bibnamefont {Jost}}, \bibinfo {author} {\bibfnamefont {C.}~\bibnamefont {Langer}}, \bibinfo {author} {\bibfnamefont {R.}~\bibnamefont {Ozeri}}, \bibinfo {author} {\bibfnamefont {S.}~\bibnamefont {Seidelin}}, \ and\ \bibinfo {author} {\bibfnamefont {D.~J.}\ \bibnamefont {Wineland}},\ }\href {\doibase 10.1103/physreva.77.012307} {\bibfield  {journal} {\bibinfo  {journal} {Physical Review A}\ }\textbf {\bibinfo {volume} {77}} (\bibinfo {year} {2008}),\ 10.1103/physreva.77.012307}\BibitemShut {NoStop}%
\bibitem [{\citenamefont {Dankert}\ \emph {et~al.}(2009)\citenamefont {Dankert}, \citenamefont {Cleve}, \citenamefont {Emerson},\ and\ \citenamefont {Livine}}]{Dankert_2009}%
  \BibitemOpen
  \bibfield  {author} {\bibinfo {author} {\bibfnamefont {C.}~\bibnamefont {Dankert}}, \bibinfo {author} {\bibfnamefont {R.}~\bibnamefont {Cleve}}, \bibinfo {author} {\bibfnamefont {J.}~\bibnamefont {Emerson}}, \ and\ \bibinfo {author} {\bibfnamefont {E.}~\bibnamefont {Livine}},\ }\href {\doibase 10.1103/physreva.80.012304} {\bibfield  {journal} {\bibinfo  {journal} {Physical Review A}\ }\textbf {\bibinfo {volume} {80}} (\bibinfo {year} {2009}),\ 10.1103/physreva.80.012304}\BibitemShut {NoStop}%
\bibitem [{\citenamefont {Magesan}\ \emph {et~al.}(2011)\citenamefont {Magesan}, \citenamefont {Gambetta},\ and\ \citenamefont {Emerson}}]{PhysRevLett.106.180504}%
  \BibitemOpen
  \bibfield  {author} {\bibinfo {author} {\bibfnamefont {E.}~\bibnamefont {Magesan}}, \bibinfo {author} {\bibfnamefont {J.~M.}\ \bibnamefont {Gambetta}}, \ and\ \bibinfo {author} {\bibfnamefont {J.}~\bibnamefont {Emerson}},\ }\href {\doibase 10.1103/PhysRevLett.106.180504} {\bibfield  {journal} {\bibinfo  {journal} {Phys. Rev. Lett.}\ }\textbf {\bibinfo {volume} {106}},\ \bibinfo {pages} {180504} (\bibinfo {year} {2011})}\BibitemShut {NoStop}%
\bibitem [{\citenamefont {Magesan}\ \emph {et~al.}(2012)\citenamefont {Magesan}, \citenamefont {Gambetta},\ and\ \citenamefont {Emerson}}]{PhysRevA.85.042311}%
  \BibitemOpen
  \bibfield  {author} {\bibinfo {author} {\bibfnamefont {E.}~\bibnamefont {Magesan}}, \bibinfo {author} {\bibfnamefont {J.~M.}\ \bibnamefont {Gambetta}}, \ and\ \bibinfo {author} {\bibfnamefont {J.}~\bibnamefont {Emerson}},\ }\href {\doibase 10.1103/PhysRevA.85.042311} {\bibfield  {journal} {\bibinfo  {journal} {Phys. Rev. A}\ }\textbf {\bibinfo {volume} {85}},\ \bibinfo {pages} {042311} (\bibinfo {year} {2012})}\BibitemShut {NoStop}%
\bibitem [{\citenamefont {Gill}\ and\ \citenamefont {Guta}(2004)}]{gill2004invitation}%
  \BibitemOpen
  \bibfield  {author} {\bibinfo {author} {\bibfnamefont {R.}~\bibnamefont {Gill}}\ and\ \bibinfo {author} {\bibfnamefont {M.}~\bibnamefont {Guta}},\ }\href@noop {} {\enquote {\bibinfo {title} {An invitation to quantum tomography},}\ } (\bibinfo {year} {2004}),\ \Eprint {http://arxiv.org/abs/quant-ph/0303020} {arXiv:quant-ph/0303020 [quant-ph]} \BibitemShut {NoStop}%
\bibitem [{\citenamefont {Banaszek}\ \emph {et~al.}(2013)\citenamefont {Banaszek}, \citenamefont {Cramer},\ and\ \citenamefont {Gross}}]{Banaszek}%
  \BibitemOpen
  \bibfield  {author} {\bibinfo {author} {\bibfnamefont {K.}~\bibnamefont {Banaszek}}, \bibinfo {author} {\bibfnamefont {M.}~\bibnamefont {Cramer}}, \ and\ \bibinfo {author} {\bibfnamefont {D.}~\bibnamefont {Gross}},\ }\href {\doibase 10.1088/1367-2630/15/12/125020} {\bibfield  {journal} {\bibinfo  {journal} {New Journal of Physics}\ }\textbf {\bibinfo {volume} {15}},\ \bibinfo {pages} {5020} (\bibinfo {year} {2013})}\BibitemShut {NoStop}%
\bibitem [{\citenamefont {Nielsen}\ \emph {et~al.}(2021)\citenamefont {Nielsen}, \citenamefont {Gamble}, \citenamefont {Rudinger}, \citenamefont {Scholten}, \citenamefont {Young},\ and\ \citenamefont {Blume-Kohout}}]{Nielsen2021gatesettomography}%
  \BibitemOpen
  \bibfield  {author} {\bibinfo {author} {\bibfnamefont {E.}~\bibnamefont {Nielsen}}, \bibinfo {author} {\bibfnamefont {J.~K.}\ \bibnamefont {Gamble}}, \bibinfo {author} {\bibfnamefont {K.}~\bibnamefont {Rudinger}}, \bibinfo {author} {\bibfnamefont {T.}~\bibnamefont {Scholten}}, \bibinfo {author} {\bibfnamefont {K.}~\bibnamefont {Young}}, \ and\ \bibinfo {author} {\bibfnamefont {R.}~\bibnamefont {Blume-Kohout}},\ }\href {\doibase 10.22331/q-2021-10-05-557} {\bibfield  {journal} {\bibinfo  {journal} {{Quantum}}\ }\textbf {\bibinfo {volume} {5}},\ \bibinfo {pages} {557} (\bibinfo {year} {2021})}\BibitemShut {NoStop}%
\bibitem [{\citenamefont {Greenbaum}(2015)}]{greenbaum2015introduction}%
  \BibitemOpen
  \bibfield  {author} {\bibinfo {author} {\bibfnamefont {D.}~\bibnamefont {Greenbaum}},\ }\href@noop {} {\enquote {\bibinfo {title} {Introduction to quantum gate set tomography},}\ } (\bibinfo {year} {2015}),\ \Eprint {http://arxiv.org/abs/1509.02921} {arXiv:1509.02921 [quant-ph]} \BibitemShut {NoStop}%
\bibitem [{\citenamefont {Blume-Kohout}\ \emph {et~al.}(2013)\citenamefont {Blume-Kohout}, \citenamefont {Gamble}, \citenamefont {Nielsen}, \citenamefont {Mizrahi}, \citenamefont {Sterk},\ and\ \citenamefont {Maunz}}]{blumekohout2013robust}%
  \BibitemOpen
  \bibfield  {author} {\bibinfo {author} {\bibfnamefont {R.}~\bibnamefont {Blume-Kohout}}, \bibinfo {author} {\bibfnamefont {J.~K.}\ \bibnamefont {Gamble}}, \bibinfo {author} {\bibfnamefont {E.}~\bibnamefont {Nielsen}}, \bibinfo {author} {\bibfnamefont {J.}~\bibnamefont {Mizrahi}}, \bibinfo {author} {\bibfnamefont {J.~D.}\ \bibnamefont {Sterk}}, \ and\ \bibinfo {author} {\bibfnamefont {P.}~\bibnamefont {Maunz}},\ }\href@noop {} {\enquote {\bibinfo {title} {Robust, self-consistent, closed-form tomography of quantum logic gates on a trapped ion qubit},}\ } (\bibinfo {year} {2013}),\ \Eprint {http://arxiv.org/abs/1310.4492} {arXiv:1310.4492 [quant-ph]} \BibitemShut {NoStop}%
\bibitem [{\citenamefont {Stark}(2014)}]{PhysRevA.89.052109}%
  \BibitemOpen
  \bibfield  {author} {\bibinfo {author} {\bibfnamefont {C.}~\bibnamefont {Stark}},\ }\href {\doibase 10.1103/PhysRevA.89.052109} {\bibfield  {journal} {\bibinfo  {journal} {Phys. Rev. A}\ }\textbf {\bibinfo {volume} {89}},\ \bibinfo {pages} {052109} (\bibinfo {year} {2014})}\BibitemShut {NoStop}%
\bibitem [{\citenamefont {Merkel}\ \emph {et~al.}(2013)\citenamefont {Merkel}, \citenamefont {Gambetta}, \citenamefont {Smolin}, \citenamefont {Poletto}, \citenamefont {Córcoles}, \citenamefont {Johnson}, \citenamefont {Ryan},\ and\ \citenamefont {Steffen}}]{Merkel_2013}%
  \BibitemOpen
  \bibfield  {author} {\bibinfo {author} {\bibfnamefont {S.~T.}\ \bibnamefont {Merkel}}, \bibinfo {author} {\bibfnamefont {J.~M.}\ \bibnamefont {Gambetta}}, \bibinfo {author} {\bibfnamefont {J.~A.}\ \bibnamefont {Smolin}}, \bibinfo {author} {\bibfnamefont {S.}~\bibnamefont {Poletto}}, \bibinfo {author} {\bibfnamefont {A.~D.}\ \bibnamefont {Córcoles}}, \bibinfo {author} {\bibfnamefont {B.~R.}\ \bibnamefont {Johnson}}, \bibinfo {author} {\bibfnamefont {C.~A.}\ \bibnamefont {Ryan}}, \ and\ \bibinfo {author} {\bibfnamefont {M.}~\bibnamefont {Steffen}},\ }\href {\doibase 10.1103/physreva.87.062119} {\bibfield  {journal} {\bibinfo  {journal} {Physical Review A}\ }\textbf {\bibinfo {volume} {87}} (\bibinfo {year} {2013}),\ 10.1103/physreva.87.062119}\BibitemShut {NoStop}%
\bibitem [{\citenamefont {Chuang}\ and\ \citenamefont {Nielsen}(1997)}]{doi:10.1080/09500349708231894}%
  \BibitemOpen
  \bibfield  {author} {\bibinfo {author} {\bibfnamefont {I.~L.}\ \bibnamefont {Chuang}}\ and\ \bibinfo {author} {\bibfnamefont {M.~A.}\ \bibnamefont {Nielsen}},\ }\href {\doibase 10.1080/09500349708231894} {\bibfield  {journal} {\bibinfo  {journal} {Journal of Modern Optics}\ }\textbf {\bibinfo {volume} {44}},\ \bibinfo {pages} {2455} (\bibinfo {year} {1997})}\BibitemShut {NoStop}%
\bibitem [{\citenamefont {Poyatos}\ \emph {et~al.}(1997)\citenamefont {Poyatos}, \citenamefont {Cirac},\ and\ \citenamefont {Zoller}}]{PhysRevLett.78.390}%
  \BibitemOpen
  \bibfield  {author} {\bibinfo {author} {\bibfnamefont {J.~F.}\ \bibnamefont {Poyatos}}, \bibinfo {author} {\bibfnamefont {J.~I.}\ \bibnamefont {Cirac}}, \ and\ \bibinfo {author} {\bibfnamefont {P.}~\bibnamefont {Zoller}},\ }\href {\doibase 10.1103/PhysRevLett.78.390} {\bibfield  {journal} {\bibinfo  {journal} {Phys. Rev. Lett.}\ }\textbf {\bibinfo {volume} {78}},\ \bibinfo {pages} {390} (\bibinfo {year} {1997})}\BibitemShut {NoStop}%
\bibitem [{\citenamefont {Fiur\'a\ifmmode~\check{s}\else \v{s}\fi{}ek}\ and\ \citenamefont {Hradil}(2001)}]{PhysRevA.63.020101}%
  \BibitemOpen
  \bibfield  {author} {\bibinfo {author} {\bibfnamefont {J.}~\bibnamefont {Fiur\'a\ifmmode~\check{s}\else \v{s}\fi{}ek}}\ and\ \bibinfo {author} {\bibfnamefont {Z.~c.~v.}\ \bibnamefont {Hradil}},\ }\href {\doibase 10.1103/PhysRevA.63.020101} {\bibfield  {journal} {\bibinfo  {journal} {Phys. Rev. A}\ }\textbf {\bibinfo {volume} {63}},\ \bibinfo {pages} {020101} (\bibinfo {year} {2001})}\BibitemShut {NoStop}%
\bibitem [{\citenamefont {Sacchi}(2001)}]{PhysRevA.63.054104}%
  \BibitemOpen
  \bibfield  {author} {\bibinfo {author} {\bibfnamefont {M.~F.}\ \bibnamefont {Sacchi}},\ }\href {\doibase 10.1103/PhysRevA.63.054104} {\bibfield  {journal} {\bibinfo  {journal} {Phys. Rev. A}\ }\textbf {\bibinfo {volume} {63}},\ \bibinfo {pages} {054104} (\bibinfo {year} {2001})}\BibitemShut {NoStop}%
\bibitem [{\citenamefont {Je\ifmmode~\check{z}\else \v{z}\fi{}ek}\ \emph {et~al.}(2003)\citenamefont {Je\ifmmode~\check{z}\else \v{z}\fi{}ek}, \citenamefont {Fiur\'a\ifmmode~\check{s}\else \v{s}\fi{}ek},\ and\ \citenamefont {Hradil}}]{PhysRevA.68.012305}%
  \BibitemOpen
  \bibfield  {author} {\bibinfo {author} {\bibfnamefont {M.}~\bibnamefont {Je\ifmmode~\check{z}\else \v{z}\fi{}ek}}, \bibinfo {author} {\bibfnamefont {J.}~\bibnamefont {Fiur\'a\ifmmode~\check{s}\else \v{s}\fi{}ek}}, \ and\ \bibinfo {author} {\bibfnamefont {Z.~c.~v.}\ \bibnamefont {Hradil}},\ }\href {\doibase 10.1103/PhysRevA.68.012305} {\bibfield  {journal} {\bibinfo  {journal} {Phys. Rev. A}\ }\textbf {\bibinfo {volume} {68}},\ \bibinfo {pages} {012305} (\bibinfo {year} {2003})}\BibitemShut {NoStop}%
\bibitem [{\citenamefont {Watrous}(2018)}]{watrous2018theory}%
  \BibitemOpen
  \bibfield  {author} {\bibinfo {author} {\bibfnamefont {J.}~\bibnamefont {Watrous}},\ }\href@noop {} {\emph {\bibinfo {title} {The theory of quantum information}}}\ (\bibinfo  {publisher} {Cambridge university press},\ \bibinfo {year} {2018})\BibitemShut {NoStop}%
\bibitem [{\citenamefont {Ostrove}\ \emph {et~al.}(2023)\citenamefont {Ostrove}, \citenamefont {Rudinger}, \citenamefont {Seritan}, \citenamefont {Young},\ and\ \citenamefont {Blume-Kohout}}]{ostrove2023nearminimalgatesettomography}%
  \BibitemOpen
  \bibfield  {author} {\bibinfo {author} {\bibfnamefont {C.}~\bibnamefont {Ostrove}}, \bibinfo {author} {\bibfnamefont {K.}~\bibnamefont {Rudinger}}, \bibinfo {author} {\bibfnamefont {S.}~\bibnamefont {Seritan}}, \bibinfo {author} {\bibfnamefont {K.}~\bibnamefont {Young}}, \ and\ \bibinfo {author} {\bibfnamefont {R.}~\bibnamefont {Blume-Kohout}},\ }\href {https://arxiv.org/abs/2308.08781} {\enquote {\bibinfo {title} {Near-minimal gate set tomography experiment designs},}\ } (\bibinfo {year} {2023}),\ \Eprint {http://arxiv.org/abs/2308.08781} {arXiv:2308.08781 [quant-ph]} \BibitemShut {NoStop}%
\bibitem [{\citenamefont {Bruzewicz}\ \emph {et~al.}(2019)\citenamefont {Bruzewicz}, \citenamefont {Chiaverini}, \citenamefont {McConnell},\ and\ \citenamefont {Sage}}]{bruzewicz2019trapped}%
  \BibitemOpen
  \bibfield  {author} {\bibinfo {author} {\bibfnamefont {C.~D.}\ \bibnamefont {Bruzewicz}}, \bibinfo {author} {\bibfnamefont {J.}~\bibnamefont {Chiaverini}}, \bibinfo {author} {\bibfnamefont {R.}~\bibnamefont {McConnell}}, \ and\ \bibinfo {author} {\bibfnamefont {J.~M.}\ \bibnamefont {Sage}},\ }\href {\doibase 10.1063/1.5088164} {\bibfield  {journal} {\bibinfo  {journal} {Applied Physics Reviews}\ }\textbf {\bibinfo {volume} {6}} (\bibinfo {year} {2019}),\ 10.1063/1.5088164}\BibitemShut {NoStop}%
\bibitem [{\citenamefont {Rudinger}\ \emph {et~al.}(2019)\citenamefont {Rudinger}, \citenamefont {Proctor}, \citenamefont {Langharst}, \citenamefont {Sarovar}, \citenamefont {Young},\ and\ \citenamefont {Blume-Kohout}}]{PhysRevX.9.021045}%
  \BibitemOpen
  \bibfield  {author} {\bibinfo {author} {\bibfnamefont {K.}~\bibnamefont {Rudinger}}, \bibinfo {author} {\bibfnamefont {T.}~\bibnamefont {Proctor}}, \bibinfo {author} {\bibfnamefont {D.}~\bibnamefont {Langharst}}, \bibinfo {author} {\bibfnamefont {M.}~\bibnamefont {Sarovar}}, \bibinfo {author} {\bibfnamefont {K.}~\bibnamefont {Young}}, \ and\ \bibinfo {author} {\bibfnamefont {R.}~\bibnamefont {Blume-Kohout}},\ }\href {\doibase 10.1103/PhysRevX.9.021045} {\bibfield  {journal} {\bibinfo  {journal} {Phys. Rev. X}\ }\textbf {\bibinfo {volume} {9}},\ \bibinfo {pages} {021045} (\bibinfo {year} {2019})}\BibitemShut {NoStop}%
\bibitem [{\citenamefont {Rebentrost}\ \emph {et~al.}(2009)\citenamefont {Rebentrost}, \citenamefont {Serban}, \citenamefont {Schulte-Herbr\"uggen},\ and\ \citenamefont {Wilhelm}}]{rebentrost2009optimal}%
  \BibitemOpen
  \bibfield  {author} {\bibinfo {author} {\bibfnamefont {P.}~\bibnamefont {Rebentrost}}, \bibinfo {author} {\bibfnamefont {I.}~\bibnamefont {Serban}}, \bibinfo {author} {\bibfnamefont {T.}~\bibnamefont {Schulte-Herbr\"uggen}}, \ and\ \bibinfo {author} {\bibfnamefont {F.~K.}\ \bibnamefont {Wilhelm}},\ }\href {\doibase 10.1103/PhysRevLett.102.090401} {\bibfield  {journal} {\bibinfo  {journal} {Phys. Rev. Lett.}\ }\textbf {\bibinfo {volume} {102}},\ \bibinfo {pages} {090401} (\bibinfo {year} {2009})}\BibitemShut {NoStop}%
\bibitem [{\citenamefont {Wallman}\ \emph {et~al.}(2016)\citenamefont {Wallman}, \citenamefont {Barnhill},\ and\ \citenamefont {Emerson}}]{wallman2016robust}%
  \BibitemOpen
  \bibfield  {author} {\bibinfo {author} {\bibfnamefont {J.~J.}\ \bibnamefont {Wallman}}, \bibinfo {author} {\bibfnamefont {M.}~\bibnamefont {Barnhill}}, \ and\ \bibinfo {author} {\bibfnamefont {J.}~\bibnamefont {Emerson}},\ }\href {\doibase 10.1088/1367-2630/18/4/043021} {\bibfield  {journal} {\bibinfo  {journal} {New Journal of Physics}\ }\textbf {\bibinfo {volume} {18}},\ \bibinfo {pages} {043021} (\bibinfo {year} {2016})}\BibitemShut {NoStop}%
\bibitem [{\citenamefont {Jing}\ and\ \citenamefont {Wu}(2018)}]{jing2018decoherence}%
  \BibitemOpen
  \bibfield  {author} {\bibinfo {author} {\bibfnamefont {J.}~\bibnamefont {Jing}}\ and\ \bibinfo {author} {\bibfnamefont {L.-A.}\ \bibnamefont {Wu}},\ }\href {\doibase 10.1038/s41598-018-19977-9} {\bibfield  {journal} {\bibinfo  {journal} {Scientific Reports}\ }\textbf {\bibinfo {volume} {8}},\ \bibinfo {pages} {1471} (\bibinfo {year} {2018})}\BibitemShut {NoStop}%
\bibitem [{\citenamefont {Piltz}\ \emph {et~al.}(2014)\citenamefont {Piltz}, \citenamefont {Sriarunothai}, \citenamefont {Varón},\ and\ \citenamefont {Wunderlich}}]{piltz2014trapped}%
  \BibitemOpen
  \bibfield  {author} {\bibinfo {author} {\bibfnamefont {C.}~\bibnamefont {Piltz}}, \bibinfo {author} {\bibfnamefont {T.}~\bibnamefont {Sriarunothai}}, \bibinfo {author} {\bibfnamefont {A.}~\bibnamefont {Varón}}, \ and\ \bibinfo {author} {\bibfnamefont {C.}~\bibnamefont {Wunderlich}},\ }\href {\doibase 10.1038/ncomms5679} {\bibfield  {journal} {\bibinfo  {journal} {Nature Communications}\ }\textbf {\bibinfo {volume} {5}},\ \bibinfo {pages} {4679} (\bibinfo {year} {2014})}\BibitemShut {NoStop}%
\bibitem [{\citenamefont {Leibfried}\ \emph {et~al.}(2003{\natexlab{a}})\citenamefont {Leibfried}, \citenamefont {DeMarco}, \citenamefont {Meyer}, \citenamefont {Lucas}, \citenamefont {Barrett}, \citenamefont {Britton}, \citenamefont {Itano}, \citenamefont {Jelenković}, \citenamefont {Langer}, \citenamefont {Rosenband},\ and\ \citenamefont {Wineland}}]{leibfried2003experimental}%
  \BibitemOpen
  \bibfield  {author} {\bibinfo {author} {\bibfnamefont {D.}~\bibnamefont {Leibfried}}, \bibinfo {author} {\bibfnamefont {B.}~\bibnamefont {DeMarco}}, \bibinfo {author} {\bibfnamefont {V.}~\bibnamefont {Meyer}}, \bibinfo {author} {\bibfnamefont {D.}~\bibnamefont {Lucas}}, \bibinfo {author} {\bibfnamefont {M.}~\bibnamefont {Barrett}}, \bibinfo {author} {\bibfnamefont {J.}~\bibnamefont {Britton}}, \bibinfo {author} {\bibfnamefont {W.~M.}\ \bibnamefont {Itano}}, \bibinfo {author} {\bibfnamefont {B.}~\bibnamefont {Jelenković}}, \bibinfo {author} {\bibfnamefont {C.}~\bibnamefont {Langer}}, \bibinfo {author} {\bibfnamefont {T.}~\bibnamefont {Rosenband}}, \ and\ \bibinfo {author} {\bibfnamefont {D.~J.}\ \bibnamefont {Wineland}},\ }\href {\doibase 10.1038/nature01492} {\bibfield  {journal} {\bibinfo  {journal} {Nature}\ }\textbf {\bibinfo {volume} {422}},\ \bibinfo {pages} {412} (\bibinfo {year} {2003}{\natexlab{a}})}\BibitemShut {NoStop}%
\bibitem [{\citenamefont {Viñas}\ and\ \citenamefont {Bermudez}(2025)}]{vinas2025microscopic}%
  \BibitemOpen
  \bibfield  {author} {\bibinfo {author} {\bibfnamefont {P.}~\bibnamefont {Viñas}}\ and\ \bibinfo {author} {\bibfnamefont {A.}~\bibnamefont {Bermudez}},\ }\href {\doibase https://doi.org/10.1038/s41534-025-00976-4} {\bibfield  {journal} {\bibinfo  {journal} {npj Quantum Information}\ }\textbf {\bibinfo {volume} {11}},\ \bibinfo {pages} {23} (\bibinfo {year} {2025})}\BibitemShut {NoStop}%
\bibitem [{\citenamefont {Ballance}\ \emph {et~al.}(2016{\natexlab{a}})\citenamefont {Ballance}, \citenamefont {Harty}, \citenamefont {Linke}, \citenamefont {Sepiol},\ and\ \citenamefont {Lucas}}]{PhysRevLett.117.060504}%
  \BibitemOpen
  \bibfield  {author} {\bibinfo {author} {\bibfnamefont {C.~J.}\ \bibnamefont {Ballance}}, \bibinfo {author} {\bibfnamefont {T.~P.}\ \bibnamefont {Harty}}, \bibinfo {author} {\bibfnamefont {N.~M.}\ \bibnamefont {Linke}}, \bibinfo {author} {\bibfnamefont {M.~A.}\ \bibnamefont {Sepiol}}, \ and\ \bibinfo {author} {\bibfnamefont {D.~M.}\ \bibnamefont {Lucas}},\ }\href {\doibase 10.1103/PhysRevLett.117.060504} {\bibfield  {journal} {\bibinfo  {journal} {Phys. Rev. Lett.}\ }\textbf {\bibinfo {volume} {117}},\ \bibinfo {pages} {060504} (\bibinfo {year} {2016}{\natexlab{a}})}\BibitemShut {NoStop}%
\bibitem [{\citenamefont {Sawyer}\ and\ \citenamefont {Brown}(2021)}]{PhysRevA.103.022427}%
  \BibitemOpen
  \bibfield  {author} {\bibinfo {author} {\bibfnamefont {B.~C.}\ \bibnamefont {Sawyer}}\ and\ \bibinfo {author} {\bibfnamefont {K.~R.}\ \bibnamefont {Brown}},\ }\href {\doibase 10.1103/PhysRevA.103.022427} {\bibfield  {journal} {\bibinfo  {journal} {Phys. Rev. A}\ }\textbf {\bibinfo {volume} {103}},\ \bibinfo {pages} {022427} (\bibinfo {year} {2021})}\BibitemShut {NoStop}%
\bibitem [{\citenamefont {Baldwin}\ \emph {et~al.}(2021)\citenamefont {Baldwin}, \citenamefont {Bjork}, \citenamefont {Foss-Feig}, \citenamefont {Gaebler}, \citenamefont {Hayes}, \citenamefont {Kokish}, \citenamefont {Langer}, \citenamefont {Sedlacek}, \citenamefont {Stack},\ and\ \citenamefont {Vittorini}}]{PhysRevA.103.012603}%
  \BibitemOpen
  \bibfield  {author} {\bibinfo {author} {\bibfnamefont {C.~H.}\ \bibnamefont {Baldwin}}, \bibinfo {author} {\bibfnamefont {B.~J.}\ \bibnamefont {Bjork}}, \bibinfo {author} {\bibfnamefont {M.}~\bibnamefont {Foss-Feig}}, \bibinfo {author} {\bibfnamefont {J.~P.}\ \bibnamefont {Gaebler}}, \bibinfo {author} {\bibfnamefont {D.}~\bibnamefont {Hayes}}, \bibinfo {author} {\bibfnamefont {M.~G.}\ \bibnamefont {Kokish}}, \bibinfo {author} {\bibfnamefont {C.}~\bibnamefont {Langer}}, \bibinfo {author} {\bibfnamefont {J.~A.}\ \bibnamefont {Sedlacek}}, \bibinfo {author} {\bibfnamefont {D.}~\bibnamefont {Stack}}, \ and\ \bibinfo {author} {\bibfnamefont {G.}~\bibnamefont {Vittorini}},\ }\href {\doibase 10.1103/PhysRevA.103.012603} {\bibfield  {journal} {\bibinfo  {journal} {Phys. Rev. A}\ }\textbf {\bibinfo {volume} {103}},\ \bibinfo {pages} {012603} (\bibinfo {year} {2021})}\BibitemShut {NoStop}%
\bibitem [{\citenamefont {Clark}\ \emph {et~al.}(2021)\citenamefont {Clark}, \citenamefont {Tinkey}, \citenamefont {Sawyer}, \citenamefont {Meier}, \citenamefont {Burkhardt}, \citenamefont {Seck}, \citenamefont {Shappert}, \citenamefont {Guise}, \citenamefont {Volin}, \citenamefont {Fallek}, \citenamefont {Hayden}, \citenamefont {Rellergert},\ and\ \citenamefont {Brown}}]{PhysRevLett.127.130505}%
  \BibitemOpen
  \bibfield  {author} {\bibinfo {author} {\bibfnamefont {C.~R.}\ \bibnamefont {Clark}}, \bibinfo {author} {\bibfnamefont {H.~N.}\ \bibnamefont {Tinkey}}, \bibinfo {author} {\bibfnamefont {B.~C.}\ \bibnamefont {Sawyer}}, \bibinfo {author} {\bibfnamefont {A.~M.}\ \bibnamefont {Meier}}, \bibinfo {author} {\bibfnamefont {K.~A.}\ \bibnamefont {Burkhardt}}, \bibinfo {author} {\bibfnamefont {C.~M.}\ \bibnamefont {Seck}}, \bibinfo {author} {\bibfnamefont {C.~M.}\ \bibnamefont {Shappert}}, \bibinfo {author} {\bibfnamefont {N.~D.}\ \bibnamefont {Guise}}, \bibinfo {author} {\bibfnamefont {C.~E.}\ \bibnamefont {Volin}}, \bibinfo {author} {\bibfnamefont {S.~D.}\ \bibnamefont {Fallek}}, \bibinfo {author} {\bibfnamefont {H.~T.}\ \bibnamefont {Hayden}}, \bibinfo {author} {\bibfnamefont {W.~G.}\ \bibnamefont {Rellergert}}, \ and\ \bibinfo {author} {\bibfnamefont {K.~R.}\ \bibnamefont {Brown}},\ }\href {\doibase 10.1103/PhysRevLett.127.130505} {\bibfield  {journal} {\bibinfo  {journal} {Phys. Rev. Lett.}\ }\textbf {\bibinfo
  {volume} {127}},\ \bibinfo {pages} {130505} (\bibinfo {year} {2021})}\BibitemShut {NoStop}%
\bibitem [{\citenamefont {Monz}()}]{comment}%
  \BibitemOpen
  \bibfield  {author} {\bibinfo {author} {\bibfnamefont {T.}~\bibnamefont {Monz}},\ }\href@noop {} {\bibinfo  {journal} {private communication}\ }\BibitemShut {NoStop}%
\bibitem [{\citenamefont {ŠAŠURA}\ and\ \citenamefont {BUŽEK}(2002)}]{vsavsura2002cold}%
  \BibitemOpen
\bibfield  {journal} {  }\bibfield  {author} {\bibinfo {author} {\bibfnamefont {M.}~\bibnamefont {ŠAŠURA}}\ and\ \bibinfo {author} {\bibfnamefont {V.}~\bibnamefont {BUŽEK}},\ }\href {\doibase 10.1080/09500340110115497} {\bibfield  {journal} {\bibinfo  {journal} {Journal of Modern Optics}\ }\textbf {\bibinfo {volume} {49}},\ \bibinfo {pages} {1593–1647} (\bibinfo {year} {2002})}\BibitemShut {NoStop}%
\bibitem [{\citenamefont {Häffner}\ \emph {et~al.}(2008)\citenamefont {Häffner}, \citenamefont {Roos},\ and\ \citenamefont {Blatt}}]{haffner2008quantum}%
  \BibitemOpen
  \bibfield  {author} {\bibinfo {author} {\bibfnamefont {H.}~\bibnamefont {Häffner}}, \bibinfo {author} {\bibfnamefont {C.}~\bibnamefont {Roos}}, \ and\ \bibinfo {author} {\bibfnamefont {R.}~\bibnamefont {Blatt}},\ }\href {\doibase https://doi.org/10.1016/j.physrep.2008.09.003} {\bibfield  {journal} {\bibinfo  {journal} {Physics Reports}\ }\textbf {\bibinfo {volume} {469}},\ \bibinfo {pages} {155} (\bibinfo {year} {2008})}\BibitemShut {NoStop}%
\bibitem [{\citenamefont {Barrett}\ \emph {et~al.}(2003)\citenamefont {Barrett}, \citenamefont {DeMarco}, \citenamefont {Schaetz}, \citenamefont {Meyer}, \citenamefont {Leibfried}, \citenamefont {Britton}, \citenamefont {Chiaverini}, \citenamefont {Itano}, \citenamefont {Jelenkovi\ifmmode~\acute{c}\else \'{c}\fi{}}, \citenamefont {Jost}, \citenamefont {Langer}, \citenamefont {Rosenband},\ and\ \citenamefont {Wineland}}]{PhysRevA.68.042302}%
  \BibitemOpen
  \bibfield  {author} {\bibinfo {author} {\bibfnamefont {M.~D.}\ \bibnamefont {Barrett}}, \bibinfo {author} {\bibfnamefont {B.}~\bibnamefont {DeMarco}}, \bibinfo {author} {\bibfnamefont {T.}~\bibnamefont {Schaetz}}, \bibinfo {author} {\bibfnamefont {V.}~\bibnamefont {Meyer}}, \bibinfo {author} {\bibfnamefont {D.}~\bibnamefont {Leibfried}}, \bibinfo {author} {\bibfnamefont {J.}~\bibnamefont {Britton}}, \bibinfo {author} {\bibfnamefont {J.}~\bibnamefont {Chiaverini}}, \bibinfo {author} {\bibfnamefont {W.~M.}\ \bibnamefont {Itano}}, \bibinfo {author} {\bibfnamefont {B.}~\bibnamefont {Jelenkovi\ifmmode~\acute{c}\else \'{c}\fi{}}}, \bibinfo {author} {\bibfnamefont {J.~D.}\ \bibnamefont {Jost}}, \bibinfo {author} {\bibfnamefont {C.}~\bibnamefont {Langer}}, \bibinfo {author} {\bibfnamefont {T.}~\bibnamefont {Rosenband}}, \ and\ \bibinfo {author} {\bibfnamefont {D.~J.}\ \bibnamefont {Wineland}},\ }\href {\doibase 10.1103/PhysRevA.68.042302} {\bibfield  {journal} {\bibinfo  {journal} {Phys. Rev. A}\ }\textbf {\bibinfo
  {volume} {68}},\ \bibinfo {pages} {042302} (\bibinfo {year} {2003})}\BibitemShut {NoStop}%
\bibitem [{\citenamefont {Schmidt}\ \emph {et~al.}(2005)\citenamefont {Schmidt}, \citenamefont {Rosenband}, \citenamefont {Langer}, \citenamefont {Itano}, \citenamefont {Bergquist},\ and\ \citenamefont {Wineland}}]{schmidt2005spectroscopy}%
  \BibitemOpen
  \bibfield  {author} {\bibinfo {author} {\bibfnamefont {P.~O.}\ \bibnamefont {Schmidt}}, \bibinfo {author} {\bibfnamefont {T.}~\bibnamefont {Rosenband}}, \bibinfo {author} {\bibfnamefont {C.}~\bibnamefont {Langer}}, \bibinfo {author} {\bibfnamefont {W.~M.}\ \bibnamefont {Itano}}, \bibinfo {author} {\bibfnamefont {J.~C.}\ \bibnamefont {Bergquist}}, \ and\ \bibinfo {author} {\bibfnamefont {D.~J.}\ \bibnamefont {Wineland}},\ }\href {\doibase 10.1126/science.1114375} {\bibfield  {journal} {\bibinfo  {journal} {Science}\ }\textbf {\bibinfo {volume} {309}},\ \bibinfo {pages} {749} (\bibinfo {year} {2005})},\ \Eprint {http://arxiv.org/abs/https://www.science.org/doi/pdf/10.1126/science.1114375} {https://www.science.org/doi/pdf/10.1126/science.1114375} \BibitemShut {NoStop}%
\bibitem [{\citenamefont {Aolita}\ \emph {et~al.}(2007)\citenamefont {Aolita}, \citenamefont {Kim}, \citenamefont {Benhelm}, \citenamefont {Roos},\ and\ \citenamefont {H\"affner}}]{PhysRevA.76.040303}%
  \BibitemOpen
  \bibfield  {author} {\bibinfo {author} {\bibfnamefont {L.}~\bibnamefont {Aolita}}, \bibinfo {author} {\bibfnamefont {K.}~\bibnamefont {Kim}}, \bibinfo {author} {\bibfnamefont {J.}~\bibnamefont {Benhelm}}, \bibinfo {author} {\bibfnamefont {C.~F.}\ \bibnamefont {Roos}}, \ and\ \bibinfo {author} {\bibfnamefont {H.}~\bibnamefont {H\"affner}},\ }\href {\doibase 10.1103/PhysRevA.76.040303} {\bibfield  {journal} {\bibinfo  {journal} {Phys. Rev. A}\ }\textbf {\bibinfo {volume} {76}},\ \bibinfo {pages} {040303} (\bibinfo {year} {2007})}\BibitemShut {NoStop}%
\bibitem [{\citenamefont {S\o{}rensen}\ and\ \citenamefont {M\o{}lmer}(1999)}]{sorensen1999quantum}%
  \BibitemOpen
  \bibfield  {author} {\bibinfo {author} {\bibfnamefont {A.}~\bibnamefont {S\o{}rensen}}\ and\ \bibinfo {author} {\bibfnamefont {K.}~\bibnamefont {M\o{}lmer}},\ }\href {\doibase 10.1103/PhysRevLett.82.1971} {\bibfield  {journal} {\bibinfo  {journal} {Phys. Rev. Lett.}\ }\textbf {\bibinfo {volume} {82}},\ \bibinfo {pages} {1971} (\bibinfo {year} {1999})}\BibitemShut {NoStop}%
\bibitem [{\citenamefont {Cirac}\ and\ \citenamefont {Zoller}(1995)}]{PhysRevLett.74.4091}%
  \BibitemOpen
  \bibfield  {author} {\bibinfo {author} {\bibfnamefont {J.~I.}\ \bibnamefont {Cirac}}\ and\ \bibinfo {author} {\bibfnamefont {P.}~\bibnamefont {Zoller}},\ }\href {\doibase 10.1103/PhysRevLett.74.4091} {\bibfield  {journal} {\bibinfo  {journal} {Phys. Rev. Lett.}\ }\textbf {\bibinfo {volume} {74}},\ \bibinfo {pages} {4091} (\bibinfo {year} {1995})}\BibitemShut {NoStop}%
\bibitem [{\citenamefont {Hilder}\ \emph {et~al.}(2022)\citenamefont {Hilder}, \citenamefont {Pijn}, \citenamefont {Onishchenko}, \citenamefont {Stahl}, \citenamefont {Orth}, \citenamefont {Lekitsch}, \citenamefont {Rodriguez-Blanco}, \citenamefont {M\"uller}, \citenamefont {Schmidt-Kaler},\ and\ \citenamefont {Poschinger}}]{PhysRevX.12.011032}%
  \BibitemOpen
  \bibfield  {author} {\bibinfo {author} {\bibfnamefont {J.}~\bibnamefont {Hilder}}, \bibinfo {author} {\bibfnamefont {D.}~\bibnamefont {Pijn}}, \bibinfo {author} {\bibfnamefont {O.}~\bibnamefont {Onishchenko}}, \bibinfo {author} {\bibfnamefont {A.}~\bibnamefont {Stahl}}, \bibinfo {author} {\bibfnamefont {M.}~\bibnamefont {Orth}}, \bibinfo {author} {\bibfnamefont {B.}~\bibnamefont {Lekitsch}}, \bibinfo {author} {\bibfnamefont {A.}~\bibnamefont {Rodriguez-Blanco}}, \bibinfo {author} {\bibfnamefont {M.}~\bibnamefont {M\"uller}}, \bibinfo {author} {\bibfnamefont {F.}~\bibnamefont {Schmidt-Kaler}}, \ and\ \bibinfo {author} {\bibfnamefont {U.~G.}\ \bibnamefont {Poschinger}},\ }\href {\doibase 10.1103/PhysRevX.12.011032} {\bibfield  {journal} {\bibinfo  {journal} {Phys. Rev. X}\ }\textbf {\bibinfo {volume} {12}},\ \bibinfo {pages} {011032} (\bibinfo {year} {2022})}\BibitemShut {NoStop}%
\bibitem [{\citenamefont {Ballance}\ \emph {et~al.}(2016{\natexlab{b}})\citenamefont {Ballance}, \citenamefont {Harty}, \citenamefont {Linke}, \citenamefont {Sepiol},\ and\ \citenamefont {Lucas}}]{Ballance_2016}%
  \BibitemOpen
  \bibfield  {author} {\bibinfo {author} {\bibfnamefont {C.~J.}\ \bibnamefont {Ballance}}, \bibinfo {author} {\bibfnamefont {T.~P.}\ \bibnamefont {Harty}}, \bibinfo {author} {\bibfnamefont {N.~M.}\ \bibnamefont {Linke}}, \bibinfo {author} {\bibfnamefont {M.~A.}\ \bibnamefont {Sepiol}}, \ and\ \bibinfo {author} {\bibfnamefont {D.~M.}\ \bibnamefont {Lucas}},\ }\href {\doibase 10.1103/physrevlett.117.060504} {\bibfield  {journal} {\bibinfo  {journal} {Physical Review Letters}\ }\textbf {\bibinfo {volume} {117}} (\bibinfo {year} {2016}{\natexlab{b}}),\ 10.1103/physrevlett.117.060504}\BibitemShut {NoStop}%
\bibitem [{\citenamefont {Srinivas}\ \emph {et~al.}(2021)\citenamefont {Srinivas}, \citenamefont {Burd}, \citenamefont {Knaack}, \citenamefont {Sutherland}, \citenamefont {Kwiatkowski}, \citenamefont {Glancy}, \citenamefont {Knill}, \citenamefont {Wineland}, \citenamefont {Leibfried}, \citenamefont {Wilson}, \citenamefont {Allcock},\ and\ \citenamefont {Slichter}}]{Srinivas_2021}%
  \BibitemOpen
  \bibfield  {author} {\bibinfo {author} {\bibfnamefont {R.}~\bibnamefont {Srinivas}}, \bibinfo {author} {\bibfnamefont {S.~C.}\ \bibnamefont {Burd}}, \bibinfo {author} {\bibfnamefont {H.~M.}\ \bibnamefont {Knaack}}, \bibinfo {author} {\bibfnamefont {R.~T.}\ \bibnamefont {Sutherland}}, \bibinfo {author} {\bibfnamefont {A.}~\bibnamefont {Kwiatkowski}}, \bibinfo {author} {\bibfnamefont {S.}~\bibnamefont {Glancy}}, \bibinfo {author} {\bibfnamefont {E.}~\bibnamefont {Knill}}, \bibinfo {author} {\bibfnamefont {D.~J.}\ \bibnamefont {Wineland}}, \bibinfo {author} {\bibfnamefont {D.}~\bibnamefont {Leibfried}}, \bibinfo {author} {\bibfnamefont {A.~C.}\ \bibnamefont {Wilson}}, \bibinfo {author} {\bibfnamefont {D.~T.~C.}\ \bibnamefont {Allcock}}, \ and\ \bibinfo {author} {\bibfnamefont {D.~H.}\ \bibnamefont {Slichter}},\ }\href {\doibase 10.1038/s41586-021-03809-4} {\bibfield  {journal} {\bibinfo  {journal} {Nature}\ }\textbf {\bibinfo {volume} {597}},\ \bibinfo {pages} {209–213} (\bibinfo {year} {2021})}\BibitemShut
  {NoStop}%
\bibitem [{\citenamefont {Sutherland}\ \emph {et~al.}(2019)\citenamefont {Sutherland}, \citenamefont {Srinivas}, \citenamefont {Burd}, \citenamefont {Leibfried}, \citenamefont {Wilson}, \citenamefont {Wineland}, \citenamefont {Allcock}, \citenamefont {Slichter},\ and\ \citenamefont {Libby}}]{Sutherland_2019}%
  \BibitemOpen
  \bibfield  {author} {\bibinfo {author} {\bibfnamefont {R.~T.}\ \bibnamefont {Sutherland}}, \bibinfo {author} {\bibfnamefont {R.}~\bibnamefont {Srinivas}}, \bibinfo {author} {\bibfnamefont {S.~C.}\ \bibnamefont {Burd}}, \bibinfo {author} {\bibfnamefont {D.}~\bibnamefont {Leibfried}}, \bibinfo {author} {\bibfnamefont {A.~C.}\ \bibnamefont {Wilson}}, \bibinfo {author} {\bibfnamefont {D.~J.}\ \bibnamefont {Wineland}}, \bibinfo {author} {\bibfnamefont {D.~T.~C.}\ \bibnamefont {Allcock}}, \bibinfo {author} {\bibfnamefont {D.~H.}\ \bibnamefont {Slichter}}, \ and\ \bibinfo {author} {\bibfnamefont {S.~B.}\ \bibnamefont {Libby}},\ }\href {\doibase 10.1088/1367-2630/ab0be5} {\bibfield  {journal} {\bibinfo  {journal} {New Journal of Physics}\ }\textbf {\bibinfo {volume} {21}},\ \bibinfo {pages} {033033} (\bibinfo {year} {2019})}\BibitemShut {NoStop}%
\bibitem [{\citenamefont {Ospelkaus}\ \emph {et~al.}(2008)\citenamefont {Ospelkaus}, \citenamefont {Langer}, \citenamefont {Amini}, \citenamefont {Brown}, \citenamefont {Leibfried},\ and\ \citenamefont {Wineland}}]{Ospelkaus_2008}%
  \BibitemOpen
  \bibfield  {author} {\bibinfo {author} {\bibfnamefont {C.}~\bibnamefont {Ospelkaus}}, \bibinfo {author} {\bibfnamefont {C.}~\bibnamefont {Langer}}, \bibinfo {author} {\bibfnamefont {J.}~\bibnamefont {Amini}}, \bibinfo {author} {\bibfnamefont {K.}~\bibnamefont {Brown}}, \bibinfo {author} {\bibfnamefont {D.}~\bibnamefont {Leibfried}}, \ and\ \bibinfo {author} {\bibfnamefont {D.}~\bibnamefont {Wineland}},\ }\href {\doibase 10.1103/physrevlett.101.090502} {\bibfield  {journal} {\bibinfo  {journal} {Physical Review Letters}\ }\textbf {\bibinfo {volume} {101}} (\bibinfo {year} {2008}),\ 10.1103/physrevlett.101.090502}\BibitemShut {NoStop}%
\bibitem [{\citenamefont {Rodriguez-Blanco}\ \emph {et~al.}(2024)\citenamefont {Rodriguez-Blanco}, \citenamefont {Shahandeh},\ and\ \citenamefont {Bermudez}}]{PhysRevA.109.052417}%
  \BibitemOpen
  \bibfield  {author} {\bibinfo {author} {\bibfnamefont {A.}~\bibnamefont {Rodriguez-Blanco}}, \bibinfo {author} {\bibfnamefont {F.}~\bibnamefont {Shahandeh}}, \ and\ \bibinfo {author} {\bibfnamefont {A.}~\bibnamefont {Bermudez}},\ }\href {\doibase 10.1103/PhysRevA.109.052417} {\bibfield  {journal} {\bibinfo  {journal} {Phys. Rev. A}\ }\textbf {\bibinfo {volume} {109}},\ \bibinfo {pages} {052417} (\bibinfo {year} {2024})}\BibitemShut {NoStop}%
\bibitem [{\citenamefont {Zhu}\ \emph {et~al.}(2006)\citenamefont {Zhu}, \citenamefont {Monroe},\ and\ \citenamefont {Duan}}]{Zhu_2006}%
  \BibitemOpen
  \bibfield  {author} {\bibinfo {author} {\bibfnamefont {S.-L.}\ \bibnamefont {Zhu}}, \bibinfo {author} {\bibfnamefont {C.}~\bibnamefont {Monroe}}, \ and\ \bibinfo {author} {\bibfnamefont {L.-M.}\ \bibnamefont {Duan}},\ }\href {\doibase 10.1103/physrevlett.97.050505} {\bibfield  {journal} {\bibinfo  {journal} {Physical Review Letters}\ }\textbf {\bibinfo {volume} {97}} (\bibinfo {year} {2006}),\ 10.1103/physrevlett.97.050505}\BibitemShut {NoStop}%
\bibitem [{\citenamefont {Leibfried}\ \emph {et~al.}(2003{\natexlab{b}})\citenamefont {Leibfried}, \citenamefont {Blatt}, \citenamefont {Monroe},\ and\ \citenamefont {Wineland}}]{RevModPhys.75.281}%
  \BibitemOpen
  \bibfield  {author} {\bibinfo {author} {\bibfnamefont {D.}~\bibnamefont {Leibfried}}, \bibinfo {author} {\bibfnamefont {R.}~\bibnamefont {Blatt}}, \bibinfo {author} {\bibfnamefont {C.}~\bibnamefont {Monroe}}, \ and\ \bibinfo {author} {\bibfnamefont {D.}~\bibnamefont {Wineland}},\ }\href {\doibase 10.1103/RevModPhys.75.281} {\bibfield  {journal} {\bibinfo  {journal} {Rev. Mod. Phys.}\ }\textbf {\bibinfo {volume} {75}},\ \bibinfo {pages} {281} (\bibinfo {year} {2003}{\natexlab{b}})}\BibitemShut {NoStop}%
\bibitem [{\citenamefont {Magnus}(1954)}]{magnus1954exponential}%
  \BibitemOpen
  \bibfield  {author} {\bibinfo {author} {\bibfnamefont {W.}~\bibnamefont {Magnus}},\ }\href {\doibase https://doi.org/10.1002/cpa.3160070404} {\bibfield  {journal} {\bibinfo  {journal} {Communications on Pure and Applied Mathematics}\ }\textbf {\bibinfo {volume} {7}},\ \bibinfo {pages} {649} (\bibinfo {year} {1954})},\ \Eprint {http://arxiv.org/abs/https://onlinelibrary.wiley.com/doi/pdf/10.1002/cpa.3160070404} {https://onlinelibrary.wiley.com/doi/pdf/10.1002/cpa.3160070404} \BibitemShut {NoStop}%
\bibitem [{\citenamefont {Blanes}\ \emph {et~al.}(2010)\citenamefont {Blanes}, \citenamefont {Casas}, \citenamefont {Oteo},\ and\ \citenamefont {Ros}}]{blanes2010pedagogical}%
  \BibitemOpen
  \bibfield  {author} {\bibinfo {author} {\bibfnamefont {S.}~\bibnamefont {Blanes}}, \bibinfo {author} {\bibfnamefont {F.}~\bibnamefont {Casas}}, \bibinfo {author} {\bibfnamefont {J.~A.}\ \bibnamefont {Oteo}}, \ and\ \bibinfo {author} {\bibfnamefont {J.}~\bibnamefont {Ros}},\ }\href {\doibase 10.1088/0143-0807/31/4/020} {\bibfield  {journal} {\bibinfo  {journal} {European Journal of Physics}\ }\textbf {\bibinfo {volume} {31}},\ \bibinfo {pages} {907} (\bibinfo {year} {2010})}\BibitemShut {NoStop}%
\bibitem [{\citenamefont {Blanes}\ \emph {et~al.}(2009)\citenamefont {Blanes}, \citenamefont {Casas}, \citenamefont {Oteo},\ and\ \citenamefont {Ros}}]{Blanes_2009}%
  \BibitemOpen
  \bibfield  {author} {\bibinfo {author} {\bibfnamefont {S.}~\bibnamefont {Blanes}}, \bibinfo {author} {\bibfnamefont {F.}~\bibnamefont {Casas}}, \bibinfo {author} {\bibfnamefont {J.}~\bibnamefont {Oteo}}, \ and\ \bibinfo {author} {\bibfnamefont {J.}~\bibnamefont {Ros}},\ }\href {\doibase 10.1016/j.physrep.2008.11.001} {\bibfield  {journal} {\bibinfo  {journal} {Physics Reports}\ }\textbf {\bibinfo {volume} {470}},\ \bibinfo {pages} {151–238} (\bibinfo {year} {2009})}\BibitemShut {NoStop}%
\bibitem [{\citenamefont {Oghittu}\ \emph {et~al.}(2024)\citenamefont {Oghittu}, \citenamefont {Safavi-Naini}, \citenamefont {Negretti},\ and\ \citenamefont {Gerritsma}}]{Oghittu_2024}%
  \BibitemOpen
  \bibfield  {author} {\bibinfo {author} {\bibfnamefont {L.}~\bibnamefont {Oghittu}}, \bibinfo {author} {\bibfnamefont {A.}~\bibnamefont {Safavi-Naini}}, \bibinfo {author} {\bibfnamefont {A.}~\bibnamefont {Negretti}}, \ and\ \bibinfo {author} {\bibfnamefont {R.}~\bibnamefont {Gerritsma}},\ }\href {\doibase 10.1103/physreva.110.063307} {\bibfield  {journal} {\bibinfo  {journal} {Physical Review A}\ }\textbf {\bibinfo {volume} {110}} (\bibinfo {year} {2024}),\ 10.1103/physreva.110.063307}\BibitemShut {NoStop}%
\bibitem [{\citenamefont {Hashim}\ \emph {et~al.}(2024)\citenamefont {Hashim}, \citenamefont {Nguyen}, \citenamefont {Goss}, \citenamefont {Marinelli}, \citenamefont {Naik}, \citenamefont {Chistolini}, \citenamefont {Hines}, \citenamefont {Marceaux}, \citenamefont {Kim}, \citenamefont {Gokhale}, \citenamefont {Tomesh}, \citenamefont {Chen}, \citenamefont {Jiang}, \citenamefont {Ferracin}, \citenamefont {Rudinger}, \citenamefont {Proctor}, \citenamefont {Young}, \citenamefont {Blume-Kohout},\ and\ \citenamefont {Siddiqi}}]{hashim2024practicalintroductionbenchmarkingcharacterization}%
  \BibitemOpen
  \bibfield  {author} {\bibinfo {author} {\bibfnamefont {A.}~\bibnamefont {Hashim}}, \bibinfo {author} {\bibfnamefont {L.~B.}\ \bibnamefont {Nguyen}}, \bibinfo {author} {\bibfnamefont {N.}~\bibnamefont {Goss}}, \bibinfo {author} {\bibfnamefont {B.}~\bibnamefont {Marinelli}}, \bibinfo {author} {\bibfnamefont {R.~K.}\ \bibnamefont {Naik}}, \bibinfo {author} {\bibfnamefont {T.}~\bibnamefont {Chistolini}}, \bibinfo {author} {\bibfnamefont {J.}~\bibnamefont {Hines}}, \bibinfo {author} {\bibfnamefont {J.~P.}\ \bibnamefont {Marceaux}}, \bibinfo {author} {\bibfnamefont {Y.}~\bibnamefont {Kim}}, \bibinfo {author} {\bibfnamefont {P.}~\bibnamefont {Gokhale}}, \bibinfo {author} {\bibfnamefont {T.}~\bibnamefont {Tomesh}}, \bibinfo {author} {\bibfnamefont {S.}~\bibnamefont {Chen}}, \bibinfo {author} {\bibfnamefont {L.}~\bibnamefont {Jiang}}, \bibinfo {author} {\bibfnamefont {S.}~\bibnamefont {Ferracin}}, \bibinfo {author} {\bibfnamefont {K.}~\bibnamefont {Rudinger}}, \bibinfo {author} {\bibfnamefont {T.}~\bibnamefont
  {Proctor}}, \bibinfo {author} {\bibfnamefont {K.~C.}\ \bibnamefont {Young}}, \bibinfo {author} {\bibfnamefont {R.}~\bibnamefont {Blume-Kohout}}, \ and\ \bibinfo {author} {\bibfnamefont {I.}~\bibnamefont {Siddiqi}},\ }\href {https://arxiv.org/abs/2408.12064} {\enquote {\bibinfo {title} {A practical introduction to benchmarking and characterization of quantum computers},}\ } (\bibinfo {year} {2024}),\ \Eprint {http://arxiv.org/abs/2408.12064} {arXiv:2408.12064 [quant-ph]} \BibitemShut {NoStop}%
\bibitem [{\citenamefont {Magnus}\ and\ \citenamefont {Neudecker}(1985)}]{magnus1985matrix}%
  \BibitemOpen
  \bibfield  {author} {\bibinfo {author} {\bibfnamefont {J.~R.}\ \bibnamefont {Magnus}}\ and\ \bibinfo {author} {\bibfnamefont {H.}~\bibnamefont {Neudecker}},\ }\href@noop {} {\bibfield  {journal} {\bibinfo  {journal} {Journal of Mathematical Psychology}\ }\textbf {\bibinfo {volume} {29}},\ \bibinfo {pages} {474} (\bibinfo {year} {1985})}\BibitemShut {NoStop}%
\bibitem [{\citenamefont {Flammia}\ and\ \citenamefont {Wallman}(2020)}]{10.1145/3408039}%
  \BibitemOpen
  \bibfield  {author} {\bibinfo {author} {\bibfnamefont {S.~T.}\ \bibnamefont {Flammia}}\ and\ \bibinfo {author} {\bibfnamefont {J.~J.}\ \bibnamefont {Wallman}},\ }\href {\doibase 10.1145/3408039} {\bibfield  {journal} {\bibinfo  {journal} {ACM Transactions on Quantum Computing}\ }\textbf {\bibinfo {volume} {1}} (\bibinfo {year} {2020}),\ 10.1145/3408039}\BibitemShut {NoStop}%
\bibitem [{\citenamefont {Scully}\ and\ \citenamefont {Zubairy}(1997)}]{scully1997quantum}%
  \BibitemOpen
  \bibfield  {author} {\bibinfo {author} {\bibfnamefont {M.~O.}\ \bibnamefont {Scully}}\ and\ \bibinfo {author} {\bibfnamefont {M.~S.}\ \bibnamefont {Zubairy}},\ }\href@noop {} {\emph {\bibinfo {title} {Quantum optics}}}\ (\bibinfo  {publisher} {Cambridge university press},\ \bibinfo {year} {1997})\BibitemShut {NoStop}%
\bibitem [{\citenamefont {Schleich}(2015)}]{schleich2015quantum}%
  \BibitemOpen
  \bibfield  {author} {\bibinfo {author} {\bibfnamefont {W.~P.}\ \bibnamefont {Schleich}},\ }\href@noop {} {\emph {\bibinfo {title} {Quantum optics in phase space}}}\ (\bibinfo  {publisher} {John Wiley \& Sons},\ \bibinfo {year} {2015})\BibitemShut {NoStop}%
\bibitem [{\citenamefont {Gardiner}\ and\ \citenamefont {Zoller}(2004)}]{gardiner2004quantum}%
  \BibitemOpen
  \bibfield  {author} {\bibinfo {author} {\bibfnamefont {C.}~\bibnamefont {Gardiner}}\ and\ \bibinfo {author} {\bibfnamefont {P.}~\bibnamefont {Zoller}},\ }\href@noop {} {\emph {\bibinfo {title} {Quantum noise: a handbook of Markovian and non-Markovian quantum stochastic methods with applications to quantum optics}}}\ (\bibinfo  {publisher} {Springer Science \& Business Media},\ \bibinfo {year} {2004})\BibitemShut {NoStop}%
\bibitem [{\citenamefont {Van~Kampen}(1992)}]{van1992stochastic}%
  \BibitemOpen
  \bibfield  {author} {\bibinfo {author} {\bibfnamefont {N.~G.}\ \bibnamefont {Van~Kampen}},\ }\href@noop {} {\emph {\bibinfo {title} {Stochastic processes in physics and chemistry}}},\ Vol.~\bibinfo {volume} {1}\ (\bibinfo  {publisher} {Elsevier},\ \bibinfo {year} {1992})\BibitemShut {NoStop}%
\bibitem [{\citenamefont {Gardiner}(1985)}]{gardiner1985handbook}%
  \BibitemOpen
  \bibfield  {author} {\bibinfo {author} {\bibfnamefont {C.}~\bibnamefont {Gardiner}},\ }\href {https://books.google.es/books?id=cRfvAAAAMAAJ} {\emph {\bibinfo {title} {Handbook of Stochastic Methods for Physics, Chemistry, and the Natural Sciences}}},\ Proceedings in Life Sciences\ (\bibinfo  {publisher} {Springer-Verlag},\ \bibinfo {year} {1985})\BibitemShut {NoStop}%
\bibitem [{\citenamefont {S\'anchez~Vel\'azquez}\ \emph {et~al.}(2025)\citenamefont {S\'anchez~Vel\'azquez}, \citenamefont {Steiner}, \citenamefont {Freund}, \citenamefont {Guevara-Bertsch}, \citenamefont {Marciniak}, \citenamefont {Monz},\ and\ \citenamefont {Bermudez}}]{PhysRevResearch.7.013008}%
  \BibitemOpen
  \bibfield  {author} {\bibinfo {author} {\bibfnamefont {J.~M.}\ \bibnamefont {S\'anchez~Vel\'azquez}}, \bibinfo {author} {\bibfnamefont {A.}~\bibnamefont {Steiner}}, \bibinfo {author} {\bibfnamefont {R.}~\bibnamefont {Freund}}, \bibinfo {author} {\bibfnamefont {M.}~\bibnamefont {Guevara-Bertsch}}, \bibinfo {author} {\bibfnamefont {C.~D.}\ \bibnamefont {Marciniak}}, \bibinfo {author} {\bibfnamefont {T.}~\bibnamefont {Monz}}, \ and\ \bibinfo {author} {\bibfnamefont {A.}~\bibnamefont {Bermudez}},\ }\href {\doibase 10.1103/PhysRevResearch.7.013008} {\bibfield  {journal} {\bibinfo  {journal} {Phys. Rev. Res.}\ }\textbf {\bibinfo {volume} {7}},\ \bibinfo {pages} {013008} (\bibinfo {year} {2025})}\BibitemShut {NoStop}%
\bibitem [{\citenamefont {Clerk}\ \emph {et~al.}(2010)\citenamefont {Clerk}, \citenamefont {Devoret}, \citenamefont {Girvin}, \citenamefont {Marquardt},\ and\ \citenamefont {Schoelkopf}}]{RevModPhys.82.1155}%
  \BibitemOpen
  \bibfield  {author} {\bibinfo {author} {\bibfnamefont {A.~A.}\ \bibnamefont {Clerk}}, \bibinfo {author} {\bibfnamefont {M.~H.}\ \bibnamefont {Devoret}}, \bibinfo {author} {\bibfnamefont {S.~M.}\ \bibnamefont {Girvin}}, \bibinfo {author} {\bibfnamefont {F.}~\bibnamefont {Marquardt}}, \ and\ \bibinfo {author} {\bibfnamefont {R.~J.}\ \bibnamefont {Schoelkopf}},\ }\href {\doibase 10.1103/RevModPhys.82.1155} {\bibfield  {journal} {\bibinfo  {journal} {Rev. Mod. Phys.}\ }\textbf {\bibinfo {volume} {82}},\ \bibinfo {pages} {1155} (\bibinfo {year} {2010})}\BibitemShut {NoStop}%
\bibitem [{\citenamefont {Varona}\ \emph {et~al.}(2024{\natexlab{a}})\citenamefont {Varona}, \citenamefont {Müller},\ and\ \citenamefont {Bermudez}}]{varona2024lindbladlikequantumtomographynonmarkovian}%
  \BibitemOpen
  \bibfield  {author} {\bibinfo {author} {\bibfnamefont {S.}~\bibnamefont {Varona}}, \bibinfo {author} {\bibfnamefont {M.}~\bibnamefont {Müller}}, \ and\ \bibinfo {author} {\bibfnamefont {A.}~\bibnamefont {Bermudez}},\ }\href {https://arxiv.org/abs/2403.19799} {\enquote {\bibinfo {title} {Lindblad-like quantum tomography for non-markovian quantum dynamical maps},}\ } (\bibinfo {year} {2024}{\natexlab{a}}),\ \Eprint {http://arxiv.org/abs/2403.19799} {arXiv:2403.19799 [quant-ph]} \BibitemShut {NoStop}%
\bibitem [{\citenamefont {Kofman}\ and\ \citenamefont {Kurizki}(2001{\natexlab{a}})}]{PhysRevLett.87.270405}%
  \BibitemOpen
  \bibfield  {author} {\bibinfo {author} {\bibfnamefont {A.~G.}\ \bibnamefont {Kofman}}\ and\ \bibinfo {author} {\bibfnamefont {G.}~\bibnamefont {Kurizki}},\ }\href {\doibase 10.1103/PhysRevLett.87.270405} {\bibfield  {journal} {\bibinfo  {journal} {Phys. Rev. Lett.}\ }\textbf {\bibinfo {volume} {87}},\ \bibinfo {pages} {270405} (\bibinfo {year} {2001}{\natexlab{a}})}\BibitemShut {NoStop}%
\bibitem [{\citenamefont {Biercuk}\ \emph {et~al.}(2011)\citenamefont {Biercuk}, \citenamefont {Doherty},\ and\ \citenamefont {Uys}}]{Biercuk_2011}%
  \BibitemOpen
  \bibfield  {author} {\bibinfo {author} {\bibfnamefont {M.~J.}\ \bibnamefont {Biercuk}}, \bibinfo {author} {\bibfnamefont {A.~C.}\ \bibnamefont {Doherty}}, \ and\ \bibinfo {author} {\bibfnamefont {H.}~\bibnamefont {Uys}},\ }\href {\doibase 10.1088/0953-4075/44/15/154002} {\bibfield  {journal} {\bibinfo  {journal} {Journal of Physics B: Atomic, Molecular and Optical Physics}\ }\textbf {\bibinfo {volume} {44}},\ \bibinfo {pages} {154002} (\bibinfo {year} {2011})}\BibitemShut {NoStop}%
\bibitem [{\citenamefont {Kofman}\ and\ \citenamefont {Kurizki}(2001{\natexlab{b}})}]{Kofman_2001}%
  \BibitemOpen
  \bibfield  {author} {\bibinfo {author} {\bibfnamefont {A.~G.}\ \bibnamefont {Kofman}}\ and\ \bibinfo {author} {\bibfnamefont {G.}~\bibnamefont {Kurizki}},\ }\href {\doibase 10.1515/zna-2001-0113} {\bibfield  {journal} {\bibinfo  {journal} {Zeitschrift für Naturforschung A}\ }\textbf {\bibinfo {volume} {56}},\ \bibinfo {pages} {83–90} (\bibinfo {year} {2001}{\natexlab{b}})}\BibitemShut {NoStop}%
\bibitem [{\citenamefont {Kofman}\ and\ \citenamefont {Kurizki}(2000)}]{Kofman2000}%
  \BibitemOpen
  \bibfield  {author} {\bibinfo {author} {\bibfnamefont {A.~G.}\ \bibnamefont {Kofman}}\ and\ \bibinfo {author} {\bibfnamefont {G.}~\bibnamefont {Kurizki}},\ }\href {\doibase 10.1038/35014537} {\bibfield  {journal} {\bibinfo  {journal} {Nature}\ }\textbf {\bibinfo {volume} {405}},\ \bibinfo {pages} {546} (\bibinfo {year} {2000})}\BibitemShut {NoStop}%
\bibitem [{\citenamefont {Kofman}\ and\ \citenamefont {Kurizki}(2004)}]{PhysRevLett.93.130406}%
  \BibitemOpen
  \bibfield  {author} {\bibinfo {author} {\bibfnamefont {A.~G.}\ \bibnamefont {Kofman}}\ and\ \bibinfo {author} {\bibfnamefont {G.}~\bibnamefont {Kurizki}},\ }\href {\doibase 10.1103/PhysRevLett.93.130406} {\bibfield  {journal} {\bibinfo  {journal} {Phys. Rev. Lett.}\ }\textbf {\bibinfo {volume} {93}},\ \bibinfo {pages} {130406} (\bibinfo {year} {2004})}\BibitemShut {NoStop}%
\bibitem [{\citenamefont {Gordon}\ \emph {et~al.}(2007)\citenamefont {Gordon}, \citenamefont {Erez},\ and\ \citenamefont {Kurizki}}]{Gordon_2007}%
  \BibitemOpen
  \bibfield  {author} {\bibinfo {author} {\bibfnamefont {G.}~\bibnamefont {Gordon}}, \bibinfo {author} {\bibfnamefont {N.}~\bibnamefont {Erez}}, \ and\ \bibinfo {author} {\bibfnamefont {G.}~\bibnamefont {Kurizki}},\ }\href {\doibase 10.1088/0953-4075/40/9/S04} {\bibfield  {journal} {\bibinfo  {journal} {Journal of Physics B: Atomic, Molecular and Optical Physics}\ }\textbf {\bibinfo {volume} {40}},\ \bibinfo {pages} {S75} (\bibinfo {year} {2007})}\BibitemShut {NoStop}%
\bibitem [{\citenamefont {Uhrig}(2008)}]{Uhrig_2008}%
  \BibitemOpen
  \bibfield  {author} {\bibinfo {author} {\bibfnamefont {G.~S.}\ \bibnamefont {Uhrig}},\ }\href {\doibase 10.1088/1367-2630/10/8/083024} {\bibfield  {journal} {\bibinfo  {journal} {New Journal of Physics}\ }\textbf {\bibinfo {volume} {10}},\ \bibinfo {pages} {083024} (\bibinfo {year} {2008})}\BibitemShut {NoStop}%
\bibitem [{\citenamefont {Cywi\ifmmode~\acute{n}\else \'{n}\fi{}ski}\ \emph {et~al.}(2008)\citenamefont {Cywi\ifmmode~\acute{n}\else \'{n}\fi{}ski}, \citenamefont {Lutchyn}, \citenamefont {Nave},\ and\ \citenamefont {Das~Sarma}}]{PhysRevB.77.174509}%
  \BibitemOpen
  \bibfield  {author} {\bibinfo {author} {\bibfnamefont {L.}~\bibnamefont {Cywi\ifmmode~\acute{n}\else \'{n}\fi{}ski}}, \bibinfo {author} {\bibfnamefont {R.~M.}\ \bibnamefont {Lutchyn}}, \bibinfo {author} {\bibfnamefont {C.~P.}\ \bibnamefont {Nave}}, \ and\ \bibinfo {author} {\bibfnamefont {S.}~\bibnamefont {Das~Sarma}},\ }\href {\doibase 10.1103/PhysRevB.77.174509} {\bibfield  {journal} {\bibinfo  {journal} {Phys. Rev. B}\ }\textbf {\bibinfo {volume} {77}},\ \bibinfo {pages} {174509} (\bibinfo {year} {2008})}\BibitemShut {NoStop}%
\bibitem [{\citenamefont {Almog}\ \emph {et~al.}(2011)\citenamefont {Almog}, \citenamefont {Sagi}, \citenamefont {Gordon}, \citenamefont {Bensky}, \citenamefont {Kurizki},\ and\ \citenamefont {Davidson}}]{Almog_2011}%
  \BibitemOpen
  \bibfield  {author} {\bibinfo {author} {\bibfnamefont {I.}~\bibnamefont {Almog}}, \bibinfo {author} {\bibfnamefont {Y.}~\bibnamefont {Sagi}}, \bibinfo {author} {\bibfnamefont {G.}~\bibnamefont {Gordon}}, \bibinfo {author} {\bibfnamefont {G.}~\bibnamefont {Bensky}}, \bibinfo {author} {\bibfnamefont {G.}~\bibnamefont {Kurizki}}, \ and\ \bibinfo {author} {\bibfnamefont {N.}~\bibnamefont {Davidson}},\ }\href {\doibase 10.1088/0953-4075/44/15/154006} {\bibfield  {journal} {\bibinfo  {journal} {Journal of Physics B: Atomic, Molecular and Optical Physics}\ }\textbf {\bibinfo {volume} {44}},\ \bibinfo {pages} {154006} (\bibinfo {year} {2011})}\BibitemShut {NoStop}%
\bibitem [{\citenamefont {Kamenev}(2011)}]{Kamenev:341673}%
  \BibitemOpen
  \bibfield  {author} {\bibinfo {author} {\bibfnamefont {A.}~\bibnamefont {Kamenev}},\ }\href {https://bib-pubdb1.desy.de/record/341673} {\emph {\bibinfo {title} {{F}ield theory of non-equilibrium systems; 1st publ.}}}\ (\bibinfo  {publisher} {Cambridge Univ. Pr.},\ \bibinfo {address} {Cambridge},\ \bibinfo {year} {2011})\ p.\ \bibinfo {pages} {341 pages}\BibitemShut {NoStop}%
\bibitem [{\citenamefont {Bermudez}\ \emph {et~al.}(2017)\citenamefont {Bermudez}, \citenamefont {Aarts},\ and\ \citenamefont {M\"uller}}]{PhysRevX.7.041012}%
  \BibitemOpen
  \bibfield  {author} {\bibinfo {author} {\bibfnamefont {A.}~\bibnamefont {Bermudez}}, \bibinfo {author} {\bibfnamefont {G.}~\bibnamefont {Aarts}}, \ and\ \bibinfo {author} {\bibfnamefont {M.}~\bibnamefont {M\"uller}},\ }\href {\doibase 10.1103/PhysRevX.7.041012} {\bibfield  {journal} {\bibinfo  {journal} {Phys. Rev. X}\ }\textbf {\bibinfo {volume} {7}},\ \bibinfo {pages} {041012} (\bibinfo {year} {2017})}\BibitemShut {NoStop}%
\bibitem [{\citenamefont {Mart\'{\i}n-V\'azquez}\ \emph {et~al.}(2022)\citenamefont {Mart\'{\i}n-V\'azquez}, \citenamefont {Aarts}, \citenamefont {M\"uller},\ and\ \citenamefont {Bermudez}}]{PRXQuantum.3.020352}%
  \BibitemOpen
  \bibfield  {author} {\bibinfo {author} {\bibfnamefont {G.}~\bibnamefont {Mart\'{\i}n-V\'azquez}}, \bibinfo {author} {\bibfnamefont {G.}~\bibnamefont {Aarts}}, \bibinfo {author} {\bibfnamefont {M.}~\bibnamefont {M\"uller}}, \ and\ \bibinfo {author} {\bibfnamefont {A.}~\bibnamefont {Bermudez}},\ }\href {\doibase 10.1103/PRXQuantum.3.020352} {\bibfield  {journal} {\bibinfo  {journal} {PRX Quantum}\ }\textbf {\bibinfo {volume} {3}},\ \bibinfo {pages} {020352} (\bibinfo {year} {2022})}\BibitemShut {NoStop}%
\bibitem [{\citenamefont {Varona}\ \emph {et~al.}(2024{\natexlab{b}})\citenamefont {Varona}, \citenamefont {Saner}, \citenamefont {Băzăvan}, \citenamefont {Araneda}, \citenamefont {Aarts},\ and\ \citenamefont {Bermudez}}]{varona2024quantumcomputingfeynmandiagrams}%
  \BibitemOpen
  \bibfield  {author} {\bibinfo {author} {\bibfnamefont {S.}~\bibnamefont {Varona}}, \bibinfo {author} {\bibfnamefont {S.}~\bibnamefont {Saner}}, \bibinfo {author} {\bibfnamefont {O.}~\bibnamefont {Băzăvan}}, \bibinfo {author} {\bibfnamefont {G.}~\bibnamefont {Araneda}}, \bibinfo {author} {\bibfnamefont {G.}~\bibnamefont {Aarts}}, \ and\ \bibinfo {author} {\bibfnamefont {A.}~\bibnamefont {Bermudez}},\ }\href {https://arxiv.org/abs/2411.05092} {\enquote {\bibinfo {title} {Towards quantum computing feynman diagrams in hybrid qubit-oscillator devices},}\ } (\bibinfo {year} {2024}{\natexlab{b}}),\ \Eprint {http://arxiv.org/abs/2411.05092} {arXiv:2411.05092 [quant-ph]} \BibitemShut {NoStop}%
\bibitem [{\citenamefont {Mart{\'{i}}nez}\ \emph {et~al.}(2024)\citenamefont {Mart{\'{i}}nez}, \citenamefont {L{\'{o}}pez},\ and\ \citenamefont {Bermudez}}]{Martinez2024thermalmasses}%
  \BibitemOpen
  \bibfield  {author} {\bibinfo {author} {\bibfnamefont {P.~V.}\ \bibnamefont {Mart{\'{i}}nez}}, \bibinfo {author} {\bibfnamefont {E.}~\bibnamefont {L{\'{o}}pez}}, \ and\ \bibinfo {author} {\bibfnamefont {A.}~\bibnamefont {Bermudez}},\ }\href {\doibase 10.22331/q-2024-07-15-1411} {\bibfield  {journal} {\bibinfo  {journal} {{Quantum}}\ }\textbf {\bibinfo {volume} {8}},\ \bibinfo {pages} {1411} (\bibinfo {year} {2024})}\BibitemShut {NoStop}%
\bibitem [{\citenamefont {Kraus}\ \emph {et~al.}(1983)\citenamefont {Kraus}, \citenamefont {B{\"o}hm}, \citenamefont {Dollard},\ and\ \citenamefont {Wootters}}]{kraus1983states}%
  \BibitemOpen
  \bibfield  {author} {\bibinfo {author} {\bibfnamefont {K.}~\bibnamefont {Kraus}}, \bibinfo {author} {\bibfnamefont {A.}~\bibnamefont {B{\"o}hm}}, \bibinfo {author} {\bibfnamefont {J.~D.}\ \bibnamefont {Dollard}}, \ and\ \bibinfo {author} {\bibfnamefont {W.}~\bibnamefont {Wootters}},\ }\href@noop {} {\emph {\bibinfo {title} {States, Effects, and Operations Fundamental Notions of Quantum Theory: Lectures in Mathematical Physics at the University of Texas at Austin}}}\ (\bibinfo  {publisher} {Springer},\ \bibinfo {year} {1983})\BibitemShut {NoStop}%
\bibitem [{\citenamefont {DeCross}\ \emph {et~al.}(2025)\citenamefont {DeCross}, \citenamefont {Haghshenas}, \citenamefont {Liu}, \citenamefont {Rinaldi}, \citenamefont {Gray}, \citenamefont {Alexeev}, \citenamefont {Baldwin}, \citenamefont {Bartolotta}, \citenamefont {Bohn}, \citenamefont {Chertkov}, \citenamefont {Cline}, \citenamefont {Colina}, \citenamefont {DelVento}, \citenamefont {Dreiling}, \citenamefont {Foltz}, \citenamefont {Gaebler}, \citenamefont {Gatterman}, \citenamefont {Gilbreth}, \citenamefont {Giles}, \citenamefont {Gresh}, \citenamefont {Hall}, \citenamefont {Hankin}, \citenamefont {Hansen}, \citenamefont {Hewitt}, \citenamefont {Hoffman}, \citenamefont {Holliman}, \citenamefont {Hutson}, \citenamefont {Jacobs}, \citenamefont {Johansen}, \citenamefont {Lee}, \citenamefont {Lehman}, \citenamefont {Lucchetti}, \citenamefont {Lykov}, \citenamefont {Madjarov}, \citenamefont {Mathewson}, \citenamefont {Mayer}, \citenamefont {Mills}, \citenamefont {Niroula}, \citenamefont {Pino}, \citenamefont
  {Roman}, \citenamefont {Schecter}, \citenamefont {Siegfried}, \citenamefont {Tiemann}, \citenamefont {Volin}, \citenamefont {Walker}, \citenamefont {Shaydulin}, \citenamefont {Pistoia}, \citenamefont {Moses}, \citenamefont {Hayes}, \citenamefont {Neyenhuis}, \citenamefont {Stutz},\ and\ \citenamefont {Foss-Feig}}]{DeCross_2025}%
  \BibitemOpen
  \bibfield  {author} {\bibinfo {author} {\bibfnamefont {M.}~\bibnamefont {DeCross}}, \bibinfo {author} {\bibfnamefont {R.}~\bibnamefont {Haghshenas}}, \bibinfo {author} {\bibfnamefont {M.}~\bibnamefont {Liu}}, \bibinfo {author} {\bibfnamefont {E.}~\bibnamefont {Rinaldi}}, \bibinfo {author} {\bibfnamefont {J.}~\bibnamefont {Gray}}, \bibinfo {author} {\bibfnamefont {Y.}~\bibnamefont {Alexeev}}, \bibinfo {author} {\bibfnamefont {C.}~\bibnamefont {Baldwin}}, \bibinfo {author} {\bibfnamefont {J.}~\bibnamefont {Bartolotta}}, \bibinfo {author} {\bibfnamefont {M.}~\bibnamefont {Bohn}}, \bibinfo {author} {\bibfnamefont {E.}~\bibnamefont {Chertkov}}, \bibinfo {author} {\bibfnamefont {J.}~\bibnamefont {Cline}}, \bibinfo {author} {\bibfnamefont {J.}~\bibnamefont {Colina}}, \bibinfo {author} {\bibfnamefont {D.}~\bibnamefont {DelVento}}, \bibinfo {author} {\bibfnamefont {J.}~\bibnamefont {Dreiling}}, \bibinfo {author} {\bibfnamefont {C.}~\bibnamefont {Foltz}}, \bibinfo {author} {\bibfnamefont {J.}~\bibnamefont {Gaebler}},
  \bibinfo {author} {\bibfnamefont {T.}~\bibnamefont {Gatterman}}, \bibinfo {author} {\bibfnamefont {C.}~\bibnamefont {Gilbreth}}, \bibinfo {author} {\bibfnamefont {J.}~\bibnamefont {Giles}}, \bibinfo {author} {\bibfnamefont {D.}~\bibnamefont {Gresh}}, \bibinfo {author} {\bibfnamefont {A.}~\bibnamefont {Hall}}, \bibinfo {author} {\bibfnamefont {A.}~\bibnamefont {Hankin}}, \bibinfo {author} {\bibfnamefont {A.}~\bibnamefont {Hansen}}, \bibinfo {author} {\bibfnamefont {N.}~\bibnamefont {Hewitt}}, \bibinfo {author} {\bibfnamefont {I.}~\bibnamefont {Hoffman}}, \bibinfo {author} {\bibfnamefont {C.}~\bibnamefont {Holliman}}, \bibinfo {author} {\bibfnamefont {R.}~\bibnamefont {Hutson}}, \bibinfo {author} {\bibfnamefont {T.}~\bibnamefont {Jacobs}}, \bibinfo {author} {\bibfnamefont {J.}~\bibnamefont {Johansen}}, \bibinfo {author} {\bibfnamefont {P.}~\bibnamefont {Lee}}, \bibinfo {author} {\bibfnamefont {E.}~\bibnamefont {Lehman}}, \bibinfo {author} {\bibfnamefont {D.}~\bibnamefont {Lucchetti}}, \bibinfo {author}
  {\bibfnamefont {D.}~\bibnamefont {Lykov}}, \bibinfo {author} {\bibfnamefont {I.}~\bibnamefont {Madjarov}}, \bibinfo {author} {\bibfnamefont {B.}~\bibnamefont {Mathewson}}, \bibinfo {author} {\bibfnamefont {K.}~\bibnamefont {Mayer}}, \bibinfo {author} {\bibfnamefont {M.}~\bibnamefont {Mills}}, \bibinfo {author} {\bibfnamefont {P.}~\bibnamefont {Niroula}}, \bibinfo {author} {\bibfnamefont {J.}~\bibnamefont {Pino}}, \bibinfo {author} {\bibfnamefont {C.}~\bibnamefont {Roman}}, \bibinfo {author} {\bibfnamefont {M.}~\bibnamefont {Schecter}}, \bibinfo {author} {\bibfnamefont {P.}~\bibnamefont {Siegfried}}, \bibinfo {author} {\bibfnamefont {B.}~\bibnamefont {Tiemann}}, \bibinfo {author} {\bibfnamefont {C.}~\bibnamefont {Volin}}, \bibinfo {author} {\bibfnamefont {J.}~\bibnamefont {Walker}}, \bibinfo {author} {\bibfnamefont {R.}~\bibnamefont {Shaydulin}}, \bibinfo {author} {\bibfnamefont {M.}~\bibnamefont {Pistoia}}, \bibinfo {author} {\bibfnamefont {S.}~\bibnamefont {Moses}}, \bibinfo {author} {\bibfnamefont
  {D.}~\bibnamefont {Hayes}}, \bibinfo {author} {\bibfnamefont {B.}~\bibnamefont {Neyenhuis}}, \bibinfo {author} {\bibfnamefont {R.}~\bibnamefont {Stutz}}, \ and\ \bibinfo {author} {\bibfnamefont {M.}~\bibnamefont {Foss-Feig}},\ }\href {\doibase 10.1103/physrevx.15.021052} {\bibfield  {journal} {\bibinfo  {journal} {Physical Review X}\ }\textbf {\bibinfo {volume} {15}} (\bibinfo {year} {2025}),\ 10.1103/physrevx.15.021052}\BibitemShut {NoStop}%
\bibitem [{\citenamefont {Pino}\ \emph {et~al.}(2021)\citenamefont {Pino}, \citenamefont {Dreiling}, \citenamefont {Figgatt}, \citenamefont {Gaebler}, \citenamefont {Moses}, \citenamefont {Allman}, \citenamefont {Baldwin}, \citenamefont {Foss-Feig}, \citenamefont {Hayes}, \citenamefont {Mayer}, \citenamefont {Ryan-Anderson},\ and\ \citenamefont {Neyenhuis}}]{Pino_2021}%
  \BibitemOpen
  \bibfield  {author} {\bibinfo {author} {\bibfnamefont {J.~M.}\ \bibnamefont {Pino}}, \bibinfo {author} {\bibfnamefont {J.~M.}\ \bibnamefont {Dreiling}}, \bibinfo {author} {\bibfnamefont {C.}~\bibnamefont {Figgatt}}, \bibinfo {author} {\bibfnamefont {J.~P.}\ \bibnamefont {Gaebler}}, \bibinfo {author} {\bibfnamefont {S.~A.}\ \bibnamefont {Moses}}, \bibinfo {author} {\bibfnamefont {M.~S.}\ \bibnamefont {Allman}}, \bibinfo {author} {\bibfnamefont {C.~H.}\ \bibnamefont {Baldwin}}, \bibinfo {author} {\bibfnamefont {M.}~\bibnamefont {Foss-Feig}}, \bibinfo {author} {\bibfnamefont {D.}~\bibnamefont {Hayes}}, \bibinfo {author} {\bibfnamefont {K.}~\bibnamefont {Mayer}}, \bibinfo {author} {\bibfnamefont {C.}~\bibnamefont {Ryan-Anderson}}, \ and\ \bibinfo {author} {\bibfnamefont {B.}~\bibnamefont {Neyenhuis}},\ }\href {\doibase 10.1038/s41586-021-03318-4} {\bibfield  {journal} {\bibinfo  {journal} {Nature}\ }\textbf {\bibinfo {volume} {592}},\ \bibinfo {pages} {209–213} (\bibinfo {year} {2021})}\BibitemShut {NoStop}%
\bibitem [{\citenamefont {Scott}(2006)}]{scott2006tight}%
  \BibitemOpen
  \bibfield  {author} {\bibinfo {author} {\bibfnamefont {A.~J.}\ \bibnamefont {Scott}},\ }\href@noop {} {\bibfield  {journal} {\bibinfo  {journal} {Journal of Physics A: Mathematical and General}\ }\textbf {\bibinfo {volume} {39}},\ \bibinfo {pages} {13507} (\bibinfo {year} {2006})}\BibitemShut {NoStop}%
\bibitem [{\citenamefont {Rudinger}\ \emph {et~al.}(2023)\citenamefont {Rudinger}, \citenamefont {Ostrove}, \citenamefont {Seritan}, \citenamefont {Grace}, \citenamefont {Nielsen}, \citenamefont {Blume-Kohout},\ and\ \citenamefont {Young}}]{rudinger2023twoqubitgatesettomography}%
  \BibitemOpen
  \bibfield  {author} {\bibinfo {author} {\bibfnamefont {K.~M.}\ \bibnamefont {Rudinger}}, \bibinfo {author} {\bibfnamefont {C.~I.}\ \bibnamefont {Ostrove}}, \bibinfo {author} {\bibfnamefont {S.~K.}\ \bibnamefont {Seritan}}, \bibinfo {author} {\bibfnamefont {M.~D.}\ \bibnamefont {Grace}}, \bibinfo {author} {\bibfnamefont {E.}~\bibnamefont {Nielsen}}, \bibinfo {author} {\bibfnamefont {R.~J.}\ \bibnamefont {Blume-Kohout}}, \ and\ \bibinfo {author} {\bibfnamefont {K.~C.}\ \bibnamefont {Young}},\ }\href {https://arxiv.org/abs/2307.15767} {\enquote {\bibinfo {title} {Two-qubit gate set tomography with fewer circuits},}\ } (\bibinfo {year} {2023}),\ \Eprint {http://arxiv.org/abs/2307.15767} {arXiv:2307.15767 [quant-ph]} \BibitemShut {NoStop}%
\bibitem [{\citenamefont {Gillespie}(1996)}]{10.1119/1.18210}%
  \BibitemOpen
  \bibfield  {author} {\bibinfo {author} {\bibfnamefont {D.~T.}\ \bibnamefont {Gillespie}},\ }\href {\doibase 10.1119/1.18210} {\bibfield  {journal} {\bibinfo  {journal} {American Journal of Physics}\ }\textbf {\bibinfo {volume} {64}},\ \bibinfo {pages} {225} (\bibinfo {year} {1996})}\BibitemShut {NoStop}%
\bibitem [{\citenamefont {Dixit}\ \emph {et~al.}(1980)\citenamefont {Dixit}, \citenamefont {Zoller},\ and\ \citenamefont {Lambropoulos}}]{PhysRevA.21.1289}%
  \BibitemOpen
  \bibfield  {author} {\bibinfo {author} {\bibfnamefont {S.~N.}\ \bibnamefont {Dixit}}, \bibinfo {author} {\bibfnamefont {P.}~\bibnamefont {Zoller}}, \ and\ \bibinfo {author} {\bibfnamefont {P.}~\bibnamefont {Lambropoulos}},\ }\href {\doibase 10.1103/PhysRevA.21.1289} {\bibfield  {journal} {\bibinfo  {journal} {Phys. Rev. A}\ }\textbf {\bibinfo {volume} {21}},\ \bibinfo {pages} {1289} (\bibinfo {year} {1980})}\BibitemShut {NoStop}%
\bibitem [{\citenamefont {Avan}\ and\ \citenamefont {Cohen-Tannoudji}(1977)}]{avan1977two}%
  \BibitemOpen
  \bibfield  {author} {\bibinfo {author} {\bibfnamefont {P.}~\bibnamefont {Avan}}\ and\ \bibinfo {author} {\bibfnamefont {C.}~\bibnamefont {Cohen-Tannoudji}},\ }\href@noop {} {\bibfield  {journal} {\bibinfo  {journal} {Journal of Physics B: Atomic and Molecular Physics}\ }\textbf {\bibinfo {volume} {10}},\ \bibinfo {pages} {155} (\bibinfo {year} {1977})}\BibitemShut {NoStop}%
\bibitem [{\citenamefont {Nielsen}\ \emph {et~al.}(2020)\citenamefont {Nielsen}, \citenamefont {Rudinger}, \citenamefont {Proctor}, \citenamefont {Russo}, \citenamefont {Young},\ and\ \citenamefont {Blume-Kohout}}]{Nielsen_2020}%
  \BibitemOpen
  \bibfield  {author} {\bibinfo {author} {\bibfnamefont {E.}~\bibnamefont {Nielsen}}, \bibinfo {author} {\bibfnamefont {K.}~\bibnamefont {Rudinger}}, \bibinfo {author} {\bibfnamefont {T.}~\bibnamefont {Proctor}}, \bibinfo {author} {\bibfnamefont {A.}~\bibnamefont {Russo}}, \bibinfo {author} {\bibfnamefont {K.}~\bibnamefont {Young}}, \ and\ \bibinfo {author} {\bibfnamefont {R.}~\bibnamefont {Blume-Kohout}},\ }\href {\doibase 10.1088/2058-9565/ab8aa4} {\bibfield  {journal} {\bibinfo  {journal} {Quantum Science and Technology}\ }\textbf {\bibinfo {volume} {5}},\ \bibinfo {pages} {044002} (\bibinfo {year} {2020})}\BibitemShut {NoStop}%
\bibitem [{\citenamefont {et. al.}(2016)}]{pygsti}%
  \BibitemOpen
  \bibfield  {author} {\bibinfo {author} {\bibfnamefont {E.~N.}\ \bibnamefont {et. al.}},\ }\href@noop {} {\enquote {\bibinfo {title} {pygsti: A python implementation of gate set tomography},}\ }\bibinfo {howpublished} {\url{https://github.com/sandialabs/pyGSTi}} (\bibinfo {year} {2016})\BibitemShut {NoStop}%
\bibitem [{\citenamefont {Aharonov}\ \emph {et~al.}(1999)\citenamefont {Aharonov}, \citenamefont {Kitaev},\ and\ \citenamefont {Nisan}}]{aharanov_kitaev}%
  \BibitemOpen
  \bibfield  {author} {\bibinfo {author} {\bibfnamefont {D.}~\bibnamefont {Aharonov}}, \bibinfo {author} {\bibfnamefont {A.}~\bibnamefont {Kitaev}}, \ and\ \bibinfo {author} {\bibfnamefont {N.}~\bibnamefont {Nisan}},\ }\href {\doibase 10.1145/276698.276708} {\bibfield  {journal} {\bibinfo  {journal} {Proc. of 30th STOC}\ } (\bibinfo {year} {1999}),\ 10.1145/276698.276708}\BibitemShut {NoStop}%
\bibitem [{\citenamefont {Degen}\ \emph {et~al.}(2017)\citenamefont {Degen}, \citenamefont {Reinhard},\ and\ \citenamefont {Cappellaro}}]{degen2017quantum}%
  \BibitemOpen
  \bibfield  {author} {\bibinfo {author} {\bibfnamefont {C.}~\bibnamefont {Degen}}, \bibinfo {author} {\bibfnamefont {F.}~\bibnamefont {Reinhard}}, \ and\ \bibinfo {author} {\bibfnamefont {P.}~\bibnamefont {Cappellaro}},\ }\href {\doibase 10.1103/revmodphys.89.035002} {\bibfield  {journal} {\bibinfo  {journal} {Reviews of Modern Physics}\ }\textbf {\bibinfo {volume} {89}} (\bibinfo {year} {2017}),\ 10.1103/revmodphys.89.035002}\BibitemShut {NoStop}%
\bibitem [{\citenamefont {Haase}\ \emph {et~al.}(2016)\citenamefont {Haase}, \citenamefont {Smirne}, \citenamefont {Huelga}, \citenamefont {Kołodynski},\ and\ \citenamefont {Demkowicz-Dobrzanski}}]{haase2016precision}%
  \BibitemOpen
  \bibfield  {author} {\bibinfo {author} {\bibfnamefont {J.~F.}\ \bibnamefont {Haase}}, \bibinfo {author} {\bibfnamefont {A.}~\bibnamefont {Smirne}}, \bibinfo {author} {\bibfnamefont {S.~F.}\ \bibnamefont {Huelga}}, \bibinfo {author} {\bibfnamefont {J.}~\bibnamefont {Kołodynski}}, \ and\ \bibinfo {author} {\bibfnamefont {R.}~\bibnamefont {Demkowicz-Dobrzanski}},\ }\href {\doibase 10.1515/qmetro-2018-0002} {\bibfield  {journal} {\bibinfo  {journal} {Quantum Measurements and Quantum Metrology}\ }\textbf {\bibinfo {volume} {5}},\ \bibinfo {pages} {13–39} (\bibinfo {year} {2016})}\BibitemShut {NoStop}%
\bibitem [{\citenamefont {Smirne}\ \emph {et~al.}(2016{\natexlab{a}})\citenamefont {Smirne}, \citenamefont {Ko\l{}ody\ifmmode~\acute{n}\else \'{n}\fi{}ski}, \citenamefont {Huelga},\ and\ \citenamefont {Demkowicz-Dobrza\ifmmode~\acute{n}\else \'{n}\fi{}ski}}]{smirne2016ultimate}%
  \BibitemOpen
  \bibfield  {author} {\bibinfo {author} {\bibfnamefont {A.}~\bibnamefont {Smirne}}, \bibinfo {author} {\bibfnamefont {J.}~\bibnamefont {Ko\l{}ody\ifmmode~\acute{n}\else \'{n}\fi{}ski}}, \bibinfo {author} {\bibfnamefont {S.~F.}\ \bibnamefont {Huelga}}, \ and\ \bibinfo {author} {\bibfnamefont {R.}~\bibnamefont {Demkowicz-Dobrza\ifmmode~\acute{n}\else \'{n}\fi{}ski}},\ }\href {\doibase 10.1103/PhysRevLett.116.120801} {\bibfield  {journal} {\bibinfo  {journal} {Phys. Rev. Lett.}\ }\textbf {\bibinfo {volume} {116}},\ \bibinfo {pages} {120801} (\bibinfo {year} {2016}{\natexlab{a}})}\BibitemShut {NoStop}%
\bibitem [{\citenamefont {Itano}\ \emph {et~al.}(1993)\citenamefont {Itano}, \citenamefont {Bergquist}, \citenamefont {Bollinger}, \citenamefont {Gilligan}, \citenamefont {Heinzen}, \citenamefont {Moore}, \citenamefont {Raizen},\ and\ \citenamefont {Wineland}}]{PhysRevA.47.3554}%
  \BibitemOpen
  \bibfield  {author} {\bibinfo {author} {\bibfnamefont {W.~M.}\ \bibnamefont {Itano}}, \bibinfo {author} {\bibfnamefont {J.~C.}\ \bibnamefont {Bergquist}}, \bibinfo {author} {\bibfnamefont {J.~J.}\ \bibnamefont {Bollinger}}, \bibinfo {author} {\bibfnamefont {J.~M.}\ \bibnamefont {Gilligan}}, \bibinfo {author} {\bibfnamefont {D.~J.}\ \bibnamefont {Heinzen}}, \bibinfo {author} {\bibfnamefont {F.~L.}\ \bibnamefont {Moore}}, \bibinfo {author} {\bibfnamefont {M.~G.}\ \bibnamefont {Raizen}}, \ and\ \bibinfo {author} {\bibfnamefont {D.~J.}\ \bibnamefont {Wineland}},\ }\href {\doibase 10.1103/PhysRevA.47.3554} {\bibfield  {journal} {\bibinfo  {journal} {Phys. Rev. A}\ }\textbf {\bibinfo {volume} {47}},\ \bibinfo {pages} {3554} (\bibinfo {year} {1993})}\BibitemShut {NoStop}%
\bibitem [{\citenamefont {Breuer}\ \emph {et~al.}(2009)\citenamefont {Breuer}, \citenamefont {Laine},\ and\ \citenamefont {Piilo}}]{PhysRevLett.103.210401}%
  \BibitemOpen
  \bibfield  {author} {\bibinfo {author} {\bibfnamefont {H.-P.}\ \bibnamefont {Breuer}}, \bibinfo {author} {\bibfnamefont {E.-M.}\ \bibnamefont {Laine}}, \ and\ \bibinfo {author} {\bibfnamefont {J.}~\bibnamefont {Piilo}},\ }\href {\doibase 10.1103/PhysRevLett.103.210401} {\bibfield  {journal} {\bibinfo  {journal} {Phys. Rev. Lett.}\ }\textbf {\bibinfo {volume} {103}},\ \bibinfo {pages} {210401} (\bibinfo {year} {2009})}\BibitemShut {NoStop}%
\bibitem [{\citenamefont {Rivas}\ \emph {et~al.}(2010)\citenamefont {Rivas}, \citenamefont {Huelga},\ and\ \citenamefont {Plenio}}]{PhysRevLett.105.050403}%
  \BibitemOpen
  \bibfield  {author} {\bibinfo {author} {\bibfnamefont {A.}~\bibnamefont {Rivas}}, \bibinfo {author} {\bibfnamefont {S.~F.}\ \bibnamefont {Huelga}}, \ and\ \bibinfo {author} {\bibfnamefont {M.~B.}\ \bibnamefont {Plenio}},\ }\href {\doibase 10.1103/PhysRevLett.105.050403} {\bibfield  {journal} {\bibinfo  {journal} {Phys. Rev. Lett.}\ }\textbf {\bibinfo {volume} {105}},\ \bibinfo {pages} {050403} (\bibinfo {year} {2010})}\BibitemShut {NoStop}%
\bibitem [{\citenamefont {Gerry}\ and\ \citenamefont {Knight}(2023)}]{gerry2023introductory}%
  \BibitemOpen
  \bibfield  {author} {\bibinfo {author} {\bibfnamefont {C.~C.}\ \bibnamefont {Gerry}}\ and\ \bibinfo {author} {\bibfnamefont {P.~L.}\ \bibnamefont {Knight}},\ }\href@noop {} {\emph {\bibinfo {title} {Introductory quantum optics}}}\ (\bibinfo  {publisher} {Cambridge university press},\ \bibinfo {year} {2023})\BibitemShut {NoStop}%
\bibitem [{\citenamefont {Eadie}\ and\ \citenamefont {Kirkpatrick}(1973)}]{eadie1973statistical}%
  \BibitemOpen
  \bibfield  {author} {\bibinfo {author} {\bibfnamefont {W.~T.}\ \bibnamefont {Eadie}}\ and\ \bibinfo {author} {\bibfnamefont {L.~D.}\ \bibnamefont {Kirkpatrick}},\ }\href@noop {} {\enquote {\bibinfo {title} {Statistical methods in experimental physics},}\ } (\bibinfo {year} {1973})\BibitemShut {NoStop}%
\bibitem [{\citenamefont {Rossi}(2018)}]{rossi2018mathematical}%
  \BibitemOpen
  \bibfield  {author} {\bibinfo {author} {\bibfnamefont {R.~J.}\ \bibnamefont {Rossi}},\ }\href@noop {} {\emph {\bibinfo {title} {Mathematical statistics: an introduction to likelihood based inference}}}\ (\bibinfo  {publisher} {John Wiley \& Sons},\ \bibinfo {year} {2018})\BibitemShut {NoStop}%
\bibitem [{\citenamefont {Ramsey}(1950)}]{PhysRev.78.695}%
  \BibitemOpen
  \bibfield  {author} {\bibinfo {author} {\bibfnamefont {N.~F.}\ \bibnamefont {Ramsey}},\ }\href {\doibase 10.1103/PhysRev.78.695} {\bibfield  {journal} {\bibinfo  {journal} {Phys. Rev.}\ }\textbf {\bibinfo {volume} {78}},\ \bibinfo {pages} {695} (\bibinfo {year} {1950})}\BibitemShut {NoStop}%
\bibitem [{\citenamefont {Ithier}\ \emph {et~al.}(2005)\citenamefont {Ithier}, \citenamefont {Collin}, \citenamefont {Joyez}, \citenamefont {Meeson}, \citenamefont {Vion}, \citenamefont {Esteve}, \citenamefont {Chiarello}, \citenamefont {Shnirman}, \citenamefont {Makhlin}, \citenamefont {Schriefl},\ and\ \citenamefont {Sch\"on}}]{PhysRevB.72.134519}%
  \BibitemOpen
  \bibfield  {author} {\bibinfo {author} {\bibfnamefont {G.}~\bibnamefont {Ithier}}, \bibinfo {author} {\bibfnamefont {E.}~\bibnamefont {Collin}}, \bibinfo {author} {\bibfnamefont {P.}~\bibnamefont {Joyez}}, \bibinfo {author} {\bibfnamefont {P.~J.}\ \bibnamefont {Meeson}}, \bibinfo {author} {\bibfnamefont {D.}~\bibnamefont {Vion}}, \bibinfo {author} {\bibfnamefont {D.}~\bibnamefont {Esteve}}, \bibinfo {author} {\bibfnamefont {F.}~\bibnamefont {Chiarello}}, \bibinfo {author} {\bibfnamefont {A.}~\bibnamefont {Shnirman}}, \bibinfo {author} {\bibfnamefont {Y.}~\bibnamefont {Makhlin}}, \bibinfo {author} {\bibfnamefont {J.}~\bibnamefont {Schriefl}}, \ and\ \bibinfo {author} {\bibfnamefont {G.}~\bibnamefont {Sch\"on}},\ }\href {\doibase 10.1103/PhysRevB.72.134519} {\bibfield  {journal} {\bibinfo  {journal} {Phys. Rev. B}\ }\textbf {\bibinfo {volume} {72}},\ \bibinfo {pages} {134519} (\bibinfo {year} {2005})}\BibitemShut {NoStop}%
\bibitem [{\citenamefont {Warren}\ \emph {et~al.}(2018)\citenamefont {Warren}, \citenamefont {Shahriar}, \citenamefont {Tripathi},\ and\ \citenamefont {Pati}}]{ramsey2}%
  \BibitemOpen
  \bibfield  {author} {\bibinfo {author} {\bibfnamefont {Z.}~\bibnamefont {Warren}}, \bibinfo {author} {\bibfnamefont {M.~S.}\ \bibnamefont {Shahriar}}, \bibinfo {author} {\bibfnamefont {R.}~\bibnamefont {Tripathi}}, \ and\ \bibinfo {author} {\bibfnamefont {G.~S.}\ \bibnamefont {Pati}},\ }\href {\doibase 10.1063/1.5008402} {\bibfield  {journal} {\bibinfo  {journal} {Journal of Applied Physics}\ }\textbf {\bibinfo {volume} {123}},\ \bibinfo {pages} {053101} (\bibinfo {year} {2018})},\ \Eprint {http://arxiv.org/abs/https://pubs.aip.org/aip/jap/article-pdf/doi/10.1063/1.5008402/14843245/053101\_1\_online.pdf} {https://pubs.aip.org/aip/jap/article-pdf/doi/10.1063/1.5008402/14843245/053101\_1\_online.pdf} \BibitemShut {NoStop}%
\bibitem [{\citenamefont {Yun}\ \emph {et~al.}(2012)\citenamefont {Yun}, \citenamefont {Zhang}, \citenamefont {Liu}, \citenamefont {Deng}, \citenamefont {You},\ and\ \citenamefont {Gu}}]{ramsey3}%
  \BibitemOpen
  \bibfield  {author} {\bibinfo {author} {\bibfnamefont {P.}~\bibnamefont {Yun}}, \bibinfo {author} {\bibfnamefont {Y.}~\bibnamefont {Zhang}}, \bibinfo {author} {\bibfnamefont {G.}~\bibnamefont {Liu}}, \bibinfo {author} {\bibfnamefont {W.}~\bibnamefont {Deng}}, \bibinfo {author} {\bibfnamefont {L.}~\bibnamefont {You}}, \ and\ \bibinfo {author} {\bibfnamefont {S.}~\bibnamefont {Gu}},\ }\href {\doibase 10.1209/0295-5075/97/63004} {\bibfield  {journal} {\bibinfo  {journal} {Europhysics Letters}\ }\textbf {\bibinfo {volume} {97}},\ \bibinfo {pages} {63004} (\bibinfo {year} {2012})}\BibitemShut {NoStop}%
\bibitem [{\citenamefont {{Haase}}\ \emph {et~al.}(2016)\citenamefont {{Haase}}, \citenamefont {{Smirne}}, \citenamefont {{Huelga}}, \citenamefont {{Ko{\l}odynski}},\ and\ \citenamefont {{Demkowicz-Dobrzanski}}}]{2016QMQM....5....2H}%
  \BibitemOpen
  \bibfield  {author} {\bibinfo {author} {\bibfnamefont {J.~F.}\ \bibnamefont {{Haase}}}, \bibinfo {author} {\bibfnamefont {A.}~\bibnamefont {{Smirne}}}, \bibinfo {author} {\bibfnamefont {S.~F.}\ \bibnamefont {{Huelga}}}, \bibinfo {author} {\bibfnamefont {J.}~\bibnamefont {{Ko{\l}odynski}}}, \ and\ \bibinfo {author} {\bibfnamefont {R.}~\bibnamefont {{Demkowicz-Dobrzanski}}},\ }\href {\doibase 10.1515/qmetro-2018-0002} {\bibfield  {journal} {\bibinfo  {journal} {Quantum Measurements and Quantum Metrology}\ }\textbf {\bibinfo {volume} {5}},\ \bibinfo {eid} {2} (\bibinfo {year} {2016})},\ \Eprint {http://arxiv.org/abs/1807.11882} {arXiv:1807.11882 [quant-ph]} \BibitemShut {NoStop}%
\bibitem [{\citenamefont {Smirne}\ \emph {et~al.}(2016{\natexlab{b}})\citenamefont {Smirne}, \citenamefont {Ko\l{}ody\ifmmode~\acute{n}\else \'{n}\fi{}ski}, \citenamefont {Huelga},\ and\ \citenamefont {Demkowicz-Dobrza\ifmmode~\acute{n}\else \'{n}\fi{}ski}}]{PhysRevLett.116.120801}%
  \BibitemOpen
  \bibfield  {author} {\bibinfo {author} {\bibfnamefont {A.}~\bibnamefont {Smirne}}, \bibinfo {author} {\bibfnamefont {J.}~\bibnamefont {Ko\l{}ody\ifmmode~\acute{n}\else \'{n}\fi{}ski}}, \bibinfo {author} {\bibfnamefont {S.~F.}\ \bibnamefont {Huelga}}, \ and\ \bibinfo {author} {\bibfnamefont {R.}~\bibnamefont {Demkowicz-Dobrza\ifmmode~\acute{n}\else \'{n}\fi{}ski}},\ }\href {\doibase 10.1103/PhysRevLett.116.120801} {\bibfield  {journal} {\bibinfo  {journal} {Phys. Rev. Lett.}\ }\textbf {\bibinfo {volume} {116}},\ \bibinfo {pages} {120801} (\bibinfo {year} {2016}{\natexlab{b}})}\BibitemShut {NoStop}%
\bibitem [{\citenamefont {Rudinger}\ \emph {et~al.}(2021)\citenamefont {Rudinger}, \citenamefont {Hogle}, \citenamefont {Naik}, \citenamefont {Hashim}, \citenamefont {Lobser}, \citenamefont {Santiago}, \citenamefont {Grace}, \citenamefont {Nielsen}, \citenamefont {Proctor}, \citenamefont {Seritan}, \citenamefont {Clark}, \citenamefont {Blume-Kohout}, \citenamefont {Siddiqi},\ and\ \citenamefont {Young}}]{PRXQuantum.2.040338}%
  \BibitemOpen
  \bibfield  {author} {\bibinfo {author} {\bibfnamefont {K.}~\bibnamefont {Rudinger}}, \bibinfo {author} {\bibfnamefont {C.~W.}\ \bibnamefont {Hogle}}, \bibinfo {author} {\bibfnamefont {R.~K.}\ \bibnamefont {Naik}}, \bibinfo {author} {\bibfnamefont {A.}~\bibnamefont {Hashim}}, \bibinfo {author} {\bibfnamefont {D.}~\bibnamefont {Lobser}}, \bibinfo {author} {\bibfnamefont {D.~I.}\ \bibnamefont {Santiago}}, \bibinfo {author} {\bibfnamefont {M.~D.}\ \bibnamefont {Grace}}, \bibinfo {author} {\bibfnamefont {E.}~\bibnamefont {Nielsen}}, \bibinfo {author} {\bibfnamefont {T.}~\bibnamefont {Proctor}}, \bibinfo {author} {\bibfnamefont {S.}~\bibnamefont {Seritan}}, \bibinfo {author} {\bibfnamefont {S.~M.}\ \bibnamefont {Clark}}, \bibinfo {author} {\bibfnamefont {R.}~\bibnamefont {Blume-Kohout}}, \bibinfo {author} {\bibfnamefont {I.}~\bibnamefont {Siddiqi}}, \ and\ \bibinfo {author} {\bibfnamefont {K.~C.}\ \bibnamefont {Young}},\ }\href {\doibase 10.1103/PRXQuantum.2.040338} {\bibfield  {journal} {\bibinfo  {journal} {PRX
  Quantum}\ }\textbf {\bibinfo {volume} {2}},\ \bibinfo {pages} {040338} (\bibinfo {year} {2021})}\BibitemShut {NoStop}%
\bibitem [{\citenamefont {Endo}\ \emph {et~al.}(2018)\citenamefont {Endo}, \citenamefont {Benjamin},\ and\ \citenamefont {Li}}]{PhysRevX.8.031027}%
  \BibitemOpen
  \bibfield  {author} {\bibinfo {author} {\bibfnamefont {S.}~\bibnamefont {Endo}}, \bibinfo {author} {\bibfnamefont {S.~C.}\ \bibnamefont {Benjamin}}, \ and\ \bibinfo {author} {\bibfnamefont {Y.}~\bibnamefont {Li}},\ }\href {\doibase 10.1103/PhysRevX.8.031027} {\bibfield  {journal} {\bibinfo  {journal} {Phys. Rev. X}\ }\textbf {\bibinfo {volume} {8}},\ \bibinfo {pages} {031027} (\bibinfo {year} {2018})}\BibitemShut {NoStop}%
\bibitem [{\citenamefont {van~den Berg}\ \emph {et~al.}(2023)\citenamefont {van~den Berg}, \citenamefont {Minev}, \citenamefont {Kandala},\ and\ \citenamefont {Temme}}]{van_den_Berg_2023}%
  \BibitemOpen
  \bibfield  {author} {\bibinfo {author} {\bibfnamefont {E.}~\bibnamefont {van~den Berg}}, \bibinfo {author} {\bibfnamefont {Z.~K.}\ \bibnamefont {Minev}}, \bibinfo {author} {\bibfnamefont {A.}~\bibnamefont {Kandala}}, \ and\ \bibinfo {author} {\bibfnamefont {K.}~\bibnamefont {Temme}},\ }\href {\doibase 10.1038/s41567-023-02042-2} {\bibfield  {journal} {\bibinfo  {journal} {Nature Physics}\ }\textbf {\bibinfo {volume} {19}},\ \bibinfo {pages} {1116–1121} (\bibinfo {year} {2023})}\BibitemShut {NoStop}%
\bibitem [{\citenamefont {Viñas}(2025)}]{myrepo}%
  \BibitemOpen
  \bibfield  {author} {\bibinfo {author} {\bibfnamefont {P.}~\bibnamefont {Viñas}},\ }\href@noop {} {\enquote {\bibinfo {title} {Context-aware-gst},}\ }\bibinfo {howpublished} {\url{https://github.com/Pablovinas/Context-aware-GST}} (\bibinfo {year} {2025})\BibitemShut {NoStop}%
\end{thebibliography}%

\end{document}